\documentclass[whitelogo]{tudelft-report}

\usepackage{graphicx} 

\usepackage{biblatex}
\usepackage{changes}

\usepackage{listings}

\usepackage{csquotes}

\usepackage{tabulary}

\usepackage{hyperref}

\usepackage{newunicodechar}

\newunicodechar{′}{'}
\newunicodechar{∧}{\wedge}

\newcommand{\stitle}[1]{\vspace{1ex}\noindent\textbf{#1}}

\newcommand\ceil[1]{\lceil#1\rceil}

\definecolor{mygreen}{rgb}{0,0.6,0}
\definecolor{mygray}{rgb}{0.5,0.5,0.5}
\definecolor{mymauve}{rgb}{0.58,0,0.82}
\definecolor{mydarkblue}{rgb}{0, 0, 0.6}

\lstdefinelanguage{TIL}{
    morekeywords={
    type, streamlet, impl, interface, in, out, import, substitute, as, prefixed, namespace, arrange, sequence, reset, process, domain, wait, test, alias,
    =, ==, --, ', -,
    true, false, Forward, Reverse, Sync, Flatten, Desync, FlatDesync,
    positive, negative, positive_hold, negative_hold, low, high, act, assert,
    Null, Bits, Group, Union, Stream},
    sensitive=true, 
    morecomment=[l]{//}, 
    morecomment=[s]{///}{///}, 
    morestring=[b]", 
    morestring=[b]\#
} %

\begin{document}

\frontmatter


\title[tudelft-white]{A Toolchain for Streaming Dataflow Accelerator Designs for Big Data Analytics}
\subtitle[tudelft-black]{Defining an IR for Composable Typed Streaming Dataflow Designs}
\author[tudelft-white]{M.\ A.\ Reukers}
\affiliation{Technische Universiteit Delft}
\coverimage{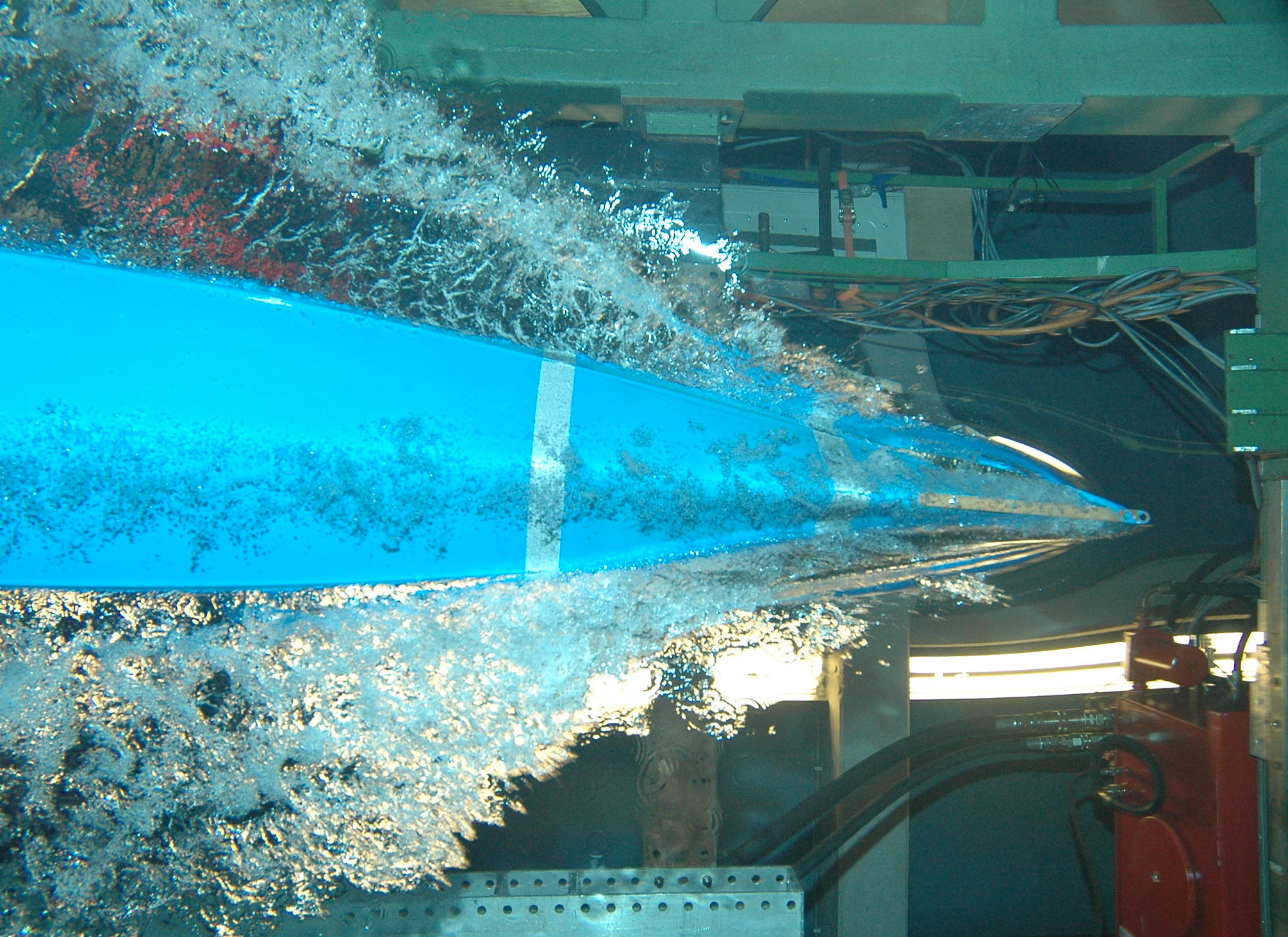}
\setpagecolor{tudelft-cyan}
\makecover[split]

\begin{titlepage}

\begin{center}


{\makeatletter
\largetitlestyle\fontsize{64}{94}\selectfont\@title
\makeatother}

{\makeatletter
\ifx\@subtitle\undefined\else
    \bigskip
   {\tudsffamily\fontsize{22}{32}\selectfont\@subtitle}    
\fi
\makeatother}

\bigskip
\bigskip

by

\bigskip
\bigskip

{\makeatletter
\largetitlestyle\fontsize{26}{26}\selectfont\@author
\makeatother}

\bigskip
\bigskip

to obtain the degree of Master of Science

at the Delft University of Technology,

to be defended publicly on Friday July 8, 2022 at 10:00 AM.

\vfill

\begin{tabular}{lll}
    Student number: & 4227220 \\
    Project duration: & \multicolumn{2}{l}{November 25, 2021 -- July 8, 2022} \\
    Thesis committee: & Prof.\ dr.\ H.\ P.\ Hofstee, & TU Delft, supervisor \\
        & Dr.\ ir.\ T.\ G.\ R.\ M.\ van Leuken, & TU Delft \\
        & Dr.\ ir.\ Z.\ Al-Ars, & TU Delft \\
        & Dr.\ ir.\ J.\ Peltenburg, & Voltron Data
\end{tabular}

\bigskip
\bigskip

\bigskip
\bigskip
An electronic version of this thesis is available at \url{http://repository.tudelft.nl/}.


\end{center}

\begin{tikzpicture}[remember picture, overlay]
    \node at (current page.south)[anchor=south,inner sep=0pt]{
        \includegraphics{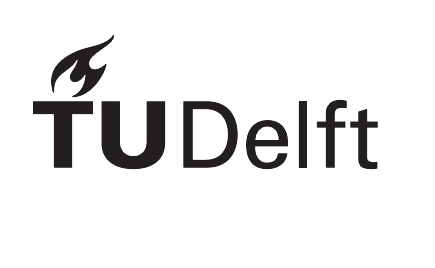}
    };
\end{tikzpicture}

\end{titlepage}


\chapter*{Abstract}
\setheader{Abstract}


Tydi is an open specification for streaming dataflow designs in digital circuits, allowing designers to express how composite and variable-length data structures are transferred over streams using clear, data-centric types. This provides a higher-level method for defining interfaces between components as opposed to existing bit- and byte-based interface specifications.
  
In this thesis, an open-source intermediate representation (IR) is introduced which allows for the declaration of Tydi's types. The IR enables creating and connecting components with Tydi Streams as interfaces, called Streamlets. It also lets backends for synthesis and simulation retain high-level information, such as documentation. Types and Streamlets can be easily reused between multiple projects, and Tydi’s streams and type hierarchy can be used to define interface contracts, which aid collaboration when designing a larger system.
  
The IR codifies the rules and properties established in the Tydi specification and serves to complement computation-oriented hardware design tools with a data-centric view on interfaces. To support different backends and targets, the IR is focused on expressing interfaces, and complements behavior described by hardware description languages and other IRs. Additionally, a testing syntax for the verification of inputs and outputs against abstract streams of data, and for substituting interdependent components, is presented which allows for the specification of behavior.
  
To demonstrate this IR, a grammar, parser, and query system have been created, and paired with a backend targeting VHDL.


\chapter*{Preface}
\setheader{Preface}


This thesis proved to be considerably challenging, and an excellent learning experience. It was very interesting to explore various technologies around hardware accelerator design, compilation, and language parsing, and learn some of the specifics of VHDL syntax while implementing the backend.

I would like to thank Prof. dr. Peter Hofstee for his continued support during my work on the thesis, and for pushing me to constantly evaluate the novelty and direction of my work. Likewise, I would like to thank Dr. Zaid Al-Ars, Yongding Tian and the entire ABS Group for the interesting meetings and support over this time. Finally, I would like to thank Johan, Matthijs Brobbel and Joost, for their ideas and the extensive discussions, for answering my questions around the Tydi specification, and for so thoroughly reviewing my paper before I submitted it to ICCAD.

Some of the contents of this thesis also appear in ``An Intermediate Representation for Composable Typed Streaming Dataflow Designs'', a paper which I submitted to ICCAD 2022. \cite{reukersintermediate2022}

\begin{flushright}
{\makeatletter\itshape
    \@author \\
    Delft, July 2022
\makeatother}
\end{flushright}

\tableofcontents

\mainmatter

\chapter{Introduction}\label{ch:introduction}


In order to transfer streaming data between components within digital circuits, designers have a choice to either design their own interfaces, or use general interface specifications such as Intel's Avalon-ST \cite{intelcorporationavalon2022} or Arm's AXI4-Stream \cite{armlimitedamba2010}. By using an interface specification, it is easier for other designers to connect components, as the signals and how they relate to data transfers are standardized. This can promote reuse, and is used by hardware design tools to provide IP (Intellectual Property) libraries and automate integration \cite{arnesenincreasing2010, jacomesurvey2001}.

The aforementioned specifications do not specify how data structures are represented, however, and as a result designers must still design, document and share these representations. Additionally, the IP integration tools are proprietary, reducing the simplicity of integrating such IPs outside of these specific tools. Addressing the first issue, Peltenburg et al. proposed Tydi (Typed dataflow interface) \cite{peltenburgtydi2020}, an open specification which allows designers to explicitly define the data which is being transferred by providing a type system for composite and variable-length data structures, in addition to defining how data elements are organized in transfers and the requirements on transfers. This thesis aims to address the second issue, by utilizing the Tydi specification as part of an IR (intermediate representation) for defining interfaces and connecting components.

The goal of the IR is not to serve as a complete hardware description language, but to provide a simple and robust way to declare Tydi's types, define interfaces and connect components which adhere to the Tydi specification, serving as part of a toolchain in order to integrate and reuse components within and across projects. To this end, the IR is not capable of directly implementing behavior, but should instead be combined with transaction-level verification to \textit{specify} intended behavior.

To demonstrate the potential of such an IR, and to explore potential approaches towards implementing a toolchain built around it, a prototype toolchain has been conceived and implemented. This consists of a query system, which tracks and computes information defined through the IR, a grammar called TIL (Tydi Intermediate Language) and parser, as a more portable, text-based way of representing designs in the IR, and finally a VHDL backend to emit designs defined in the IR.

\section{Problem Statement}\label{sec:problem_statement}



While much research is focused on developing and accelerating algorithms for streaming data in both hardware \cite{nowatzkistreamdataflow2017, plavecstream2010} and software \cite{isahsurvey2019}, many designs for low-level hardware still have to transfer streams over interfaces which are either custom or based on generic, bit- and/or byte-oriented specifications such as AXI4-Stream \cite{armlimitedamba2010} and Avalon-ST \cite{intelcorporationavalon2022}. As a result, higher-level information about data structures and how streams of data are organized over transfers must be devised and implemented by designers, and are not reflected by the declaration of the interface in a traditional HDL.

Some of this design effort can be alleviated through the use of high-level synthesis: tools such as Vivado HLS can be employed to leverage C, C++ or SystemC combined with IP-blocks using \textit{ap\_fifo} or AXI4-Stream to handle data streams \cite{advancedmicrodevicesinc.interfaces2022}, while synthesizing compilers such as Optimus \cite{hormatioptimus2008} have been developed in the past to leverage StreamIt \cite{thiesstreamit2002}, a language specifically for streaming applications. At the same time, many researchers are working on improved hardware description languages and IRs, such as Chisel \cite{bachrachchisel2012}, FIRRTL \cite{izraelevitzreusability2017} and LLHD \cite{schuikillhd2020}.

These are not suitable replacements for a higher-level interface specification, however: HLS tools either obfuscate the interfaces between low-level hardware and/or use proprietary IRs and tools to connect components, making reuse more difficult. While the HDLs and IRs mentioned are aimed at more general hardware designs, so still require custom interfaces for streaming data transfers.

As such, the aim of this thesis is to develop a free, open-source IR for defining high-level streaming dataflow interfaces mapped onto hardware and for connecting these interfaces. This would complement existing HDLs and IRs which describe behavior, and enable components designed in higher-level front-end languages for HLS to propagate more type information to the resulting interfaces.

Long-term, the Accelerated Big Data Systems group aims to develop a toolchain for streaming dataflow accelerator designs for big data analytics. The work done for this thesis is part of such a toolchain, providing a grammar and parser, query system for the IR, and a compiler to VHDL. At the same time, Yongding Tian has been working on a front-end language for his thesis and as another component of this toolchain, enabling designers to express behavior as well.




\section{Methodology}\label{sec:methodology}


As the aim of this thesis is not only to define an intermediate representation, but contribute to a toolchain, there was an increased focus on \textit{implementing} such tools. Essentially, by creating and iterating on a ``vertical slice'' of a partial toolchain, it is possible to evaluate the effectiveness of the IR and the feasibility of the proposed toolchain overall.

An ideal vertical slice would have the following properties and components:

\begin{enumerate}
    \item A (partial) specification for the intermediate representation; i.e., what (additional) concepts should it be able to express.
    \item A means to integrate a compiler, using one or both of:
    \begin{enumerate}
        \item A grammar and a parser, taking a text-based representation and allowing a subsequent backend to interpret the results.
        \item A \textit{query system} not unlike the one employed by the \textit{Rust} compiler \cite{rustcompilerteamqueries2021}, which would allow a backend to perform queries to retrieve and/or compute information from a definition in the IR, reducing the need for separate optimizing passes. (As such, the information stored in the system does not necessarily need to be the result of a (single) parser, but can be input programmatically.)
    \end{enumerate}
    \item A backend for emitting designs defined in the IR as a conventional hardware description language suitable for simulation and synthesis. Due to familiarity and broad support, this language will be VHDL(-93). The backend should be capable of as many of the following as possible:
    \begin{enumerate}
        \item Emit Streamlets with structural implementations; i.e., Streamlets which contain and connect other Streamlets.
        \item Link behavioral implementations.
        \item Emit a testbench based on high-level assertions defined in the IR.
    \end{enumerate}
\end{enumerate}

Based on interim progress towards these features and results of finished (prototype) implementations, the next step would be to:
\begin{itemize}
    \item Continue working on and/or expanding specific features. (The initial implementation is successful and/or promising, or the feature requires further evaluation.)
    \item Revise goals and/or the IR specification. (The feature is not feasible, or an alternative appears more effective.)
    \item Omit them and instead recommend their implementation as future work. (The feature is feasible and promising, but cannot be implemented satisfactorily within the time frame of this thesis.)
\end{itemize}

As an example, if a concept expressed in the IR is impossible or very difficult to express in VHDL (or any target HDL), the solution would be to either revise the IR to include more information (i.e., the concept is possible to express, but requires more/different input), or to summarize these findings in this thesis and remove it or recommend it as future work.

The results of this methodology are described in Chapter \ref{ch:evaluation}.

\section{Contributions}

The contributions of this thesis can be summarized as follows, with references to the relating chapters and sections:

\begin{itemize}
    \item \textbf{An intermediate representation for composition and linking behavior} --- This thesis proposes an intermediate representation for composing streaming dataflow designs, using the Tydi specification. It features means to define Streams and the data they carry, to create interfaces with clock and reset domains tied to specific ports, create components (Streamlets) and connect them, and link to implementations of behavior. Sections \ref{sec:types_interfaces} and \ref{sec:component_implementation}
    \item \textbf{Recommendations for further improvements to the intermediate representation} --- After evaluation of the intermediate representation, it was determined certain language and compiler-oriented features would improve the intermediate representation's ability to describe designs and aid backends in emitting them to a target language. Specifically, the addition of type parameters (\ref{subsec:type_parameters}), the inclusion of code generation constructs (\ref{subsec:generation}), the addition of limited, behavioral intrinsic functions (\ref{subsec:intrinsics}), the inclusion of annotations for backends (\ref{subsec:annotations}) and changes to the existing representation to improve the readability of the output (\ref{sec:readability}). This thesis also discusses potential difficulties when implementing them, as features need to translate well to many potential target languages.
    \item \textbf{Proposals for (and partial, preliminary implementation of) a high-level testing framework} --- As the intermediate representation primarily exposes typed interfaces, tests can be performed as high-level assertions against transfers of typed data (Sections \ref{sec:assertions} and \ref{sec:spec_poc}), while the inclusion of substitutions helps when testing more complex or incomplete dependencies (\ref{subsec:substitution}). This thesis also discusses the potential problems (and some solutions) when setting up (resetting) subject components in Section \ref{sec:setup}.
    \item \textbf{A complete toolchain as proof of concept, from intermediate representation to target language} --- As part of the work on this thesis, a query system for the intermediate representation (\ref{sec:query_system}), a VHDL backend (\ref{sec:vhdl_backend}), and a text-based grammar (Tydi Intermediate Language, TIL) and parser (\ref{sec:grammar_parser}) were implemented. These are provided in a free, open-source repository along with a simple example application which allows a user to compile a TIL file to VHDL, described in Section \ref{sec:impl_example}.
    \item \textbf{Validating the Tydi interface specification} --- By implementing the Tydi interface specification programmatically, a number of unaddressed and contradictory situations were brought to light, for which the specification should be amended. Section \ref{sec:tydi_spec}
    \item \textbf{Evaluation of the intermediate representation's ability to describe interfaces and connections} --- The intermediate representation was evaluated for its ability to produce human-readable and traceable output in Section \ref{sec:readability}. The evaluation of its effectiveness in representing streaming interfaces and connections between them is described in Section \ref{sec:effort}, using the existing AXI4 and AXI4-Stream standards as a reference point.
\end{itemize}

\chapter{Background}\label{ch:background}

\section{Stream Processing}\label{sec:background_stream_processing}


\subsection{Data Streams}

Stream processing refers to means of processing data which is produced or consumed incrementally, rather than a set of data which is known and stored ahead of time on the system. The order of the data and rate at which it arrives cannot necessarily be controlled, and the number of elements is potentially unbounded, requiring the system to process elements as they arrive and before the next element does.

Examples of practically unbounded data streams would be analyzing real-time weather events or human behavior, but even limited sets of data can be treated as streams when timing is critical to performance, such as when encrypting and decrypting data to and from a storage medium. As a result, stream processing has been actively researched for over 20 years, with software paradigms and hardware acceleration being worked on in parallel in attempts to improve performance and establish effective data and execution models \cite{fragkoulissurvey2020, isahsurvey2019}.

\stitle{Software models} In software, stream processing has been approached in many different ways to various ends. More recent examples of stream processing include Kafka Streams, which is a stream processing library of Apache Kafka \cite{saxapache2018}, and Spark Streaming \cite{zahariadiscretized2012, apachesoftwarefoundationspark2022} (now Structured Streaming \cite{apachesoftwarefoundationstructured2022}). Both aim to provide a useful subset of high-level functions for processing data streams, mapped onto their existing domains. There are also more wholesale approaches, such as StreamIt \cite{thiesstreamit2002}, which is a language specifically designed for streaming applications. 

\stitle{Hardware Acceleration} In hardware acceleration, the constraints of stream processing are less uncommon; hardware designs are already heavily constrained by timing, and do not necessarily have a notion of state. More specifically addressing recent needs of stream processing, there are a number of frameworks such as Fleet \cite{thomasfleet2020} and S2FA (Spark-to-FPGA-Accelerator) \cite{yus2fa2018} which use FPGAs to accelerate streaming operations which conventional processors may struggle with.





\subsection{Interface Specifications}\label{subsec:background_interfaces}




When designing digital circuits for stream processing hardware accelerators, internal communication will likewise take the form of unbounded streams of messages between sources and sinks. There exists a number of interface specifications to ensure these streams of data are correctly transferred and represented: For example, ARM devised the AXI4-Stream protocol \cite{armlimitedamba2010}, and Intel defines the Avalon-ST interface specification \cite{intelcorporationavalon2022}.

Both protocols are able to optionally organize sequences of data into packets over transfers; AXI4-Stream uses the \textit{last} signal to indicate that a transfer is the last in a sequence making up a packet, while Avalon-ST uses the \textit{startofpacket} and \textit{endofpacket} signals to do the same. Likewise, both interface specifications incorporate means to indicate whether data is being transferred from a source (using a \textit{valid} signal) and can be transferred to a sink (using a \textit{ready} signal), and allows for valid transfers to be indicated as entirely or partially empty. These properties ensure that sequences of data can be transferred over time and without needing to account for the rate at which individual elements arrive, nor for the total size of a sequence.

Additionally, by using a standard and well-defined interface, designers can not only ensure that transfers are consistent within a project, but can share components between different projects and even across organizations. For example, this is employed by AMD/Xilinx \cite{advancedmicrodevicesinc.intellectual2022} and Intel \cite{intelcorporationintroduction2021} to establish libraries of IP cores for designs implemented on their respective FPGAs. Finally, when representing streams of data in high-level synthesis, such standards ensure data can be consumed from or produced for such IP cores; e.g., Vivado HLS uses AXI4-Stream and \textit{ap\_fifo} for this purpose \cite{advancedmicrodevicesinc.interfaces2022}.





\section{Tydi}\label{sec:background_tydi}

The Tydi specification and type system was introduced by Peltenburg et al. \cite{peltenburgtydi2020} and defines an abstract way to describe data structures transferred over hardware streams. Tydi promises to reduce the design effort of creating hardware for streaming dataflow computing, by providing clear and intuitive ways to map composite, variable-length data structures onto a hardware streaming protocol. An open-source repository and documentation \cite{vanstratenintroduction2021} expanding on the specification and providing example code for mapping Tydi's streams onto VHDL component ports is now available.

The specification defines five \textit{logical types}: the stream-manipulating Stream type, and the element-manipulating Null, Bits, Group and Union types.

\subsection{Element-manipulating Types}

Element-manipulating types are how Tydi represents kinds of data; that is to say, these types represent arbitrary data as well as specific data structures.

\begin{figure}[ht]
  \centering
\includegraphics[width=0.95\linewidth]{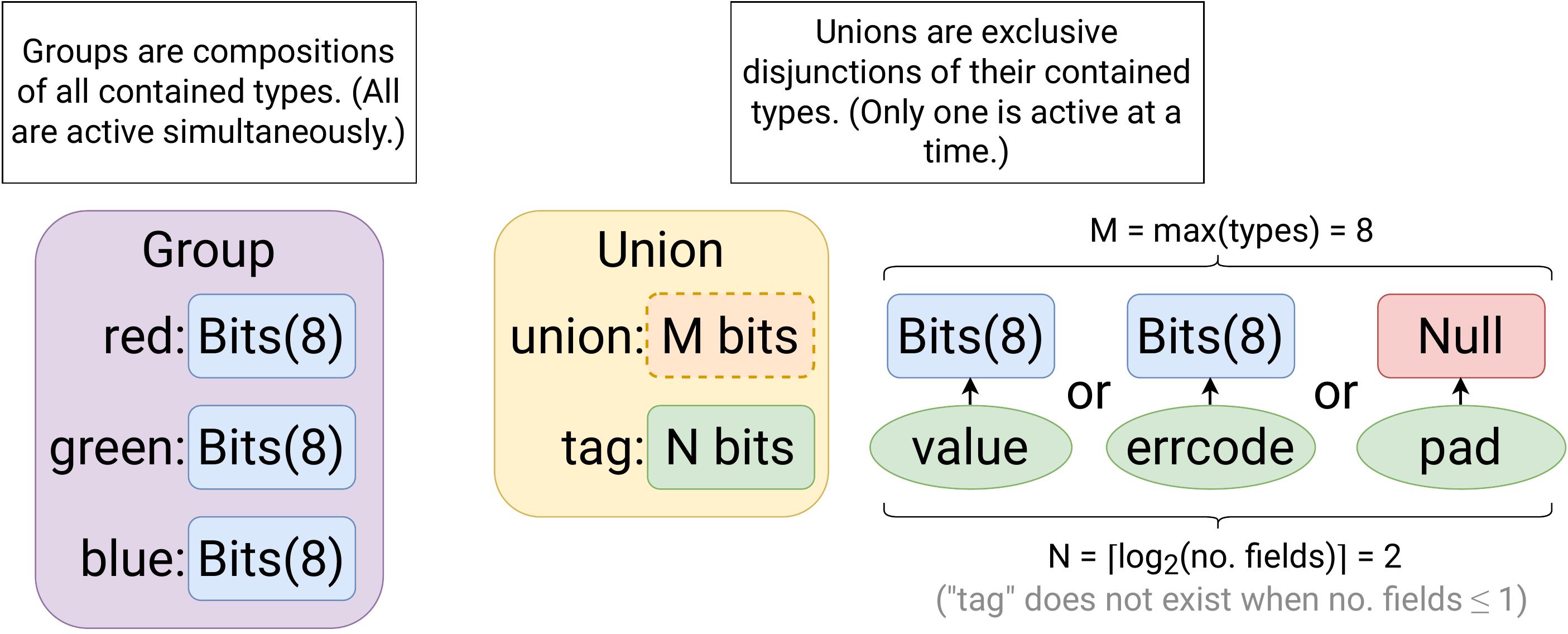}
  \caption{An illustration of the difference between Group and Union types.}
  \label{fig:tydi_union_vs_group}
\end{figure}

\begin{itemize}
    \item The \textit{Null} type is for transfers of one-valued data, its only valid value is null. This can be used to indicate (part of) a transfer being valid and active, but no data being transferred.
    \item The \textit{Bits(N)} type represents a data signal of N bits. It is used to transfer arbitrary data.
    \item The \textit{Group} type contains named \textit{fields}, which are themselves any logical type. Groups are compositions, and represent all fields being active simultaneously.
    \item The \textit{Union} type is comparable to the Group type, in that it contains named \textit{fields} of logical types. Unlike Groups, Unions are exclusive disjunctions; only one field may be active at a time. Figure \ref{fig:tydi_union_vs_group} further illustrates the difference.
\end{itemize}

Field \textit{Name}s of Groups and Unions consist of (ASCII) letters, digits and/or underscores \cite{vanstratenlogical2021}, and may not contain two or more consecutive underscores. Names within a Group or Union must be unique, cannot start or end with an underscore, and cannot start with a digit. The latter constraints ensure broad compatiblity with potential target HDLs, while consecutive underscores are reserved for use in \textit{Path Name}s, discussed in Section \ref{subsec:physical_streams}.

The element-manipulating types alone can represent many data structures, for example:
\begin{itemize}
    \item \textit{Bits(N)} can be used to transfer primitive data types such as numbers, booleans and characters.
    \item A \textit{Union} of \textit{Null} and another type can indicate optional data.
    \item \textit{Group}s can be used to directly represent records of data.
\end{itemize}

However, as Streams are also logical types, Groups, Unions and Streams themselves can carry further nested logical Streams, each with their own data and properties.

\subsection{Streams}

The Stream type adds a further layer of flexibility to these element-manipulating types. It does not only represent the physical stream and signals carrying the element-manipulating types, but also features properties for further describing data structures. Notably, Streams have a \textit{dimensionality} property, which indicates whether the data being transferred is part of a sequence. In hardware, this is translated to a ``last'' signal; when this signal is driven high, it indicates that the data being transferred is the last element in a sequence, and a Stream with a higher dimensionality will have multiple last bits, to indicate nested sequences.

\begin{figure}[ht]
  \centering
\includegraphics[width=0.95\linewidth]{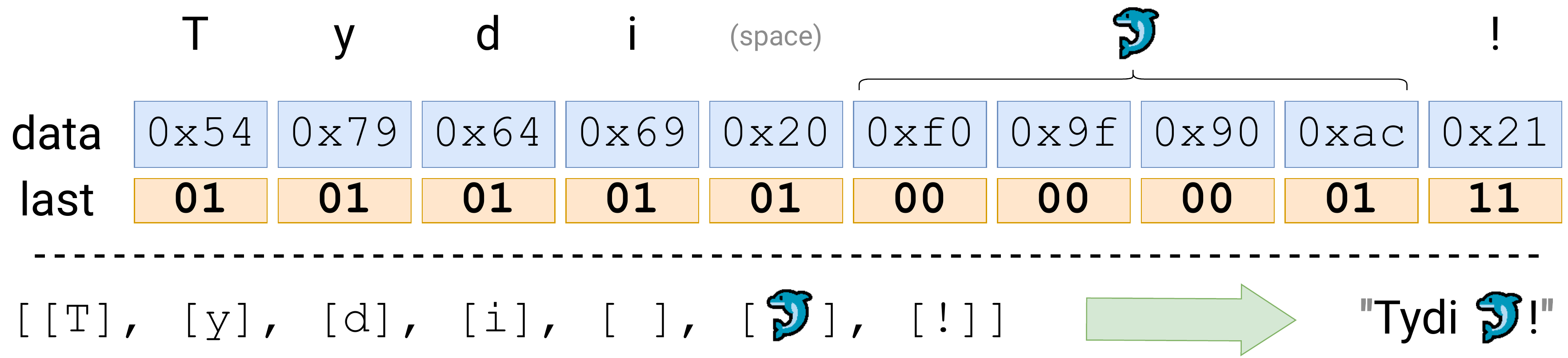}
  \caption{Using multiple last bits to transfer a UTF-8 encoded string as bytes, where the inner dimension is used to easily distinguish groups of bytes making up single characters.}
  \label{fig:tydiutf8}
\end{figure}

A \textit{last} signal is typically used to reflect sequences or other kinds of variable-length data. Both AXI4-Stream and Avalon-ST lack Tydi's ability to assign multiple \textit{last} bits to a transfer or element, however. This gives Tydi interfaces more flexibility in naturally reflecting different data types, or combining multiple variable-length data structures. Figure \ref{fig:tydiutf8} illustrates how this can be used to better reflect a UTF-8 encoded string transferred as bytes; the inner dimension is used to represent UTF-8 characters, which can be between 1 and 4 bytes long, and the outer dimension is used to represent the string as a whole. Of course, UTF-8 itself already encodes whether a byte is part of a group making up a character, but this simplifies processing downstream, and alleviates the need for similar encoding on other data types. For instance, when representing a video with an arbitrary resolution, three dimensions can be used to indicate the end of a row, the end of a frame, and the end of the video overall respectively.

\begin{figure}[ht]
  \centering
\includegraphics[width=0.95\linewidth]{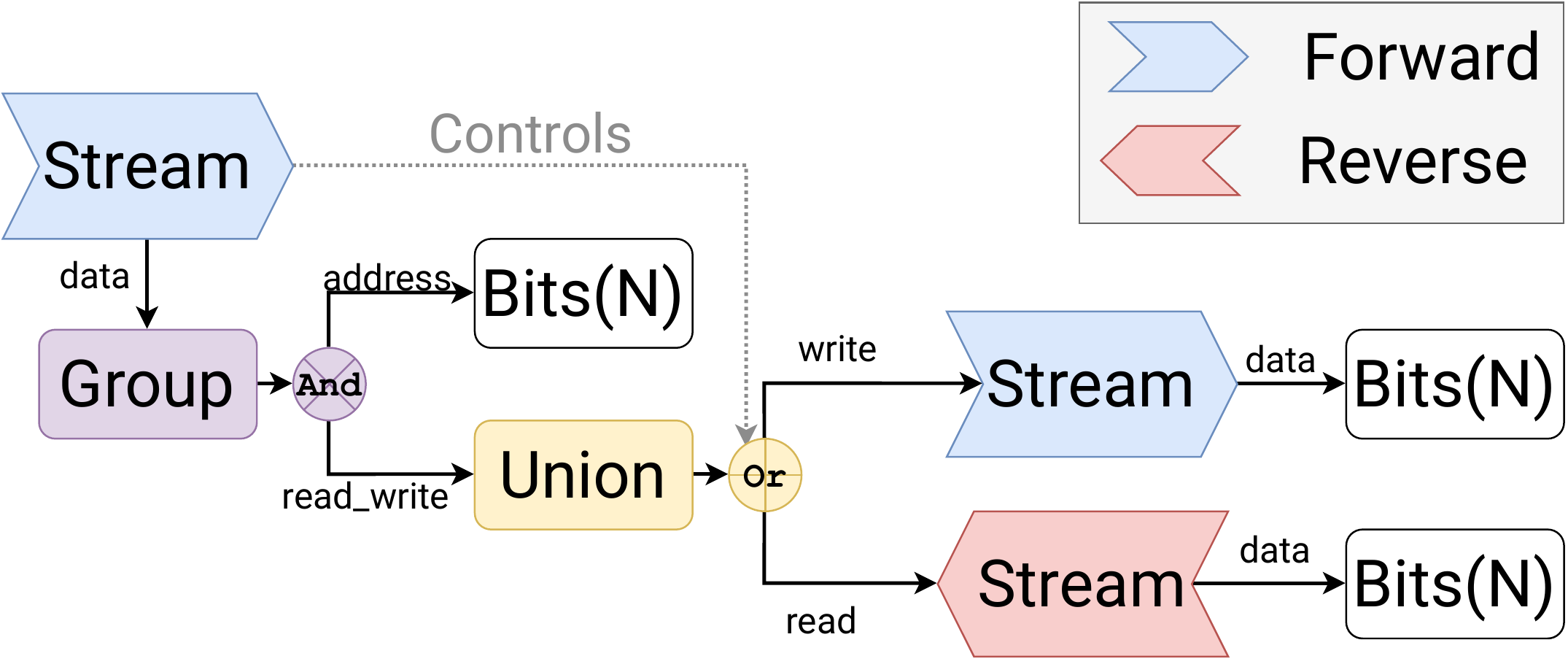}
  \caption{A Union of Streams with opposite directions allows a parent Stream to control whether data should be read or written.}
  \label{fig:tydirw}
\end{figure}

Another noteworthy property is \textit{direction}, which indicates whether a Stream flows in the same direction as its parent, or in reverse. This allows designers to express that certain Streams have a relation: As an example, a Group can have both a ``Forward'' and ``Reverse'' Stream to indicate that interdependent data is transferred between the sink and source, such as a memory address and the data retrieved from that address. Giving a parent and child Stream different directions can also be used to indicate that one Stream (direction) directly controls another; expanding on the previous example, consider a Stream which can be used to read and write to memory, but which prevents reading and writing simultaneously. One way to implement this is as follows: The parent Stream carries a Group with the fields ``address'' and ``read\_write'', ``address'' is simply a Bits(N) type, but ``read\_write'' is a Union carrying both a Forward ``write'' and Reverse ``read'' field. As illustrated by figure \ref{fig:tydirw}, this allows the parent Stream to control whether read or write is active by setting the Union's \textit{tag}.

In addition to dimensionality and direction, Streams have properties for describing how transfers should be organized in space and time, the specifics of their implementation will be described in the next section:

\begin{itemize}
    \item \textit{Throughput} is a positive, rational number indicating how many elements are expected to be transferred per individual handshake, or relative to its parent Stream. The number of element lanes is a Stream's \textit{throughput} multiplied by that of all parent Streams, rounded up to a natural number.
    \item \textit{Synchronicity} refers to how strong the relation between a child Stream and its parents are with regards to dimensional information. ``Sync'' indicates that for each element transferred on the parent, the child has a matching transfer, while ``Desync'' indicates that the child may have transfers of arbitrary size. Both options also have a ``Flat'' variant, which results in redundant last signals on the child being omitted.
    \item \textit{Complexity} is a number which encodes guarantees on how elements of a sequence are transferred. In brief, lower complexities place more constraints on \textit{source} streams, such as by requiring that transfers of sequences occur over consecutive cycles.
    \item A \textit{keep} property can be used to ensure a \textit{logical} Stream is synthesized into physical signals, as nested Streams may otherwise be combined into a single physical stream.
\end{itemize}

Finally, in the event these properties are insufficient for a use-case, Streams can also have a \textit{user} signal carrying an element-manipulating type. This user signal can be used to provide additional information independent from transfers or clock cycles.

\subsection{Physical Streams}\label{subsec:physical_streams}

Many of the properties described in the previous section have no impact on the kinds of data being transferred, but instead affect \textit{how} it will be transferred. These changes are reflected in the \textit{physical streams} \cite{vanstratenphysical2021} resulting from a logical Stream definition. A physical stream ``canonically'' consists of some variation on the following signals:

\begin{itemize}
    \item \textit{ready}: 1 bit, when driven high, this indicates that the sink device is prepared to accept transfers.
    \item \textit{valid}: 1 bit, when driven high, this indicates that the source device is transferring valid data.
    \item \textit{data}: $E \times N$ bits, carries an element-manipulating logical type (of size $0 \leq E < \infty$ bits), and may be composed of one or more data lanes ($1 \leq N < \infty$).
    \item \textit{last}: $D \lor D \times N$ bits, indicates whether a particular transfer, or element, represents the end of one or more sequences. If complexity $C \geq 8$, there is a last signal/slice per element lane, otherwise, this signal refers to the entire transfer. Its size is equal to dimensionality $D$.
    \item \textit{endi}: $0 \lor \ceil{{log}_2(N)}$ bits, the ``end index'': Only exists when the number of element lanes $N > 1$, and complexity $C \geq 5$ or dimensionality $D \geq 1$. Indicates the end of all \textit{active} element lanes in a valid transfer.
    \item \textit{stai}: $0 \lor \ceil{{log}_2(N)}$ bits, the ``start index'': Only exists when the number of element lanes $N > 1$, and complexity $C \geq 6$. Indicates the start of all \textit{active} element lanes in a valid transfer. Can be combined with \textit{endi} to form a range of active elements, but cannot be used to mark an entire transfer as inactive. (The value of \textit{stai} must be smaller than or equal to that of \textit{endi}.)
    \item \textit{strb}: $0 \lor 1 \lor N$ bits, the ``strobe'' signal: When dimensionality $D \geq 1$, can be used to mark all of a transfer's elements inactive, as a single bit. When complexity $C \geq 7$, all element lanes $N$ have an individual \textit{strb} bit, allowing for individual element lanes to be marked inactive, rather than the ranges supported by the start- and end indices.
    \item \textit{user}: $U$ bits, this signal carries the element-manipulating type defined in the Stream's \textit{user} property, its properties are entirely user-defined.
\end{itemize}

Any signals sized 0 are omitted entirely, and Tydi allows for the \textit{ready} and \textit{valid} signals to be omitted when the physical stream is always ready or always valid, respectively. Physical streams are not necessarily directly equivalent to logical Streams; this is a result of Tydi making Streams themselves logical types, allowing for nested Streams in a Stream's data property. As the \textit{data} signal itself cannot represent a Stream, such logical Streams will be split into multiple physical streams.

The exact procedures for converting logical types into physical streams are defined in the Tydi specification as the \textit{split}, \textit{fields} and \textit{synthesis} functions. In brief:
\begin{itemize}
    \item Logical Streams are \textit{split} into a list of named physical streams. Names of physical streams are based on potential field \textit{Name}s of Groups and Unions, which are concatenated hierarchically as \textit{Path Name}s; Path Names are emitted as Names joined by two underscores. As a root Stream or directly nested Stream is not part of a field, split Stream names may be empty. When two Streams are directly nested, they may be \textit{flattened}, combining their properties into a single Stream (by for example multiplying their \textit{throughput}s and determining an absolute \textit{direction}).
    \item Non-Stream logical types are converted into \textit{fields}, which are lists of named bitfields, based on their size. These fields eventually make up the \textit{data} signal. As before, Names of Group and Union fields are used to determine these names, and are concatenated into Path Names. Likewise, a field name may be empty if it is directly part of a Stream's data or user property.
    \item The \textit{synthesis} function converts all \textit{split} Streams into a list of named ${PhysicalStream}(E, N, D, C, U)$ definitions:\begin{itemize}
        \item $E$ is the \textit{element content}, derived from the \textit{fields} function on a split Stream's \textit{data} property.
        \item $N$ is the number of \textit{element lanes}, which is equal to the split Stream's \textit{throughput} $\ceil{t}$.
        \item $D$ is the physical stream's \textit{dimensionality}, equal to that of the split Stream.
        \item $C$ is the physical stream's \textit{complexity}, equal to that of the split Stream.
        \item $U$ is the \textit{user content}, derived from the \textit{fields} function on a split Stream's \textit{user} property.
    \end{itemize}
\end{itemize}

The \textit{synthesis} function also accounts for defining separate, user-defined signals which flow in parallel to a physical stream (in addition to the \textit{user} signal), but this is not currently relevant to the use-case of specifically generating Tydi-compliant interfaces. It is also worth noting that the output of the \textit{split} function is not discarded after \textit{synthesis}: The \textit{direction} property is not part of physical streams, so must be retrieved from their respective Stream definition.

The Tydi specification also permits \textit{alternative representations} of physical streams, bundling element types into aggregate/record types in the target language. For example, in VHDL, rather than simply using a bit vector to represent a Group's data over multiple lanes, it is possible to instead use a record type with field names corresponding to those of the Group's, and then using an array of this record type to represent the data signal overall. However, it recommends that any ``outer'' interfaces still use the ``canonical'' representation described at the start of this section, to ensure interoperability between potential IP blocks.

The \textit{number of element lanes} $N$ applies to the data signal, multiplying the total bit width defined by \textit{element content} $E$. It does not apply to the \textit{user} signal.

The \textit{complexity} $C$ affects how elements and sequences are organized over element lanes and consecutive lanes. Overall, a lower complexity imposes more restrictions on a source, in the inverse, this results in a higher complexity making it more difficult to implement a sink. As an example, a complexity of $\leq2$ requires that elements of an inner sequence are transferred over consecutive cycles by a source, while higher complexities allow it to stall independently from the sink. The specification currently defines 8 levels of complexity \cite{vanstratenphysical2021}. Table \ref{tab:complexity_levels} illustrates the cumulative changes between complexity levels.

\begin{table}[ht]
\includegraphics[width=\linewidth]{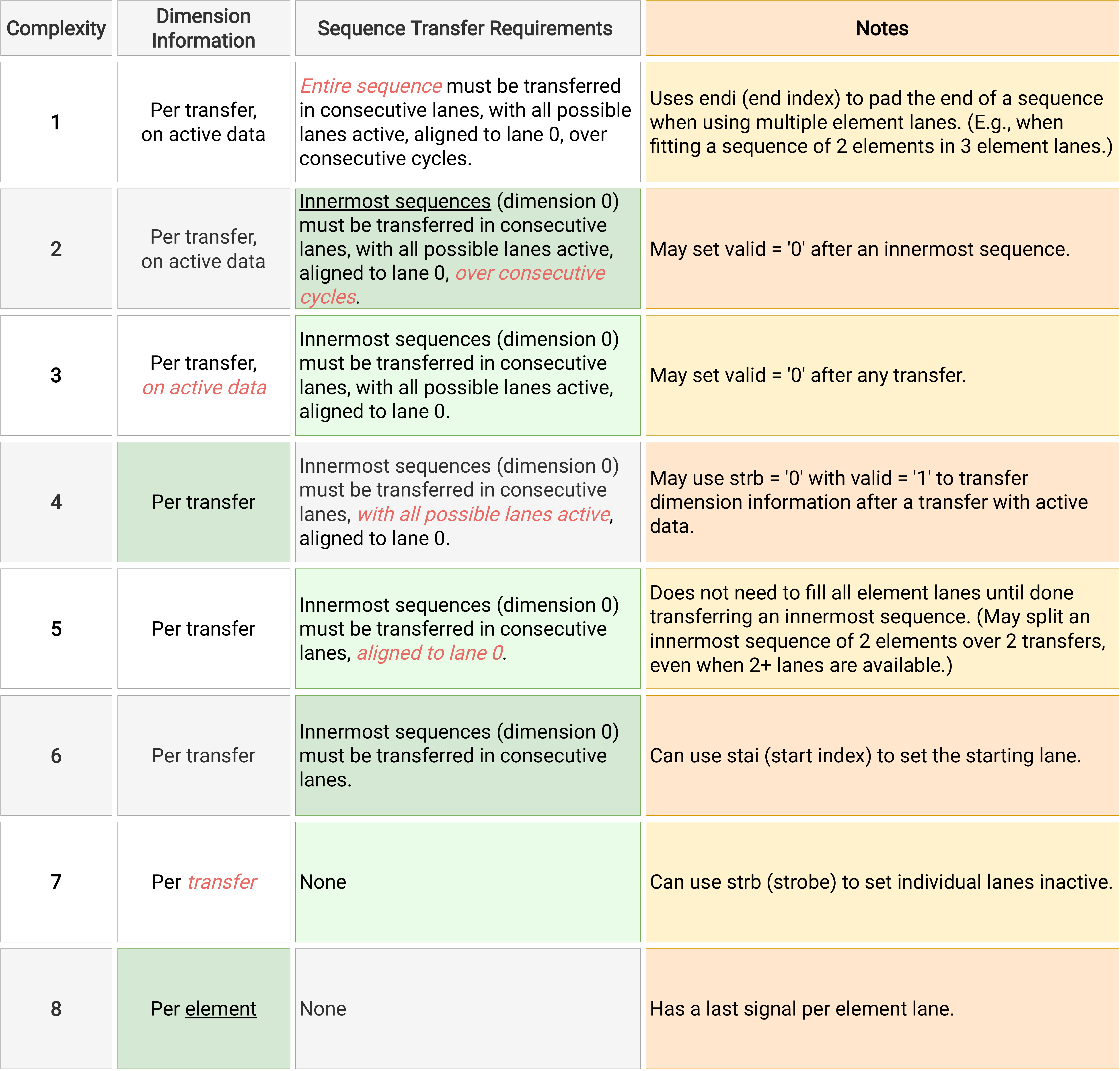}
    \caption{The changes between Tydi's \textit{complexity} levels. For clarity, constraints which will be omitted next complexity level are \textit{italicized} and marked red, and any potential replacement constraint or property is \underline{underlined}.}
    \label{tab:complexity_levels}
\end{table}

\begin{figure}[ht]
  \centering
\includegraphics[width=0.9\linewidth]{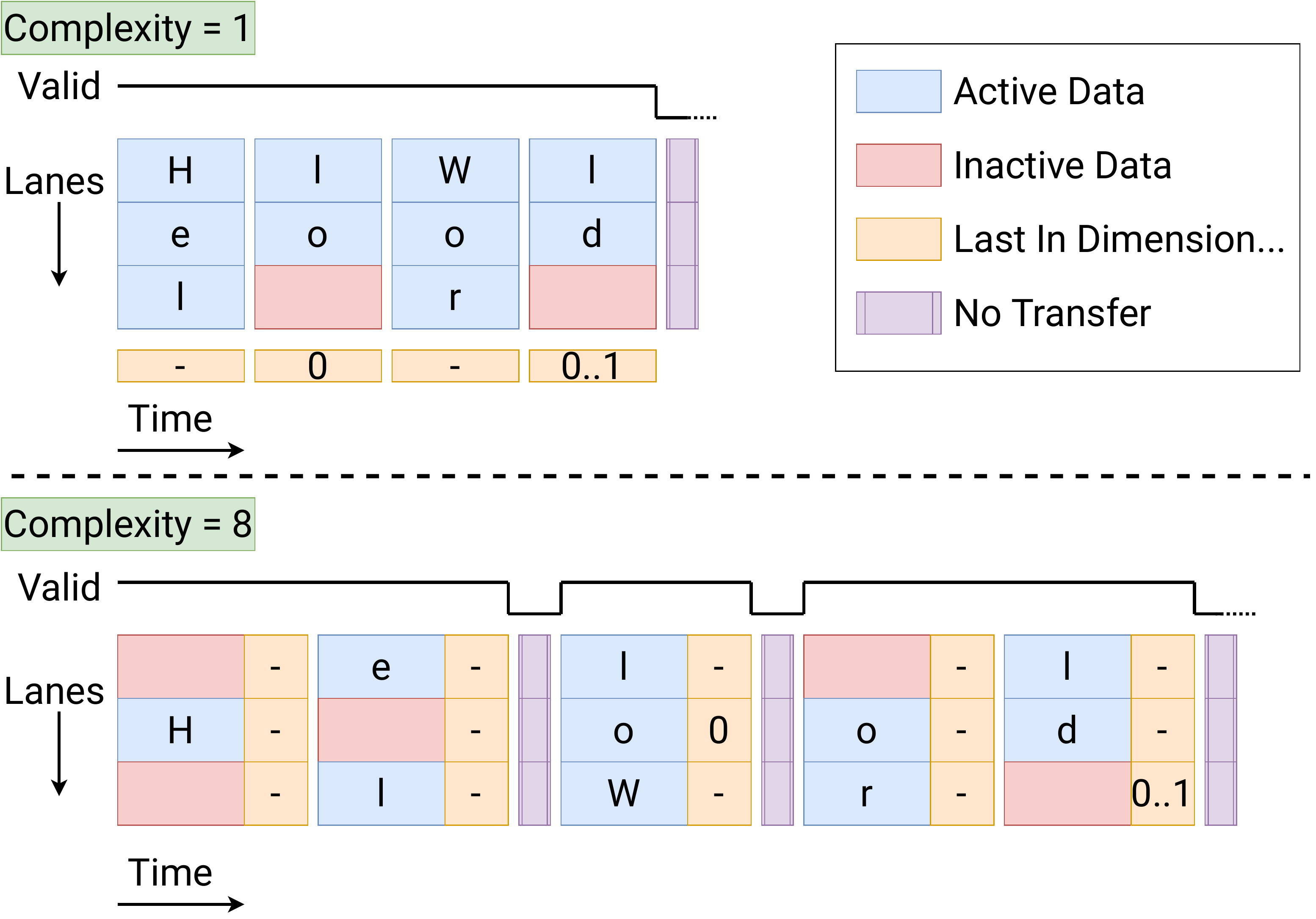}
  \caption{Streams determine which signals are used and valid to organize elements in transfers, and how transfers are organized over time.}
  \label{fig:streamproperties}
\end{figure}

Figure \ref{fig:streamproperties} illustrates how a higher \textit{complexity} allows for transfers to be organized differently. When transferring \texttt{[[H, e, l, l, o], [W, o, r, l, d]]}, at \textit{complexity} $=1$ all elements must be aligned to the first lane, \textit{last} data is asserted per transfer, and all data must be transferred over consecutive cycles and lanes. At \textit{complexity} $=8$, there are no requirements for how elements are aligned, transfers may be postponed (asserting \textit{valid} low), and \textit{last} data is asserted per lane, and may be postponed (using an inactive lane to assert \textit{last} for a previous lane or transfer).

The \textit{synchronicity} and \textit{keep} properties of the original logical Stream type are not directly reflected by the physical stream, but do affect how they are synthesized and how transfers are expected to behave. As mentioned in the previous section, \textit{synchronicity} $s$ indicates whether the transfers of a child Stream are constrained by its parent Stream, and only applies if the child Stream's \textit{dimensionality} $d>0$.
\begin{itemize}
    \item If $s = {Sync}$, for each element in the parent Stream, one sequence must be transferred on the child Stream. E.g., both the parent and child Stream have \textit{dimensionality} $d=1$, and the parent Stream features a \textit{Group(a: Bits(1), b: Stream(data: Bits(1), ...)}, where \textit{b} is this child Stream. To transfer the sequence \texttt{[ (a: 1, b: [1, 0, 1] ), (a: 0, b: [0, 0]) ]}, the parent Stream will use its \textit{data} signal to transfer the value of \textit{a}, while the child Stream will transfer the sequence on \textit{b} independently, but before the next transfer on the parent Stream. The child physical stream will have $d=2$, and duplicate the \textit{last} information of the parent transfer. (Resulting in a transfer \texttt{[ (a: 1), (a: 0) ]} on the parent, and \texttt{[ [1, 0, 1], [0, 0] ]} on the child.)
    \item If $s = {Flatten}$, the child Stream will behave as \textit{Sync}, but the child physical stream will not duplicate the the (redundant) \textit{last} information of the parent physical stream. (Hence, the previous examples transfer would become \texttt{[1, 0, 1], [0, 0]} for the child.)
    \item If $s = {Desync}$, the child Stream will still inherit the parent's dimensionality (i.e., if the parent and child have $d=1$, the resulting child physical stream will still have $D=2$), but its transfers will not be constrained by those on its parent Stream. I.e., its transfers will still be shaped as \texttt{[ [ ... ], [ ... ] ]}, but they will have no (apparent) relation to the parent Stream's transfers. This does not preclude designers from using the \textit{user} signal to provide context, instead.
    \item $s = {FlatDesync}$ is the equivalent of \textit{Flatten} for \textit{Desync}, in that it does not duplicate the parent Stream's dimensionality for the child Stream. In effect, the child Stream now has no apparent relation to the parent Stream.
\end{itemize}

In the event two Streams are directly nested, they are \textit{flattened} by default in the \textit{split} function: Rather than a parent Stream with no \textit{data} signal, whose only purpose is to transfer the outer \textit{last} bits, the function will instead produce a single Stream combining the dimensionalities of parent and child. This process will only occur when both the \textit{data} and \textit{user} signals of the parent would otherwise be empty. The \textit{keep} property mentions before prevents a Stream from being \textit{flattened} regardless of its data and user signals being empty. This partly avoids issues with the parent and child Stream not having unique names (which would otherwise be derived from the Group or Union field name for the child Stream), requiring one replace the other, though the use of the \textit{keep} and \textit{user} properties can nonetheless introduce this issue, as discussed later in Section \ref{subsec:issue_nested}.

\section{Alternatives}\label{sec:background_alternatives}

This thesis aims to simplify development of streaming dataflow accelerator designs for big data analytics by improving reusability and making it easier to connect different components through the use of the Tydi specification. Outside of the application of Tydi, however, these are existing problems for which solutions are already being developed, and which can ostensibly be further adapted to suit more specific needs. This section lists alternative solutions which have been considered, and how they align with the goals of the IR and toolchain.

\subsection{Hardware Description Languages}\label{subsec:alternatives_languages}


Strictly speaking, it would be possible to implement hardware accelerators with Tydi interfaces in existing hardware description languages manually, rather than building a toolchain to emit them. It would also be possible to limit the scope of the toolchain, by directly compiling from a more abstract front-end language to an existing HDL, or by providing a standard library of components, or by generating templates in a target HDL. In particular, there are a number of languages which promote reuse and/or are suited to expressing streaming data processing efficiently.


Lime \cite{auerbachlime2010}, StreamIt \cite{thiesstreamit2002} and HPVM \cite{kotsifakouhpvm2018} are able to express streaming data processing, with StreamIt in particular being designed for this purpose alone. StreamIt and HPVM are not explicitly designed as hardware description languages, but nontheless map very well to hardware (using a suitable synthesizing compiler in StreamIt's case \cite{hormatioptimus2008}).

Chisel \cite{bachrachchisel2012}, FIRRTL \cite{izraelevitzreusability2017} and LLHD \cite{schuikillhd2020} are more general HDLs which aim to simplify expression of hardware and hardware interfaces, and promote reuse. All have since been incorporated into the CIRCT (Circuit IR Compilers and Tools) project \cite{llvmcommunitycirct2022} as different parts of an overall toolchain.

The IR is, first and foremost, an extension of the existing Tydi interface specification. In that it codifies the rules for designing and connecting interfaces, how to define data types, and how to transfer data. As such, the goal is not to outright replace any of the aforementioned languages, but serve a complementary role by expressing Tydi streams and Streamlets as efficiently as possible. It also aims to propagate high-level information down to the languages a backend might emit, including documentation. By intentionally limiting the IR's scope compared to conventional HDLs, it should also serve as an intermediary for very different kinds of front-ends. For instance, its focus on composable interfaces can also be applied to more visually-oriented design tools, such as Vivado's ``block design'' view discussed in the next section.








\subsection{Design Tools and High-Level Synthesis}


At the same time, there are multiple ongoing efforts to improve the tools used for designing such hardware accelerators, in the form of new hardware description languages \cite{bachrachchisel2012, izraelevitzreusability2017}, intermediate representations \cite{schuikillhd2020} and compilers \cite{llvmcommunitycirct2022}, high-level synthesis based on software programming languages \cite{nanesurvey2016}, and more general program representations for heterogeneous systems \cite{plavecstream2010, kotsifakouhpvm2018}.

\textbf{High-level synthesis} can help programmers who are unfamiliar with HDLs and hardware design in general to more quickly implement their ideas. This is especially relevant when the goal is to use a hardware accelerator to speed up an algorithm which was previously implemented in software, as it allows for these ideas to be translated more easily. However, the ideas expressed in HLS rarely propagate very far to the resulting hardware descriptions and simulations, making it more difficult to perform verification and analyze issues and targets for optimization from the same perspective. Some of this can be addressed by also building simulators for the high-level language, and introducing additional directives and macros to better match hardware, as in SystemC \cite{accellerasystemsinitiativesystemc2022}.

\begin{figure}[ht]
    \centering
    \includegraphics[width=0.8\linewidth]{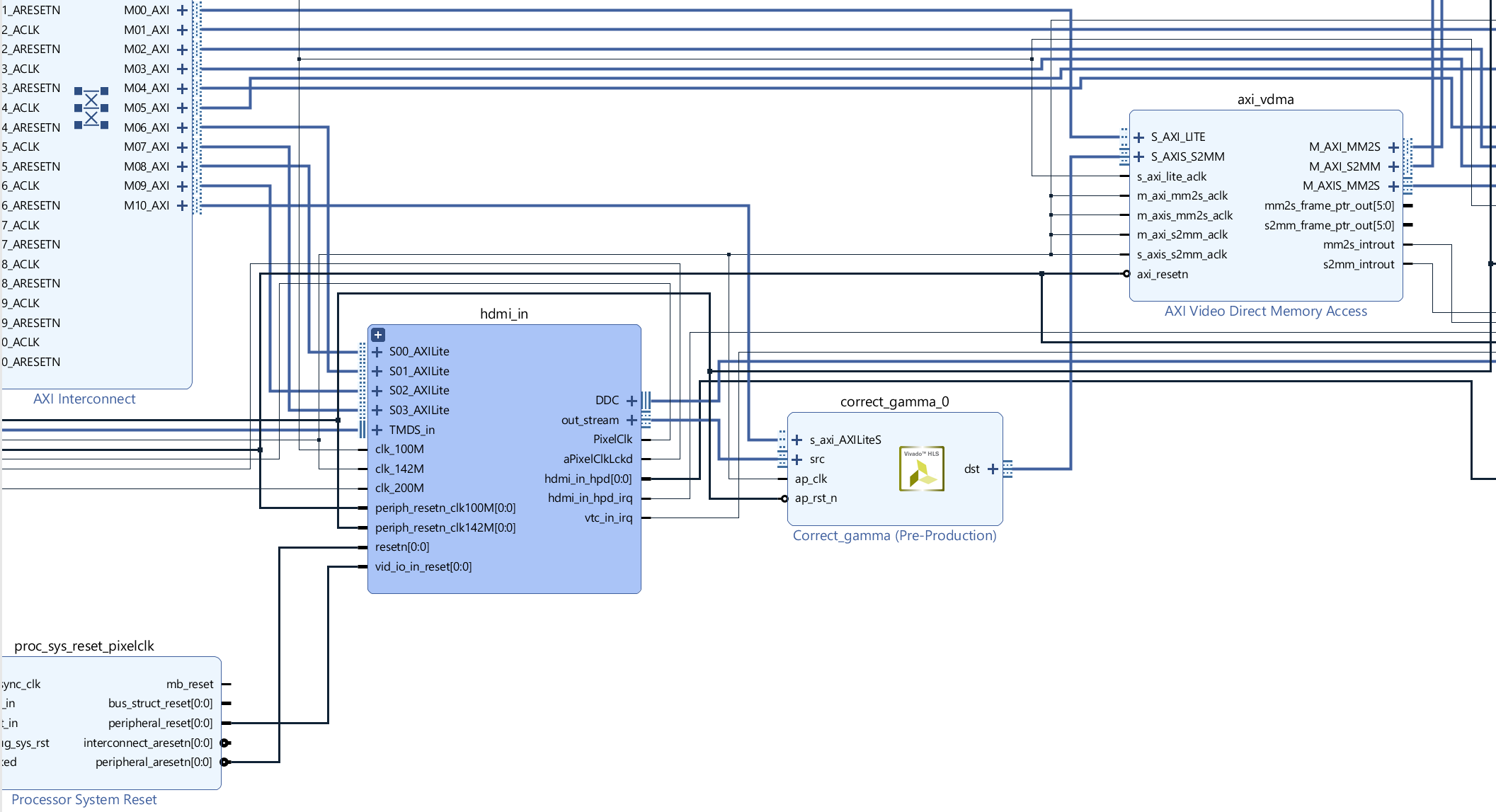}
    \caption{Using Vivado's ``block design'' interface to connect individual IP blocks using, among others, AXI4-Lite interfaces.}
    \label{fig:block_design_connect}
\end{figure}

Comprehensive \textbf{design tools} may also incorporate ways to encourage and improve avenues for reuse. For instance, Vivado includes a ``block design'' view which allows for individual components (IP blocks) to be connected using standard AXI4(-Lite/-Stream) or \textit{ap\_fifo} interfaces, as shown in Figure \ref{fig:block_design_connect}. This is combined with Xilinx's existing HLS tools and IP block library \cite{advancedmicrodevicesinc.intellectual2022} to allow for integrating these components as well, such as through pragmas (directives) indicating particular parameters or variables in a high-level language should correspond to a given interface type \cite{advancedmicrodevicesinc.interfaces2022}.

\begin{figure}[ht]
    \centering
    \includegraphics[width=0.8\linewidth]{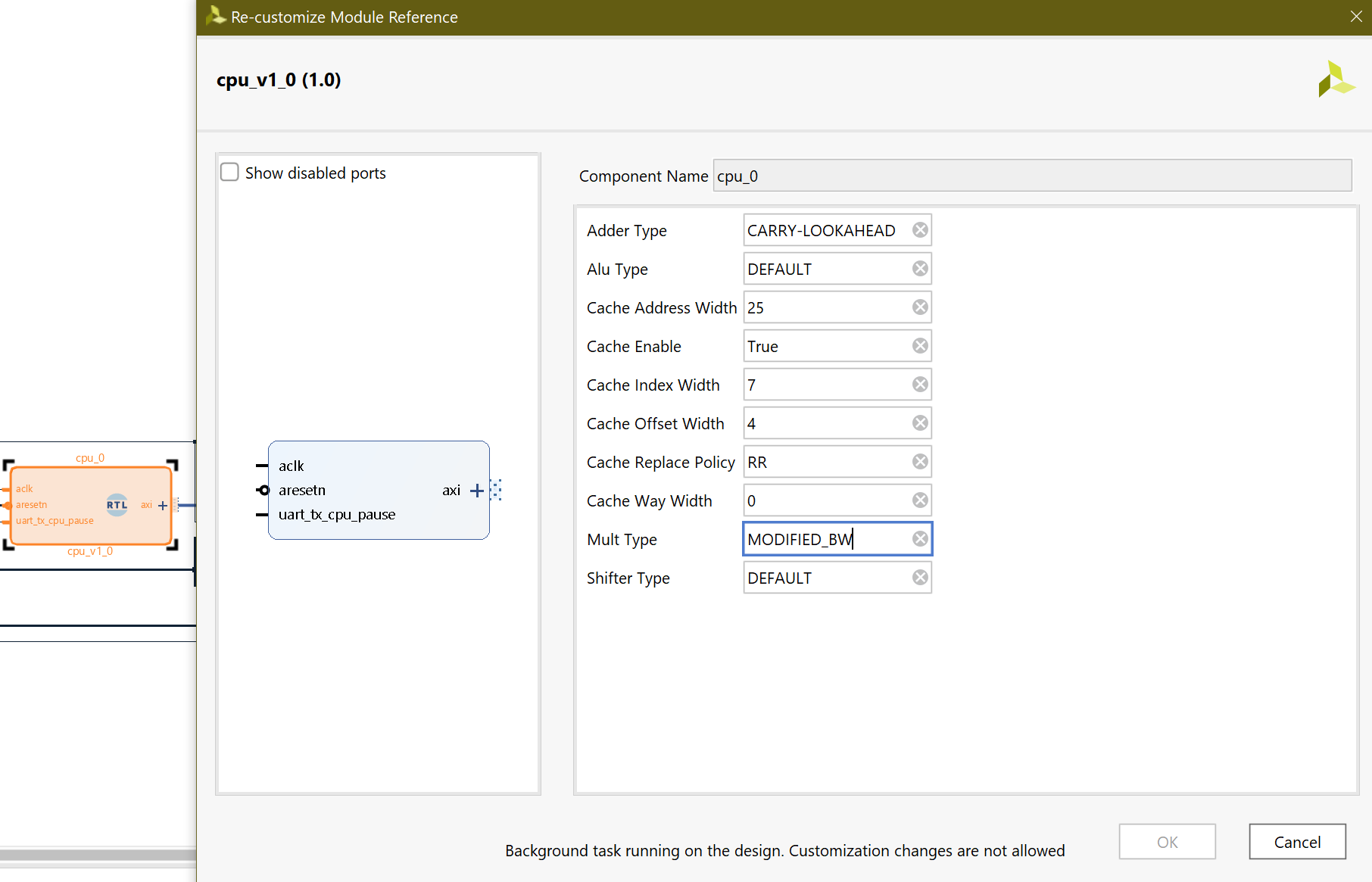}
    \caption{Using Vivado's ``block design'' interface to configure properties of certain IP blocks.}
    \label{fig:block_design_configure}
\end{figure}

Such tools are undoubtedly easy to use, especially when IP blocks surface configuration items to allow them to be modified from the same visual block design interface, as in Figure \ref{fig:block_design_configure}. The reusability enabled by such tools is less clear, however; they are proprietary, and any components designed for them must adhere to the constraints set by Vivado to enable the most desirable features, such as easily connected interfaces and exposed configuration properties. It can very much promote reuse, but only within a closed ecosystem.

\subsection{Frameworks and Embedded Domain-Specific Languages}

One approach to building hardware accelerators for a specific domain is to instead surface hardware-oriented language and/or methods within an existing language. For instance, Fleet \cite{thomasfleet2020}, S2FA \cite{yus2fa2018} and Melia \cite{wangmelia2016} all promise to easily integrate FPGA accelerators into existing software data processing environments. These improve ease-of-use by bringing the accelerator designs closer to their intended targets, and enable reuse of accelerator ``kernels'' throughout a project or over different projects. This approach can be quite effective, but is too specific to benefit hardware accelerator design reuse across different domains and potential processing frameworks.




\chapter{Intermediate Representation: Composition}\label{ch:ir}

The primary purpose of the intermediate representation is to define Tydi Streams and Interfaces, and use these to compose Streamlets. This chapter describes the ways this functionality is implemented, its use-cases, and the considerations which have gone into the IR's design overall.

To illustrate the various IR concepts described in this chapter, there are listings in TIL (Tydi Intermediate Language) a grammar for the IR which was designed (and can be parsed) as part of the overall prototype toolchain. For more details on TIL and its implementation, see Section \ref{sec:grammar_parser}.

\section{Type Declarations and Interface Design}\label{sec:types_interfaces}

\subsection{Type Declarations}\label{subsec:type_declarations}

As described in Section \ref{sec:background_tydi}, Tydi features 5 ``logical types'', with Groups, Unions and Streams themselves having fields or properties containing these logical types. The IR must be able to represent definitions of all types, account for being able to nest types, and enable comparison between types to ensure compatibility between interfaces. Listing \ref{lst:element_manip_expr_til} demonstrates expressions for the four ``element-manipulating'' types.

\begin{lstlisting}[basicstyle=\ttfamily\small,caption={Expressions for element-manipulating logical types in TIL},label={lst:element_manip_expr_til}]
Null

Bits(7)

Group (
  field_name1: Bits(2),
  field_name2: Bits(7),
)

Union (
  field_name1: Bits(7),
  field_name2: Group (
    field_name1: Bits(2),
    field_name2: Bits(3),
  ),
  field_name3: Null,
)
\end{lstlisting}

The ``stream-manipulating'' logical type, \textit{Stream}, is defined in a similar way. As it is the only type with explicit properties, however, it also features a number of \textit{default} values for some of these properties when they are omitted, as explained in the comments in Listing \ref{lst:stream_def_expr_til}. Note that Streams can be used in the exact same way as any other logical type, in that it can be used as a Group or Union's field, or another Stream's \textit{data} property; the only exception is the Stream's \textit{user} property, which may only contain element-manipulating types.

\begin{lstlisting}[basicstyle=\ttfamily\small,caption={Expression the stream-manipulating logical type in TIL},label={lst:stream_def_expr_til}]
Stream (
  data: Bits(8),
  throughput: 2.0,    // 1.0 by default
  dimensionality: 0,
  synchronicity: Sync,
  complexity: 4,
  direction: Reverse, // Forward by default
  user: Bits(2),      // Null by default
  keep: true,         // false by default
)
\end{lstlisting}

As tracking deeply nested types can become convoluted for a compiler emitting to the IR, and make the output hard to read, the IR also features the ability to declare types within a \textit{namespace} and give them a unique identifier to track them by. Listing \ref{lst:type_decls_til} showcases how these identifiers can be used; identifiers serve as an alternative to explicit type definition expressions.

\begin{lstlisting}[basicstyle=\ttfamily\small,caption={Statements declaring logical types in TIL},label={lst:type_decls_til}]
namespace namespace_name {
  type bits_type_name = Bits(7);
  
  type group_type_name = Group (
    field_name1: Bits(2),
    field_name2: bits_type_name,
  );
  
  type union_type_name = Union (
    field_name1: bits_type_name,
    field_name2: group_type_name,
    field_name3: Null,
  );
  
  type stream_type_name = Stream (
    data: union_type_name,
    dimensionality: 0,
    synchronicity: Sync,
    complexity: 4,
  );
  
  type parent_stream_type_name = Stream (
    data: stream_type_name,
    dimensionality: 1,
    synchronicity: Sync,
    complexity: 4,
  );
}
\end{lstlisting}

So as not to diverge from the Tydi specification, which does not feature identifiers as a property of logical types, these identifiers only exist as a property of the namespace, and should not affect the output of a compiler for the IR. That is to say, a \textit{Bits(8)} is always the same as any other \textit{Bits(8)}, regardless of the identifier it was given, and whether it was given an identifier at all. The merits and demerits of this approach are elaborated on in Sections \ref{subsec:interface_compat} and \ref{sec:readability}.

In order to represent each type definition in the query system, each type definition is \textit{interned}: Each distinct type is stored as an immutable entry in memory and tracked using a unique identifier. This has a number of advantages:
\begin{itemize}
    \item It reduces the amount of data stored by the query system. (No need for multiple entries in memory for identical type definitions.)
    \item Types which contain nested types as fields or properties only need to contain the identifier, instead of a copy of the definition or a direct reference to memory.
    \item Comparison between types is trivial, as it is only necessary to compare intern identifiers: If types are different, their identifiers will be different, as well.
    \item The query system will not have to track different kinds of type expressions: Whether they are namespace identifiers or direct definitions, each type ultimately becomes an intern identifier.
\end{itemize}

\subsection{Interfaces as Contracts}\label{subsec:interfaces}

As Section \ref{sec:background_tydi} would suggest, Tydi's types can convey a significant amount of information; not just what data is transferred, but also how it is transferred, and how sequences of elements relate to one another. In effect, a sufficiently detailed Stream definition can be treated as a \textit{contract} between components (and in a sense, designers) on how a stream of data will be implemented.

The intermediate representation builds on this when declaring \textit{Interfaces}. In its simplest form, an Interface represents a collection of ports on a component (Streamlet), each of which carries a logical Stream either into or out of the component. Any streamlet must have an interface; as a result, all streamlet definitions can be subsetted into interfaces, as shown in Listing \ref{lst:iface_decls_til}. By default, backends are not expected to emit interface declarations which are not part of streamlet definitions, and the names of interface declarations should not have any effect on the resulting output. It is however allowed to define a streamlet without any implementation, consisting only of an interface - defining an interface without any ports is also allowed.

\begin{lstlisting}[basicstyle=\ttfamily\small,caption={Statements declaring interfaces and streamlets in TIL},label={lst:iface_decls_til}]
interface my_interface = (a: in stream, b: out stream);

streamlet my_streamlet = my_interface;

streamlet my_impl_streamlet = (
  a: in other_stream,
  b: out stream
) {
  impl: ...
};

streamlet subsetted_streamlet = my_streamlet {
  impl: ...
};
\end{lstlisting}

However, each Interface and its ports may also feature \textit{documentation}. Distinct from comments on a grammar, documentation is an actual property of a port or interface, and is expected to be implemented by a backend, typically by generating matching comments on the related output. Documentation being propagated from higher-level descriptions to the actual computation-oriented design tools that the IR complements is primarily useful when either implementing a component based on an interface template, or when trying to identify how physical signals relate to their abstract definition.

While Tydi's Streams assume a single clock and reset signal, which together make up their clock and reset domain, regardless of how many physical streams they are composed of, the ports of an Interface do not need to rely on the same clock and reset signals. Instead, an Interface may have one or more uniquely named \textit{domains} which represent a clock and reset signal, each of which is associated with one or more of the Interface's ports.

Subsequently, while the intermediate representation does not feature the ability to define a specific clock or how a reset signal should be handled, designers can use these domains to ensure multiple clock and reset signals are available on a component, and that ports which belong to different domains are not directly connected. In the event no domain is specified on the Interface, a default domain is instead created and assigned to all ports, as Tydi currently only defines Streams in the context of a clock.

It is worth noting, as a recommendation for future work, that the use of ready-valid signals \textit{should} make it possible to represent fully asynchronous (clock-less) micropipelines \cite{sutherlandmicropipelines2007} using Tydi. Even if the specification currently assumes the existence of a clock, many of the timing constraints enforced through it can be replaced by the ready and valid signals serving as events for forward and reverse propagation.


\subsection{Compatibility}\label{subsec:interface_compat}


The ports of Interfaces are compatible with one another when they have the same logical type, appropriate directions (for each physical stream, there is a source and matching sink), and the same (clock) domain.

A domain in Tydi and the IR consists of a clock and reset signal; while the reset signal does not have any specific constraints to the clock signal, the reset behavior, requiring the \textit{valid} and \textit{ready} signals to be driven low during a reset, is constrained by it. The Tydi specification generally assumes a single clock and reset signal, but this only applies in the context of a Stream and its compatibility with other Streams. Therefor, it is possible to surmise that a (clock) domain and a Stream are intrinsically linked; the compatibility of two interfaces using Tydi Streams is contingent on them being part of the same clock domain.

To reflect these properties, the IR assumes a single ``default'' domain, but allows for the definition of additional/alternative domains and for linking them to specific interface Streams; the actual clock speed is irrelevant to compatibility, only whether a designer indicates something is a different domain.

\begin{lstlisting}[basicstyle=\ttfamily\small,label={lst:domain_compat}]
// As no domain has been defined, the "default" is assigned
(a: in stream, b: out stream) {
  impl: {
    // Sharing one domain, these are compatible
    a -- b;
  }
}

// Declaring new domains removes the default domain, and
// requires that they are assigned to individual ports.
<'a_domain, 'b_domain>(
  a: in stream 'a_domain,
  b: out stream 'b_domain,
) {
  impl: {
    // As these now have different domains, these are incompatible
    a -- b;
  }
}
\end{lstlisting}


Note that while types in the IR may be defined with identifiers, these identifiers are not a property of the logical type in question, and only exist within the namespace. This choice was made to restrict the IR to properties defined in the Tydi specification.

As a result, types with different names but otherwise identical properties are fully compatible; on an abstract level, this can be interpreted as a kind of implicit casting between types. Although when evaluating this with respect to readability of backend output, discussed in Section \ref{sec:readability}, and in light of the potential added value of a stricter type system, this approach may need to be reconsidered in the future. An alternative approach might make identifiers an intrinsic property of types, and separately support type aliases for functionality similar to the current behavior - depending on the language being targeted, such aliases could even be propagated to the backend.

However, while type identifiers are not currently relevant to compatibility, \textit{field} identifiers are an actual property of the Group and Union types. Hence, a \texttt{Group(a: Null)} is not compatible with a \texttt{Group(b: Null)}, regardless of whether they are physically identical.

\begin{lstlisting}[basicstyle=\ttfamily\small,label={lst:type_compat}]
type bits8 = Bits(8);
type byte = Bits(8);
// bits8 and byte are compatible

type a_group = Group(a: bits8);
type b_group = Group(b: bits8);
type group_a = Group(a: byte);
// a_group and group_a are compatible, but neither are compatible with b_group
\end{lstlisting}

Finally, while \textit{complexity} is a property of the Stream type, the Tydi specification does conditionally allow Streams with different complexities but otherwise identical properties to be connected. Specifically, a physical \textit{source} stream may be connected to a \textit{sink} if its complexity is equal to or lower than that of the sink. Note however that this applies to physical streams: logical Streams do not have a notion of sinks and sources, and may contain child Streams which flow in reverse directions, resulting in them containing both sink and source physical streams.

As such, the IR considers the Streams of ports incompatible when their complexity is not identical. While the process of connecting compatible physical streams can be optimistically automated to improve reuse, as discussed later in Section \ref{subsec:intrinsics}, designers should generally strive for a shared, normalized complexity between Streams.



\section{Component Composition and Implementation}\label{sec:component_implementation}


In addition to Interfaces, the IR introduces the ability to declare components, referred to as \textit{Streamlets}. These Streamlets consist of an Interface and optionally an Implementation. In effect, there are two different kinds of Implementation for a Streamlet: a \textit{structural} implementation, which can be used to combine instances of streamlets into a larger design, and a \textit{link} to an implementation of behavior in the target language or format.

Streamlets are the intended output of a project; Types, Interfaces and Implementations are not expected to be included in a backend's emissions unless they are part of a Streamlet, but can be shared between IR projects.

As Streamlets always have an Interface, they can be \textit{subsetted} to Interfaces, which can be used to express alternate implementations of the same component, e.g. when versioning a component or when substituting one for the purposes of testing as described in Section \ref{subsec:substitution}.

\subsection{Structural Composition}

As the goal of both Tydi and the IR is to improve compatibility and reuse of primitive components, the IR features the ability to connect Streamlets to one another. The IR refers to this as a \textit{Structural} implementation.

Structural implementations can contain \textit{instances} of Streamlets and connections between ports of Streamlets. Instances consist of a local name and a reference to a Streamlet declaration, the ports of their interfaces are assigned separately through connections. If the parent interface has named domains, these must also be assigned to the Streamlet instance.

\begin{lstlisting}[basicstyle=\ttfamily\small,caption={Statements for instantiating instances of Streamlets in TIL.},label={lst:structural_instance_til}]
// Creating an instance with a default domain
instance_name = streamlet_name;

// Creating an instance and assigning domains
instance_name = streamlet_name<'parent_domain_name>;
// Or:
instance_name = streamlet_name<'streamlet_domain_name = 'parent_domain_name>;
\end{lstlisting}

Connections can be created between the ports of both Streamlet instances and the containing Streamlet which is being implemented, and require both ports to have identical types and clock domains (for the reasons described in Section \ref{subsec:interface_compat}). Connections are explicitly not ``assignments'', as the direction of a port is already known, and there is not necessarily one overall direction for a Stream type due to the possibility to define Streams which are \textit{Reversed} (such as when representing request and response streams). Hence, the \textit{source} and \textit{sink} between two ports of a connection is determined during lowering for each resulting Physical Stream.

\begin{lstlisting}[basicstyle=\ttfamily,caption={Statements for connecting ports in TIL.},label={lst:structural_connection_til}]
instance_name.instance_port1 -- instance_name.instance_port2;
instance_name.instance_port -- parent_port;
parent_port1 -- parent_port2;
\end{lstlisting}

By default, the IR requires that each port of each Streamlet is connected to exactly one other port. Leaving ports unconnected is against the Tydi specification, which requires that a default signal is driven for omitted signals \cite{vanstratenphysical2021}. While HDLs such as VHDL and Verilog support one-to-many and many-to-one connections at a signal-level, these are not allowed by the IR due to the fact that ports represent Streams with handshake signals, which would need to be combined.

\begin{lstlisting}[basicstyle=\ttfamily\scriptsize,caption={A full Structural implementation in TIL.},label={lst:structural_implementation_til}]
streamlet example_streamlet = <
  'parent_domain1,
  'parent_domain2,
> (
  parent_port1: in stream 'parent_domain1,
  parent_port2: out stream 'parent_domain1,
) {
    impl: {
        parent_port1 -- parent_port2;
        
        // dom_example has two domains, one for ports a and d, and one for port b and c
        different_domains = dom_example<'parent_domain1, 'parent_domain2>;
        different_domains.a -- different_domains.d;
        different_domains.b -- different_domains.c;
        
        // By assigning them the same domain, a and b and c and d can nonetheless be connected
        same_domains = dom_example<'parent_domain1, 'parent_domain1>;
        same_domains.a -- same_domains.b;
        same_domains.c -- same_domains.d;
        
        // For clarity, when assigning domains it's also possible to specify
        // which domain of the instance is being assigned to, rather than using their order.
        explicit_doms = blank_doms<'c = 'parent_domain1, 'a = 'parent_domain2, 'b ='parent_domain2>;
        
        // It's also possible to mix named assignments with ordered assignments,
        // provided the named assignments succeed all ordered assignments.
        mixed_assignments = blank_doms<'parent_domain2, 'c = 'parent_domain1, 'b ='parent_domain2>;
    }
};
\end{lstlisting}

While combining the \textit{ready} signals of multiple sinks could be achieved with simple logical \textit{and} expressions for a one-to-many connection, combining multiple transfers in a many-to-one connection has no clear, universally applicable, solution. Even the aforementioned one-to-many implementation is not universal, as some designs may call for only one of the many to alternately accept transfers. Finally, as a connection does not necessarily have a single direction, a one-to-many connection between ports may well contain physical many-to-one transfers.

Instead, the solution to unconnected and one-to-many ports would be to explicitly define their behavior. In the current implementation, that means designing specific Streamlets for this purpose; but as there is a common subset of expected behavior (drive default, sink and ignore input, transmit transfers to all in a one-to-many source-to-sink configuration, etc.), the ultimate goal would be to provide intrinsics which automatically implement this behavior, as described in Section \ref{subsec:intrinsics}.






\subsection{Linked Implementations}\label{subsec:linked_impls}

The intermediate representation intentionally omits expressions for implementing or simulating arbitrary behavior of components. Designing a language or set of expressions for functional hardware design and simulation is a difficult problem which is already being addressed by many researchers and organizations, as explained in Sections \ref{sec:background_stream_processing} and \ref{subsec:alternatives_languages}. Instead, ``behavioral implementations" in the IR exist only as \textit{links} to directories, which contain the relevant code in languages more suited for expressing behavior.

How these links are used is left up to the backend, though a simple use-case would be to create or copy a file in the target output language based on the Streamlet's name. As these are directories, multiple such files can exist side-by-side for different targets, and implementations do not need to be constrained to a single file; a linked directory could even be used to refer to a project or library consisting of multiple files, provided this exposes the Interface of the Streamlet being implemented.

It is worth noting that linked implementations are not to be treated like \textit{imports}: A linked implementation is still fundamentally part of the IR project, and should be included with its sources/output. A link should not refer to a common library directory, and linked implementations should use relative paths (relative to the project root), rather than absolute paths. Provided front-end languages use these same constraints, this has the added benefit of ensuring projects can be easily shared between developers and tracked in version control systems.

In TIL's grammar, links are simply written as path strings enclosed in double quotes. Whether a path uses valid formatting and characters is determined by the query system, though the query system will not verify whether the directory actually exists; it is up to the connected backend (and its potential configuration) to decide whether to treat a non-existent directory as a failure condition, or to instead create the directory if it does not exist.

\begin{lstlisting}[basicstyle=\ttfamily\scriptsize,caption={A linked implementation in TIL.},label={lst:linked_implementation_til}]
streamlet example_streamlet = <
  'parent_domain1,
  'parent_domain2,
> (
  parent_port1: in stream 'parent_domain1,
  parent_port2: out stream 'parent_domain1,
) {
    impl: "./path/to/a/directory"
};
\end{lstlisting}

Note that, as shown in Listing \ref{lst:linked_implementation_til}, linked implementations still require a complete Streamlet definition, consisting of a name and an interface in addition to its implementation. As mentioned before, the Streamlet name is used to let the backend determine which file or set of files to use from a directory. It also ensures that its interface definition can be included in the target language's project structure, and instances can be created inside structural implementations. E.g., when emitting to VHDL, the interface definition and name are used to create a component definition in the emitted project's \textit{package} file(s), and these components are used inside the emitted architectures of structural implementations. The interface definition and name can also be used to automatically generate a correctly named and structured \textit{template} in the linked directory, if the target does not already exist.

\begin{figure}[h]
  \centering
  \includegraphics[width=\linewidth]{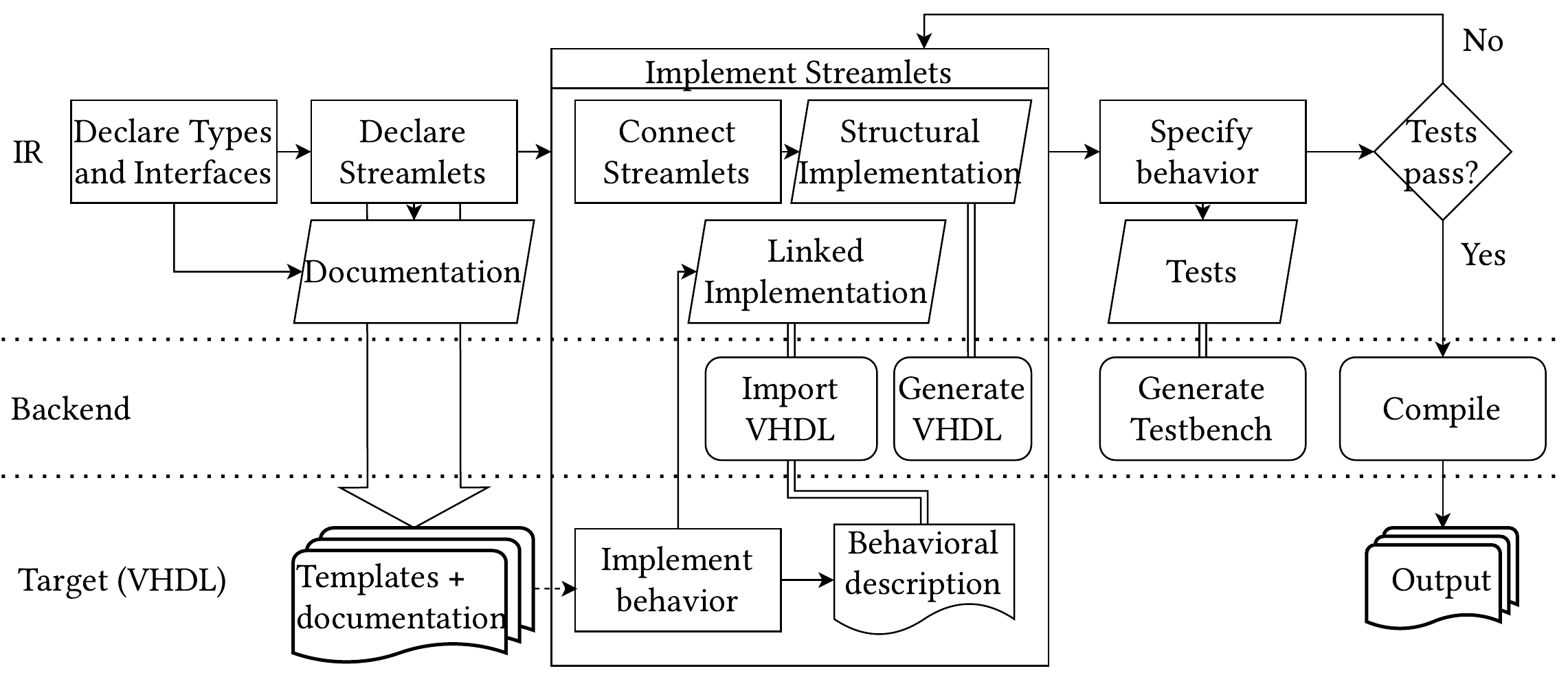}
  \caption{An example workflow, demonstrating how Streamlets are implemented using the IR, a suitable backend, and behavior defined in the target language.}
  \label{fig:irworkflow}
\end{figure}

Figure \ref{fig:irworkflow} illustrates how linked implementations fit within a partial toolchain and workflow, consisting of Streamlets, structural implementations and tests defined in the IR, combined with behavior defined in a target language (VHDL, in this example) by a suitable backend. Not pictured are tools for simulating the testbenches produced by the backend, further passes on the output, nor any potential frontend language.


\section{Recommendations for Language and Compiler Features}\label{sec:proposed_features}


While the previous sections cover the primary functionality of the IR, which is to describe types, interfaces, Streamlets, and implementations of Streamlets, there are still a number of language features which can improve or simplify the IR's ability to describe designs overall. To preface these language features, however, it is important to note that these are not (solely) intended to make the IR easier to write or read by humans; as this is an IR, it will primarily be emitted as output of a more ergonomic, front-end language and read by a compiler via the query system. As such, language features in the IR should not be ``syntactic sugar'', which is ultimately different styles of expression for existing language constructs. New features should meaningfully translate to something a compiler can directly implement, and should not expand to existing IR constructs (except potentially as a fallback). By extension, these features should be possible to implement in common HDLs.

\subsection{Type Parameters}\label{subsec:type_parameters}


Type parameters allow for the creation of variable types (also known as \textit{generic types}), by parameterizing specific properties and allowing variations upon the type to be instantiated as needed. Listing \ref{lst:generics_til} shows what such type parameters might look like in TIL. Such type parameters can be useful in simplifying and organizing the expression of related interfaces which share many properties, for example, if a certain (collection of) Streams represents a memory interface, its bus width can be parameterized.

\begin{lstlisting}[basicstyle=\ttfamily\small,caption={Theoretical grammar for generic types in TIL.},label={lst:generics_til}]
type generic_stream<
  a: bitcount,
  b: type,
  c: complexity,
  d: throughput,
  e: keep,
  f: direction,
> = Stream(
  data: Group(a: Bits(a), a2: Bits(a), b: b),
  complexity: c,
  throughput: d,
  keep: e,
  synchronicity: Sync,
  dimensionality: 0,
  direction: f,
);

type concrete_stream = generic_stream<6, Bits(5), 4, 3.14, false, Forward>;
\end{lstlisting}

The omission of type parameters was not purely due to implementation time constraints, but due to the properties of such parameters being subject to debate, relating to the quality described in the preface: Language features should be possible to translate by a backend to their target language. There are two distinct ways to implement generic types in the IR:

\begin{enumerate}
    \item Generic types are evaluated before compilation, either by the parser or by the query system before a compiler requests a definition. Only concrete types exist in the resulting output.
    \item Type parameters are stored as properties on the IR, and can be used during compilation to translate to equivalent language features.
\end{enumerate}

The first option is easiest to implement; as suggested, evaluating generic types to concrete types does not even need to be a feature of the IR itself, but can be handled by the parser for TIL. This makes them similar to identifiers on the namespace: A feature for tracking and reusing types, rather than something to be propagated to a compiler's output. For instance, one could save time on implementing multiple Streams with the same complexity, synchronicity and dimensionality by making the relevant properties type parameters.

However, like type identifiers, these generics are not \textit{strictly} required to be part of the IR or even TIL: A front-end language and compiler can define generics themselves, and simply expand these to concrete types for the IR.

The second option provides significantly more added value, but would also be significantly more difficult to implement. After all, while many (hardware description) languages which could be targeted may feature similar type parameters, not all languages do, and not all type parameters are equally flexible. For example, while VHDL(-93) has type parameters for determining the size of certain arrays, and newer versions (2019) even include support for types as parameters (like the ``b'' parameter in Listing \ref{lst:generics_til}), there are no parameters which omit ports altogether (when passing a Tydi Null, changing the \textit{complexity}, or setting the \textit{keep} property from true to false) or change the direction of ports (the ``f'' parameter in the example Listing).

There are a few potential ways around these limitations:
\begin{enumerate}
    \item The IR can be limited in which properties can be parameterized (e.g., only \textit{bitcount} and \textit{throughput} can be parameterized).
    \item A compiler using the query system can selectively request certain parameters to be evaluated ahead of time.
    \item Require that compilers for the Tydi IR must support all generic parameters.
\end{enumerate}

Each solution is ultimately flawed in some way, in that the first solution does not account for languages which lack any type parameters, and the third solution omits a large number of HDLs. The second solution's flaws are less obvious, this solution means that functionality is at least equivalent to the alternative interpretation of generics (always evaluating to concrete types), but also greatly limits their potential; it requires that it \textit{must} be possible to evaluate any generic type ahead of time.

This constraint means that designers cannot define generic Streamlets or Interfaces, or cannot use them effectively as instances. Additionally, to convert otherwise entirely generic types (and/or Streamlets) into concrete versions, the IR will have to generate unique names for each variant. E.g., an instance may simply be defined as \texttt{streamlet\_name<a = 3, b = 4>}, where \textit{b} is a bitcount and supported, but \textit{a} is the complexity level and not supported: To resolve this, the query system returns a \texttt{streamlet\_name\_\_a\_is\_3} with a bitcount type parameter, and any number of other combinations.

Though with that said, it is still possible to make a recommendation for maximum functionality:
\begin{itemize}
    \item Type parameters should be a part of the IR, and not evaluated beforehand. As the minimal possible outcome is one which matches functionality with pre-compilation evaluation.
    \item All properties, and types themselves, should be possible to use as parameters. As different target languages have different limitations.
    \item Compilers should be able to indicate which parameters they support, at which point the query system should do one of the following, potentially based on further configuration: \begin{enumerate}
        \item Fail if the project contains type parameters which the compiler/target language does not support.
        \item Only return types and Streamlets which feature the supported type parameters. (If no type parameters are supported, only return concrete types and Streamlets.)
        \item Generate uniquely named variants of types and Streamlets which are concrete (have fully-defined parameters), but feature unsupported type parameters.
    \end{enumerate}
\end{itemize}

\subsection{Generation}\label{subsec:generation}
As a more general language feature for structural implementations, the IR could expose forms of \textit{generation}. Specifically, generating multiple instances of the same Streamlet definition in \textit{arrays}, and generating connections between such instances in \textit{for loops}. For example, rather than expressing a number of instances with unique names, such generation would leverage the target language's ability to generate similar arrays:

\begin{minipage}{0.95\textwidth}
  \centering
  \begin{minipage}[t]{.45\textwidth}
    \centering
    \begin{lstlisting}[basicstyle=\ttfamily]
a_1 = streamlet_decl_a;
a_2 = streamlet_decl_a;
a_3 = streamlet_decl_a;
b_1 = streamlet_decl_b;
b_2 = streamlet_decl_b;
b_3 = streamlet_decl_b;
a_1.out_port -- b_1.in_port;
a_1.in_port -- b_1.out_port;
a_2.out_port -- b_2.in_port;
a_2.in_port -- b_2.out_port;
a_3.out_port -- b_3.in_port;
a_3.in_port -- b_3.out_port;
    \end{lstlisting}
  \end{minipage}
  \begin{minipage}[t]{.45\textwidth}
    \centering
    \begin{lstlisting}[basicstyle=\ttfamily]
a[3] = streamlet_decl_a;


b[3] = streamlet_decl_b;


for i in 0..3 {
  a[i].out_port -- b[i].in_port;
  a[i].in_port -- b[i].out_port;
}
    \end{lstlisting}
  \end{minipage}
  \captionsetup{type=lstlisting}
  \caption{Arrays of instances and loops for connections (right) compared to equivalent, explicit instances and connections (left).}
  \label{fig:twoalg}
\end{minipage}

These forms of generation are relatively safe to include as part of the IR, because many common HDLs already support equivalent functionality:
\begin{itemize}
    \item VHDL provides \texttt{for ... generate} to generate multiple \textit{port map}s and connect arrays of signals based on an index. \cite[Section~11.8]{ieee2009}
    \item Verilog likewise allows for modules to be instantiated in \texttt{generate for ...} loops. \cite[Section~12.4]{ieee2006}
\end{itemize}

By adding support for such loops, it is possible to generate more readable output: If a front-end allows for the expression of (for) loops, propagating these loops to the IR and the eventual target language better reflects the designer's intentions.

In the event that a language does not feature compatible constructs, it is possible to safely expand these loops to explicit instantiations and connections in the IR. This can be performed as a fallback function of the query system, rather than requiring a compiler to implement it, and can simply generate instatiations with index numbers as part of a ``Path Name'' (which normally reflect namespaces or nested fields of logical types). To expand on why this is safe: The Tydi specification (and by extension the IR) does not allow for \textit{Name}s to start with numbers, and normally requires \textit{Path Name}s to consist of valid \textit{Name}s; by using a number as part of a \textit{Path Name} for this expansion, we are guaranteed unique output names. (Conversely, directly appending numbers to \textit{Name}s could result in conflicts with otherwise valid \textit{Name}s.)

Combining generation with type (and specifically Streamlet definition-bound) parameters can also be incorporated into the IR, but does not necessarily feature equivalent constructs in potential target languages, and may not be safely expanded for the same reasons as expressed in the previous section.

\subsection{Intrinsics}\label{subsec:intrinsics}

While the intermediate representation does not support expressing \textit{arbitrary} functionality, there is arguably a subset of functionality useful for implementing Tydi-based components and streaming dataflow designs in general. For general design purposes, small components which aid with building pipelines and ensuring consistent parallel operation, such as slices, buffers and synchronizers, are relatively simple in their implementation and commonly used. In the specific context of Tydi, there are are number of limitations enforced by the specification and IR which can be addressed somewhat easily; for instance, all ports of an interface must be connected - to address this, a designer needs to drive a default or constant value to this port, or simply indicate it is unused (by driving ready and/or valid low).

Hence, there is cause to establish a minimal, portable set of intrinsic functions, or \textit{intrinsics}, to be implemented by any backend. Specifically, intrinsics should only cover commonly used, simple functionality which cannot be implemented by a library of fixed component designs; as an example, slices are commonly used and simple in both their functionality and implementation, but a fixed library cannot address each possible interface design. The same applies to driving default values to physical streams, which are specified and relatively easy to apply, but too variable to incorporate into a library. Furthermore, neither functionality would commonly be implemented as a Streamlet, but instead incorporated into one (as a lower-level component, or directly).

Intrinsics will primarily be useful in the context of structural implementations, but may also be useful when defining types and interfaces. For example, in the absence of type parameters, or to provide a more explicit way of expressing it, an intrinsic which \textit{reverses} a given Stream type could be implemented.

To include such intrinsics in TIL, accounting for the different scopes they may be applicable in, and avoiding an excess of keywords potentially conflicting with identifiers, it would make sense to preface intrinsics with a control character. For the purposes of the next examples, this control character will be \texttt{!}, as it is otherwise unused, and not valid as part of any identifier, as shown in Listing \ref{lst:excl_intrinsic_til}.

\begin{lstlisting}[basicstyle=\ttfamily,caption={Using \texttt{!} as a control character for intrinsic functions in TIL},label={lst:excl_intrinsic_til}]
type my_rev_stream = !reverse(stream_identifier);
...
instance.port_a -- !default;
...
instance.port_a -- !buffer(3) -- parent_port;
\end{lstlisting}

Listing \ref{lst:excl_intrinsic_til} also demonstrates that intrinsics benefit from being able to modify operators and produce different kinds of statements, by overriding what a port may be connected to (port connecting to \texttt{!default}), and producing different kinds of connections (a \texttt{!buffer} between a connection). As intrinsics will be part of the IR, and not user-definable, this will be possible to account for and implement.

As a recommendation for future work and summary, the following intrinsic functions are likely to be suitable for the IR:
\begin{enumerate}
    \item \textit{Slices} and (FIFO) \textit{buffers}, to break up combinatorial paths and/or account for different operations taking a variable number of cycles.
    \item A \textit{synchronizer}, which attempts to combine the ready/valid control signals of multiple input and/or output Streams.
    \item A \textit{parallelizer}, which converts a single source Stream/port and attempts to split elements or outer sequences into separate Streams. (I.e., this only applies to Streams without dimensionality, or selects the cut-off point based on the outer dimension. For simplicity, this should only apply to Streams which are represented as single physical streams.)
    \item A \textit{serializer}, which converts multiple matching source Streams into a single Stream, based on their outer (or lack of) dimension. The inverse of the parallelizer above.
    \item A throughput \textit{reshaper}, which results in a Stream with a different number of element lanes, either combining multiple transfers into one, or splitting up transfers into multiple.
    \item An intrinsic which automatically drives \textit{default} values to the physical streams that make up a given Stream, as defined by the Tydi specification \cite{vanstratenphysical2021}.
    \item A way to explicitly mark a port as \textit{unconnected}, circumventing the IR's validation against unconnected ports, and optionally driving ready/valid low.
    \item A \textit{constant value} source, based on the assertion system described in Section \ref{sec:assertions}, allowing designers to quickly stub connections beyond driving their default values.
    \item An optimistic source-to-sink \textit{complexity bridging connection}: Provided a port's source Stream(s) have a lower complexity and otherwise perfectly match another port's sink Stream(s), these ports can be connected according to the Tydi specification. (Note that this will need to account for a port's parent (source) Stream potentially containing reversed child (sink) Streams, which cannot have lower complexities than their counterparts.)
    \item A configurable \textit{complexity downshifter}: Using sufficiently deep buffers, transfers from a higher-complexity source Stream may be converted to match the constraints of lower complexities. (A buffer is required because lower complexities can impose timing constraints on transfers, and may not allow transfers with inactive lanes.)
    \item A \textit{reverse} function for Stream expressions, which takes an existing Stream type and simply switches its direction property between Forward and Reverse. (This functionality can potentially be implemented on the query system or TIL parser, rather than by a compiler.)
    \item Different, configurable intrinsics for \textit{N-to-M connections} based on parallelizers and serializers, and potentially by simply sending transfers from a source Stream to multiple sinks by synchronizing their ready/valid signals.
\end{enumerate}

This is not an exhaustive list, and more specific functions may arise based on actual use, but these should serve as relevant, minimal examples of which functions could be incorporated into the IR directly.

These proposed intrinsic functions also illustrate another property to account for: Not every intrinsic will necessarily succeed, and it may not be possible to determine their successfulness based on static evaluation. For example, the complexity bridging connection will simply fail on static evaluation if it is determined that the connection includes a source with a higher complexity than its sink counterpart. Conversely, whether the ``complexity downshifter'' will succeed depends entirely on whether the buffer is sufficiently large to concatenate all incoming transfers to match the target complexity's constraints; complexity $1$ requires that \textit{all} transfers occur over consecutive clock cycles, requiring either the buffer to be sufficiently deep for all expected sequence lengths, or for the source to not postpone transfers often enough to let a smaller buffer run out. The latter case can only be evaluated by the designer or potentially in simulation.

Finally, there will likely be a number of potential intrinsic functions which are broadly applicable, but can only be used in a subset of scenarios and/or target languages. For example, a simple clock generator for simulation purposes could be expressed as an intrinsic, but would not be possible or useful to synthesize. To this end, such subsets should have a hierarchical naming scheme to indicate they are not generally available and/or expected to be implemented by all backends. The clock generator example could be expressed as \texttt{!sim.clockgen(...)}, for example.





\subsection{Annotations}\label{subsec:annotations}

One final suggested addition to the language as a whole is support for \textit{annotations}, which are syntactic metadata to be interpreted by compilers. Up until now, the additional language features discussed were intended to be generally applicable between most compilers and potential target languages. By contrast, annotations are intended for including information which is specific to a target language, compiler, or hardware platform, or may be interpreted (very) differently between them. As an example of the latter scenario: A ``deprecated'' annotation could produce warnings, be ignored altogether, or outright prevent compilation depending on the target, compiler, and configuration.

The IR should not predicate the properties or implementation of annotations, and it should be possible to add annotations to any language construct: That is to say, annotations can be used to provide metadata for any expression, statement, or intrinsic, and may be combined with other annotations. To avoid unintentional conflicts of annotations, to ensure they can be parsed as text in TIL and can be stored in the query system, and to avoid incompatibility between IR projects, the following constraints and properties are recommended:

\begin{enumerate}
    \item An IR project being valid should \textit{never} be contingent on the existence of annotations. I.e., a compiler must be able to ignore all annotations and produce valid (albeit not necessarily correct or desirable) output. This can be enforced by the TIL parser and IR query system not making use of the metadata defined in annotations. Note that compilers may still fail on \textit{incorrect} annotations, based on their interpretation.
    \item Use a unique control character as delimiters for an annotation in TIL. For example, the \texttt{@} symbol, which is otherwise unused and unsupported by the IR.
    \item All annotations must have a ``namespace'': Namespaces can be used to indicate certain annotations belong to specific compilers, languages, or operations. This ensures a compiler can easily query only relevant annotations within a project. To prevent further conflicts within a specific namespace (such as ones for a target language which is served by different compilers), a common forum to establish their meaning will be necessary. \begin{itemize}
        \item E.g.: \texttt{@namespace.sub\_namespace.property@}
        \item This also allows for the organization of multiple properties based on namespaces, as shown in Listing \ref{lst:annotation_collection_til}.
    \end{itemize}
    \item An annotation can be one of the following kinds of properties: \begin{enumerate}
        \item A \textit{flag}; their existence implies a boolean \textit{true}. E.g.: \texttt{.property}
        \item An \textit{assignment}; a name, followed by a single value being assigned. E.g.: \texttt{.property = value}
        \item A \textit{constructor}; a name, followed by multiple properties of that name being assigned a value. E.g.: \texttt{.property(a = value, b = value)}
    \end{enumerate}
    \item Annotations can be assigned one of the following kinds of values: \begin{itemize}
        \item A \textit{number}: These can be positive, negative or floating point, but will not be evaluated by the parser and simply stored as strings. E.g.: \texttt{1, -1, 1.0, -1.0}
        \item A \textit{unit value}: An arbitrary sequence of (non-control or otherwise conflicting) characters preceding and/or succeeding a number. Stored as an optional string, a number (also a string) and another optional string. E.g.: \begin{itemize}
            \item \texttt{\$ 100.00}
            \item \texttt{40 GHz}
            \item \texttt{after 10 ns}
        \end{itemize}
        \item A \textit{string}: Arbitrary sequences of characters, enclosed by double-quotes. Control characters should be escaped, but it may be possible to provide a ``raw string'' syntax in TIL, as well.
        \item A \textit{name}: Sequences of (non-control or otherwise conflicting) characters, can be used to imply constants or values of enumerations, are stored as strings. E.g.: \texttt{.target\_hw = HAL9000}
        \item An \textit{object}: Arbitrary collections of (unique) names and nested values, stored as maps using a string (name) as key. E.g.: \texttt{\{a: value, b: value\}}
        \item A \textit{list}: A sequence of values, the parser and query system will not enforce they are the same kind of value. E.g.: \texttt{["string", 1.0, name]}
    \end{itemize}
\end{enumerate}

\begin{lstlisting}[basicstyle=\ttfamily\small,caption={A collection of annotated properties},label={lst:annotation_collection_til}]
@namespace.namespace2 {
  namespace3.namespace4.flagvalue,
  namespace5 {
    constructorvalue(a = 2, b = 3),
    flagvalue2,
    namespace6 {
       flagvalue3,
       assignvalue = "a string",
    }
  }
}@
\end{lstlisting}

Combined, these properties should allow for the clear, readable expression of virtually any kind of metadata. Note that annotations overall simply amount to collections of (tagged) strings, ensuring they do not burden the query system or TIL parser with evaluating specific constraints. By extension, any kinds of values not addressed (e.g., various non-decimal representations of numbers) can simply be passed directly as strings.

\section{Project Structure and Reusability}\label{sec:project_structure}


In order to support organizing information over different files, and to provide further options for configuration, the IR should also feature some form of ``projects''; definitions of which files in a given directory or set of directories belong to one another. This also opens the way for \textit{imports}, not just of namespaces within a file or project, but between different projects altogether.

At this time, the TIL parser does not support using multiple files, does not allow for imports between namespaces, and the only configuration is the input file and output directory through the example application described Section \ref{sec:impl_example}. Additionally, the starting position of relative paths is defined relative to the directory from which the example application is run. Many of these issues can be addressed through a project definition; the project file would define which TIL files are part of the project, what the desired output directory is, and its location could serve as the root location for relative paths (alternatively, the file could make this configurable).

The notion of projects, and that of imports between both projects and namespaces already exists to an extent on the query system; all internable structures even implement a \texttt{MoveDb} trait, which as the name implies allows for them to be moved (or copied) between query system databases, generating new identifiers as needed. This is necessary as some structures will themselves contain identifiers which would otherwise refer to their old database. Imports generally will not need to be reflected in a target language, meaning the query system does not need to reflect any new structures. As such, there are two aspects to define: What information should a ``project'' describe, and how should importing declarations between namespaces and projects behave?

\subsection{Project Properties}

It is important to distinguish which properties are relevant to the TIL \textit{parser}, and which properties are relevant to the query system's notion of a project. As such, the query system's project should contain the following information:

\begin{itemize}
    \item A name for the project.
    \item The location to use as root for any relative paths.
    \item Any configuration relevant to the backend, such as the desired output directory, and potentially backend-specific configuration items.
    \item Potential configuration relevant to the query system, such as how to handle specific type parameters, as discussed in Section \ref{subsec:type_parameters}.
\end{itemize}

In addition to these properties, the \textit{parser}'s project (file) should perform the following functions:

\begin{itemize}
    \item Serve as the root for relative paths, or configure it.
    \item Contain the configurations described above, or point towards configuration files for these properties.
    \item Configure parser-specific properties; e.g., should there be a way to optimize the parse speed by not creating and emitting an error report, this should be configurable.
\end{itemize}

\subsection{Import Behavior}

There are two main questions to answer about how imports should be have:

\begin{enumerate}
    \item How should imported declarations be identified?
    \item Which declarations can be imported?
\end{enumerate}

\stitle{Identifiers} The first question is relevant to situations where multiple identifiers overlap, either between imports, or between an import and the namespace being imported to. The query system simply provides an \texttt{import\_as} function, which allows the name of the import to be set afterwards. This does not specify any constraints as to how they should be named, however. One of, or more likely some combination of, the following methods can be employed:

\begin{enumerate}
    \item Duplicate identifiers should result in an error.
    \item Imports should always be prefixed by their namespace, and further prefixed by their project name if imported from another project. (Note that multiple projects may have different names, however.)
    \item Imports are arbitrarily aliased, determined by whichever means emits TIL.
    \item Imports are arbitrarily prefixed, determined by whichever means emits TIL.
\end{enumerate}

\stitle{Declarations to import} As the use-cases for imports have not been fully realized at this point, future work on the IR or similar projects should consider the following questions:

\begin{itemize}
    \item Should it be possible to selectively import declarations from a namespace?
    \item Would declarations in a namespace benefit from a public/private distinction, or similar?
    \item May certain namespaces be excluded from being imported altogether?
\end{itemize}


\subsection{Notes on Reusability}

While being able to import declarations between projects can aid reusability, this ultimately depends on whether these imports are actually possible from the front-end emitting the IR. A front-end language can also establish its own project structure and import behavior, and simply emit ``flat'' TIL. However, namespaces are designed to be able to reflect any project structure a front-end may utilize, and imports should be designed to do the same. The ideal outcome would be for importing IR projects generated by different front-ends to be feasible, by making it as easy as possible to map between the IR and any structures the front-end may use.


\chapter{Intermediate Representation: Specification}\label{ch:specification}

While the intermediate representation lacks the ability to completely implement behavior, it can nonetheless allow for the specification of behavior through tests.

Unlike the previous chapter, the listings shown in this chapter feature a theoretical, suggested grammar, as work on tests at the query system- and VHDL compiler-level did not advance far enough to warrant implementation of a parser.


\section{High-level Assertions}\label{sec:assertions}

As the IR is used to represent ports consisting of Streams carrying logical types, it is best suited for transaction-level verification. Inputs and outputs should be verified against abstract streams of data, upon which the IR combined with a backend will generate the necessary signalling behaviour and assertions. This enables designers to verify the behaviour of components and correctness of their interfaces without needing to concern themselves with the target language.

There are two key properties to consider when designing and generating tests for Interfaces based on transactions:
\begin{enumerate}
    \item Ports of an Interface are not required to be interdependent or synchronized with one another.
    \item A port's Stream does not necessarily have a single direction, as child Streams can be \textit{Reversed}.
\end{enumerate}

To address these, the recommended testing grammar has the following properties:
\begin{enumerate}
    \item Transaction verification on ports should be assumed to happen in parallel by default, rather than in the sequence assertions are declared. Discussed in Section \ref{subsec:parallelbydefault}.
    \item Rather than explicit \textit{assign} and \textit{compare} methods, the IR should automatically determine whether physical streams are sinks or sources. Discussed in Sections \ref{subsec:assertingequality} and \ref{subsec:issues_explicit}.
\end{enumerate}

These properties by themselves will still not allow high-level assertions to cover every possible use of Tydi interfaces, however. Section \ref{subsec:substitution} describes how to mitigate these limitations.

\subsection{Parallel by Default}\label{subsec:parallelbydefault}

Within a test scope, all statements should be assumed to occur in parallel, rather than being executed sequentially. If the input of one or more streams are required before an output can be produced, this should be implemented  through the \textit{ready} and \textit{valid} signal(s) of the Streamlet. For example, implementing a Streamlet which adds two inputs could be represented as follows, assuming the output ``result'' does not assert \textit{valid} until it has received and added two inputs:

\begin{lstlisting}[basicstyle=\ttfamily]
test test_name {
  adder = adder_def;
  
  adder.result = "010";
  adder.in1 = "01";
  adder.in2 = "01";
}
\end{lstlisting}

As assertions occur in parallel, it is not possible to perform multiple separate assertions on a port in the same scope. However, if the Streamlet being tested also implements proper \textit{ready} signalling and/or buffers to ensure one set of inputs ("in1" and "in2") corresponds to one output, the following can be asserted:

\begin{lstlisting}[basicstyle=\ttfamily]
adder.result = ("010", "001", "011");
adder.in1 = ("01", "01", "10");
adder.in2 = ("01", "00", "01");
\end{lstlisting}

Where \lstinline{("01", "00", "01")} represents a series of \lstinline{Bits(2)} to be transferred over a Stream without dimensionality. This is to be transferred depending on throughput; e.g., one port could support two elements per transfer and require only two transfers, while another might only support one element per transfer and require three. In this proposed syntax, square brackets would be used to indicate dimensionality: \lstinline{[["1", "0"], ["0"]]} represents a Stream with data \lstinline{Bits(1)} and dimensionality 2.

\subsection{Sequences}\label{subsec:sequences}

While transactions on ports are not \textit{necessarily} interdependent, it is reasonable to expect that they will be in many cases regardless. While stateless behavior can be tested in parallel, as each transfer still requires a valid handshake, components which do observe state require that transactions on ports can be asserted in a specific sequence. For example, a counter which accumulates based on input transfers and always drives its output with its current value, or an instruction for a state machine, require that the transfer on the input succeeds before the value on the output is tested.

To this end, the proposed testing grammar also includes \textit{sequences} of explicit stages, each with their own scope; the assertions within each stage still happen in parallel, but each stage must successfully pass before the assertions in the next stage are performed:

\begin{lstlisting}[basicstyle=\ttfamily]
sequence sequence_name {
  initial_state {
    counter.count = "0000";
  }, increment {
    counter.increment = "1";
  }, result_state {
    counter.count = "0001";
  },
};
\end{lstlisting}

In simulation, such stages could be implemented by creating specific flags for each assertion in that scope, then requiring that the normally parallel processes wait until each flag in that stage is set.

For the purposes of tracking their progress and for giving flags descriptive names, sequences \textit{must} have a unique \textit{Name} (in their scope), while individual stages \textit{may} have unique names. Stages are propagated as \textit{Path Names}, using the enclosing scope (the sequence) as their root; if a stage does not have a name, \textit{stage\#} is used instead, with \textit{\#} being the number of that stage (starting from 1, and counting stages which do have names). The flag names of individual assertions within a scope are to be determined by the backend; a descriptive example would be to simply use the name of the instance and port being asserted on, as these are guaranteed to be unique within that scope.

It is possible to specify multiple sequences in a single test, which will occur in parallel. Likewise, nesting sequences is allowed, in which case the nested sequence will occur in parallel with any other assertions in that scope (including other sequences). Note that \textit{any} port asserted on in a sequence is thereby excluded from being asserted on in the sequence's parent scope.

\subsection{Descriptive Errors}\label{subsec:assertion_labels}

For organization purposes and to emit descriptive error messages when tests fail, (optional) labels on assertions, sequences and sequence stages can be added to tests. By specifying intended error messages for the backend, it is possible to better reflect the high-level assertions described in the IR. The proposed syntax for labels/messages on assertions is as follows:

\begin{lstlisting}[basicstyle=\ttfamily]
"overall test label": test test_name {
  ...

  "this is an assertion label": component.port = "1010";
  
  "sequence label": sequence sequence_name {
    "stage label": stage_with_name {
      ...
    },
    "stage label": { // No stage name
      ...
    }
  }
}
\end{lstlisting}

That is to say, labels will consistently use the <string><colon> syntax, behaving similarly to documentation. Test labels may be combined with their parent labels for further clarity; ergo, the query system should provide an ordered list of all parent labels.

In the event no label was supplied for an assertion, it will be up to the backend to generate messages/names if necessary, based on other properties of that assertion. Regardless of the existence of labels, it will be prudent for the query system to also store ancillary information about all assertions, for the backend to optionally use. For instance, if the tests were parsed from a TIL file, line numbers and character spans of assertions and stages can be propagated to the backend to provide more descriptive errors, even when labels are provided. By extension, it is also possible for a frontend emitting the IR to use labels to propagate such (line number and character span) information from its own test definitions.


\subsection{Asserting Equality}\label{subsec:assertingequality}


The IR should automatically determine whether physical streams are sinks or sources, rather than requiring explicit language to drive or compare a signal. The latter property means that something closer to mathematical equality is implemented; ``the transaction on port \textit{a} is equal to data \textit{x}'', whereupon it is automatically determined whether \textit{x} should be driven, or observed and compared.

While this has very little effect on the examples in the previous section, other than potentially removing unnecessary keywords or operators:

\begin{lstlisting}[basicstyle=\ttfamily]
adder.result == ("010", "001", "011");
adder.in1 = ("01", "01", "10");
adder.in2 = ("01", "00", "01");
// OR:
assert adder.result = ("010", "001", "011");
act adder.in1 = ("01", "01", "10");
act adder.in2 = ("01", "00", "01");
\end{lstlisting}

It greatly simplifies assertions on nested and reversed child streams. For example, we can use the same adder concept described before, but combine its ports into a single Stream and port with a Reversed child Stream to indicate a response:

\begin{lstlisting}[basicstyle=\ttfamily]
add: in Stream(
  data: Group(
    in1: Stream(
      data: Bits(2),
      direction: Forward,
      ...
    ),
    in2: Stream(
      data: Bits(2),
      direction: Forward,
      ...
    ),
    result: Stream(
      data: Bits(3),
      direction: Reverse,
      ...
    ),
  ),
  direction: Forward,
  ...,
  keep: false,
)
\end{lstlisting}

Subsequently, the assertion can be represented as follows:

\begin{lstlisting}[basicstyle=\ttfamily,caption={Representing the \textit{in} and \textit{result} Streams as independent.},label={lst:group_independent}]
adder.add = {
  in1: ("01", "01", "10"),
  in2: ("01", "00", "01"),
  result: ("010", "001", "011"),
};
\end{lstlisting}

Or as follows, to emphasize the relation between transfers on the Streams:

\begin{lstlisting}[basicstyle=\ttfamily,caption={Representing the \textit{in} and \textit{result} Streams as interdependent.},label={lst:group_sync}]
adder.add = ({
  in1: "01",
  in2: "01",
  result: "010",
}, {
  in1: "01",
  in2: "00",
  result: "001",
}, {
  in1: "10",
  in2: "01",
  result: "011",
});
\end{lstlisting}

In the above examples, the parent ``add'' stream will have been flattened into ``in1'' and ``in2'', but this physical implementation detail has little bearing on the assertion itself. Furthermore, if we decide to turn ``in1'' and ``in2'' into simple \textit{Bits(2)} fields rather than \textit{Stream}s, the assertion in Listing \ref{lst:group_sync} will remain unchanged.

It is worth noting that Listing \ref{lst:group_independent} and Listing \ref{lst:group_sync} should produce identical results, with independent, parallel transfers on the \textit{in} and \textit{result} Streams attempting to match the specified throughput (and physical element lanes). Synchronicity across multiple child Streams relative to a parent Stream at an element level can only be respected if the parent Stream has dimensionality or transfers elements itself. This is why the parent ``add'' Stream can be flattened into its child ``in'' Streams.

It is of course still possible for a Streamlet to be implemented such that it enforces synchronicity through ready-valid signals on the child Streams, or by setting \textit{keep = true} on the parent Stream. Conversely, it is allowed for the \textit{adder} example to have a higher throughput on the ``result'' Stream, and buffer results (or inputs) to provide them all as multiple elements in a single transfer.

An example of a naturally synchronous set of Streams without the parent having dimensionality is a \textit{Union} of Streams, in which the parent Stream transfers the \textit{tag} (indicating the active field) over its data signal. This can be asserted as follows, and should result in sequential assertions on the child Streams, synchronized to transfers on the parent Stream:

\begin{lstlisting}[basicstyle=\ttfamily,caption={Asserting on a Stream carrying a Union with an alternating input (request) and output (response) Stream. The parent (query) Stream controls the \textit{tag} determining which Stream should be active.},label={lst:union_assert}]
storage.query = ({
  request: ["0001", "0100"], // derive tag value based on the field name
}, {
  response: ["11001010", "10010000", "00110001", ...],
}, {
  request: ...
}, ...)
\end{lstlisting}

\subsection{Issues with Explicit Assignment and Comparison}\label{subsec:issues_explicit}

Conversely, consider a way to represent the initial scenario from Listing \ref{lst:group_independent} using \textit{=} and \textit{==} operators:

\begin{minipage}{0.95\textwidth}
  \centering
  \begin{minipage}[t]{.45\textwidth}
    \centering
\begin{lstlisting}[basicstyle=\ttfamily\small]
adder.add = {
  in1: ("01", "01", "10"),
  in2: ("01", "00", "01"),
  result: ("010", "001", "011"),
};
\end{lstlisting}
  \end{minipage}
  \begin{minipage}[t]{.45\textwidth}
    \centering
\begin{lstlisting}[basicstyle=\ttfamily\small]
adder.add = {
  in1 = ("01", "01", "10"),
  in2 = ("01", "00", "01"),
  result == ("010", "001", "011"),
};
\end{lstlisting}
  \end{minipage}
  \captionsetup{type=lstlisting}
  \caption{Two approaches to using explicit assign and compare operators.}
  \label{lst:assert_compare_approaches}
\end{minipage}

The left example of Listing \ref{lst:assert_compare_approaches} is clearly incorrect, as it is using a ``assign'' operator, but actually ``comparing'' \textit{result}. The right example is more subtly wrong, however: The parent ``add'' Stream does not actually exist, making the initial \texttt{add =} statement irrelevant. If we were to make ``add'' an \textit{out} port, and reverse \textit{in1} and \textit{in2} instead of \textit{result}, we may instead assert it as follows:

\begin{lstlisting}[basicstyle=\ttfamily]
adder.add == {
  in1 = ("01", "01", "10"),
  in2 = ("01", "00", "01"),
  result == ("010", "001", "011"),
};
\end{lstlisting}

However, the physical implementation of either design is identical, as is the actual assertion. The outer \texttt{add <operator>} has no bearing on the inner scope; it requires work to track the direction of the parent Stream to no practical benefit. So alternatively, we may split up all statements as follows:

\begin{lstlisting}[basicstyle=\ttfamily]
adder.add.in1 = ("01", "01", "10");
adder.add.in2 = ("01", "00", "01");
adder.add.result == ("010", "001", "011");
\end{lstlisting}

This requires knowledge of which Streams are converted to physical streams, however. Moreover this would not work when the parent stream does exist (as the Group may contain element types in addition to Streams), and could not be used to perform the Union assertion in Listing \ref{lst:union_assert}.

Finally, in order for compilers to emit (or designers to write) these statements in the IR, they will need to track the specific direction of each physical stream, based on the logical types and interfaces. As this process is necessary regardless, it is more effective to offload it to the query system (or compiler) consuming the IR, which already concerns itself with converting Interfaces and logical Streams to physical streams. This also ensures the front-end compilers (or designers) will only need to concern themselves with the abstract, logical definitions of transactions, rather than their exact physical implementation.

\section{Proof of Concept}\label{sec:spec_poc}


\subsection{Physical Transfers}

Work on high-level assertions had begun as part of this thesis, though they could not be implemented in full. Assertions were developed bottom-up, in that the focus was on the results to be emitted by the VHDL backend, with the intent of raising the level abstractions to match the proposed syntax from there.

Specifically, there is support within the query system and the VHDL backend for assertions on \textit{physical streams}, with a \texttt{PhysicalTransfers} trait being able to arbitrarily switch between driving signals and asserting them against a given transfer. The \texttt{PhysicalTransfers} trait is implemented automatically by any object implementing the \texttt{PhysicalSignals} trait, which features methods for returning the direction of a physical stream, and for automatically driving or comparing the signals (\textit{data}, \textit{endi}, etc.) based on this direction.

Combined with a set of \texttt{handshake} (driving ready or valid, and waiting for and/or asserting the inverse) methods, a \textit{sequence} of elements can be easily transferred over multiple cycles. To further allow for the verification of various timing constraints, the handshake signals can be driven to either be held high (resulting in a transfer over consecutive cycles), or to be driven low after a cycle. The \texttt{PhysicalTransfers} trait reflects this by having separate \texttt{open\_transfer} and \texttt{close\_transfer} methods, along with a ``test\_staggered'' parameter on \texttt{transfer}.

The transfers themselves are called \texttt{PhysicalTransfer}, featuring properties for how the transfer should behave derived from the physical stream being driven, such as which lanes may be inactive, and whether the transfer needs to occur over consecutive cycles. The contents of the transfer are taken from a more free-form \texttt{LogicalTransfer}, which is either explicitly an ``empty sequence'', or an iterator of simple \textit{elements}.

Each logical element contains an optional \textit{data} field which is either Null, Bits, or a Group or Union of further \textit{element data}, it also contains an optional \textit{last} property to indicate that the element represents the end of one or more dimensions in a sequence. If \textit{last} is not set, it simply means the \textit{last} signal should not be driven. However, if \textit{data} is not set, that indicates that the data lane itself is inactive.

When a \texttt{LogicalTransfer} is set on a \texttt{PhysicalTransfer}, it is verified whether this transfer is possible given the physical stream's constraints. For instance, it will fail if the elements contain data types which do not match the physical stream, or if it attempts to drive \textit{last} for multiple elements despite the physical stream's complexity being $C<8$.

\subsection{Demonstration}

The results of this work are demonstrated by \texttt{process\_tests}\footnote{\url{https://github.com/matthijsr/til-vhdl/blob/main/crates/til_vhdl/tests/process_tests.rs}}, as such:

Taking the following ``physical transfers'', representing the sequence: \texttt{[ [ [ 11, -, 11, 10 ], [ 01, 00, 10 ] ], - ], -} (With \texttt{-} representing inactive lanes.)

\begin{lstlisting}[basicstyle=\ttfamily\scriptsize,language=scala]
let transfer_1 =
    PhysicalTransfer::new(Complexity::new_major(8), Positive::new(3).unwrap(), 2, 3, 3)
        .with_logical_transfer(([Some("11"), None, Some("11")], "101"))?; // [[[11, -, 11
let transfer_2 =
    PhysicalTransfer::new(Complexity::new_major(8), Positive::new(3).unwrap(), 2, 3, 3)
        .with_logical_transfer([("01", Some(0..0)), ("10", None), ("00", None)])?; // 10], [01, 00
let transfer_3 =
    PhysicalTransfer::new(Complexity::new_major(8), Positive::new(3).unwrap(), 2, 3, 3)
        .with_logical_transfer([("01", Some(0..1)), ("-", Some(2..2)), ("-", None)])?; // 10]], -], -
\end{lstlisting}

An input physical stream can be addressed as follows:
\begin{lstlisting}[basicstyle=\ttfamily\scriptsize,language=scala]
drive_stream.open_transfer()?;
drive_stream.transfer(transfer_1.clone(), false, "test message drive 1")?;
drive_stream.transfer(transfer_2.clone(), false, "test message drive 2")?;
drive_stream.transfer(transfer_3.clone(), false, "test message drive 3")?;
drive_stream.close_transfer()?;
\end{lstlisting}

Producing the following VHDL, driving its ports with the correct values and timings:

\begin{lstlisting}[basicstyle=\ttfamily\scriptsize,language=vhdl]
process is
begin
  a__x_valid <= '1';
  a__x_data(1 downto 0) <= "11";
  a__x_data(5 downto 4) <= "11";
  a__x_last(2 downto 0) <= (others => '0');
  a__x_last(5 downto 3) <= (others => '0');
  a__x_last(8 downto 6) <= (others => '0');
  a__x_strb <= "101";
  a__x_user(2 downto 0) <= "101";
  wait until rising_edge(clk) and a__x_ready = '1';
  a__x_data(1 downto 0) <= "10";
  a__x_data(3 downto 2) <= "01";
  a__x_data(5 downto 4) <= "00";
  a__x_last(2 downto 0) <= "001";
  a__x_last(5 downto 3) <= (others => '0');
  a__x_last(8 downto 6) <= (others => '0');
  a__x_strb <= "111";
  a__x_stai <= std_logic_vector(to_unsigned(0, 2));
  a__x_endi <= std_logic_vector(to_unsigned(2, 2));
  wait until rising_edge(clk) and a__x_ready = '1';
  a__x_data(1 downto 0) <= "10";
  a__x_last(2 downto 0) <= "011";
  a__x_last(5 downto 3) <= "100";
  a__x_last(8 downto 6) <= (others => '0');
  a__x_strb <= "100";
  wait until rising_edge(clk) and a__x_ready = '1';
  a__x_valid <= '0';
  wait until rising_edge(clk);
end process a__x;
\end{lstlisting}

While an output physical stream is addressed in the same way:
\begin{lstlisting}[basicstyle=\ttfamily\scriptsize,language=scala]
compare_stream.open_transfer()?;
compare_stream.transfer(transfer_1.clone(), false, "test message compare 1")?;
compare_stream.transfer(transfer_2.clone(), false, "test message compare 2")?;
compare_stream.transfer(transfer_3.clone(), false, "test message compare 3")?;
compare_stream.close_transfer()?;
\end{lstlisting}

But is automatically converted to comparisons in VHDL:

\begin{lstlisting}[basicstyle=\ttfamily\scriptsize,language=vhdl]
process is
begin
  wait until rising_edge(clk) and a__y_valid = '1';
  assert a__y_data(1 downto 0) = "11" report "test message compare 1";
  assert a__y_data(5 downto 4) = "11" report "test message compare 1";
  assert a__y_last(2 downto 0) = (others => '0') report "test message compare 1";
  assert a__y_last(5 downto 3) = (others => '0') report "test message compare 1";
  assert a__y_last(8 downto 6) = (others => '0') report "test message compare 1";
  assert a__y_strb = "101" report "test message compare 1";
  assert a__y_user(2 downto 0) = "101" report "test message compare 1";
  a__y_ready <= '1';
  wait until rising_edge(clk) and a__y_valid = '1';
  assert a__y_data(1 downto 0) = "10" report "test message compare 2";
  assert a__y_data(3 downto 2) = "01" report "test message compare 2";
  assert a__y_data(5 downto 4) = "00" report "test message compare 2";
  assert a__y_last(2 downto 0) = "001" report "test message compare 2";
  assert a__y_last(5 downto 3) = (others => '0') report "test message compare 2";
  assert a__y_last(8 downto 6) = (others => '0') report "test message compare 2";
  assert a__y_strb = "111" report "test message compare 2";
  assert a__y_stai = std_logic_vector(to_unsigned(0, 2)) report "test message compare 2";
  assert a__y_endi = std_logic_vector(to_unsigned(2, 2)) report "test message compare 2";
  a__y_ready <= '1';
  wait until rising_edge(clk) and a__y_valid = '1';
  assert a__y_data(1 downto 0) = "10" report "test message compare 3";
  assert a__y_last(2 downto 0) = "011" report "test message compare 3";
  assert a__y_last(5 downto 3) = "100" report "test message compare 3";
  assert a__y_last(8 downto 6) = (others => '0') report "test message compare 3";
  assert a__y_strb = "100" report "test message compare 3";
  a__y_ready <= '1';
  wait until rising_edge(clk) and a__y_valid = '1';
  a__y_ready <= '0';
  wait until rising_edge(clk);
end process a__y;
\end{lstlisting}

\subsection{Results and Future Work}\label{subsec:assertions_future}

While work did not complete within the span of the thesis, this proof of concept does demonstrate that high-level assertions such as those described in Section \ref{sec:assertions} are possible. Notably, even this reduced version greatly improves the ergonomics of performing transfer-level assertions, as they require fewer lines of code (scaffolding around the query system required for the test notwithstanding) and better represent the high-level intentions.

To expand on this, the next steps would be to:
\begin{enumerate}
    \item Use the query system to convert an arbitrary (but type-appropriate) sequence into multiple transfers automatically.
    \item Use the query system to convert assertions on logical Streams such as those in \ref{sec:assertions} into multiple assertions on the corresponding physical streams.
\end{enumerate}

Note that these steps do not require further input on the backend; provided the backend implements the requisite \texttt{PhysicalTransfers} trait, all further logic can be implemented for \textit{all} possible backends on the query system itself. And when this is successful, one may create a minimal grammar and parser, to reduce the amount of scaffolding when testing these functions.



\section{Complex Test Cases}\label{subsec:substitution}

\subsection{Limitations of High-Level Assertions}\label{subsec:assertion_limitations}


Of course, not all behavior can be tested through transfer-based, high-level assertions. There are a few specific properties which make a Streamlet difficult to test, or make a test scenario difficult to implement:

\begin{itemize}
    \item \textbf{A \textit{user} signal} --- The Tydi specification allows for Streams to have a \textit{user} signal, which exists specifically to address use-cases not covered by Tydi's transfer specification. There are very few constraints on the user signal, other than it being a \textit{signal}, and not its own (physical) stream. The user signal can be driven independently from transfers and clock cycles, what constraints there are to its behavior are entirely determined by the designer, and so cannot be translated to the assertion system described in the previous sections. Hence, the user signal will be omitted from the assertion system, and designers of highly custom interfaces should implement tests manually.
    \item \textbf{Testing large ranges of inputs} --- The assertions described before use \textit{constants}, making more exhaustive testing difficult to implement. For example, when testing a 8-bit adder, one would expect it to work for any combination of inputs in that 8-bit range. This \textit{can} be implemented through an exhaustive series of assertions, but is better served by a (random) number generator, or another external source of inputs and expected outputs.
    \item \textbf{Testing against randomness} --- In the same vein as the previous limitation, the use of constants in assertions makes accounting for Streamlets which itself produces random outputs more difficult. The most simple example of this would be the Streamlet itself being a random number generator, with the test attempting to assert that its output is sufficiently statistically random. (A more complex scenario could involve a Streamlet using randomness for the purposes of encryption.)
    \item \textbf{Unimplemented dependencies} --- If a design is made up of multiple Streamlets, assertions can only be performed against the completed product and against its individual components. If one or more pieces are not yet implemented, the tests cannot succeed.
    \item \textbf{Verifying the use of dependencies} --- On a more abstract level, it is not possible to verify that a composite design actually employs its intended dependencies. E.g., does the ``encryption'' Streamlet actually use the verified ``random number generator'' Streamlet, or does it implement its own (potentially incorrect) random number generation?
    \item \textbf{Creating predictable dependencies} --- Conversely, a dependency may have been implemented, but not be conducive to testing. In the earlier example of an ``encryption'' Streamlet depending on a random number generator, it may be desirable to test to the Streamlet with non-random numbers for more predictable assertions. In effect, the goal is to isolate only one Streamlet's functionality.
\end{itemize}



\subsection{Using Test Streamlets for Verification}

In scenarios where high-level assertions cannot serve as a useful source of inputs, and/or cannot properly verify outputs, it is still possible to use the IR's ability to link and compose Streamlets to instead create ``test Streamlets'', and connect these to the subject's input and/or output.

For instance, in the previously described scenario of wanting to test a range of inputs against an adder, it is possible to create a Streamlet which generates numbers and verifies outputs (either against a known-good adder, or by drawing from an external source), and connect it to the adder. The same principle can be used to create a Streamlet which tests a random number generator's randomness, or one which drives the user signal.

These applications are obviously not very different from manually creating a testbench in the target language, but they ensure that the tests remain organized through the IR, and make it easier to reuse certain solutions. Additionally, these ``test Streamlets'' can be combined with high-level assertions for mixed test scenarios, or to use the assertions as configuration or verification on the test Streamlet.

Such ``test Streamlets'' can already be implemented in the IR through simple namespaces, but should be declared in tests or test files instead for better organization of both the IR and the backend's output, e.g:

\begin{lstlisting}[basicstyle=\ttfamily\scriptsize]
streamlet general_test_streamlet = <definition>;

test test_name {
  streamlet very_specific_test_streamlet = <definition>;
  streamlet test_tld = (correct: out test_result) {
    impl: {
      tester = very_specific_test_streamlet;
      subject = actual_streamlet;
      
      subject.input -- tester.output;
      subject.output -- tester.input;
      tester.correct -- correct;
    }
  };
  
  test_subject = test_tld;
  
  test_subject.correct = "1";
}
\end{lstlisting}

\subsection{Substitution}

In order to isolate a composite Streamlet's functionality from its dependencies, it will be helpful for the IR to provide some way of redefining or \textit{substituting} Streamlet definitions. The test Streamlets in the previous sections are helpful when the dependencies are internal, or part of a structural implementation; when Streamlets are embedded in a behavioral implementation, there are no such options.

Provided the behavioral implementation depends on Streamlets tracked by and generated from the IR, it should be possible to redefine their implementation when creating the test, or test project. E.g., in VHDL, it would include different architecture definitions in the workspace of the testbench.

This can be used for the following purposes:
\begin{itemize}
    \item Performing proper ``unit tests'', by removing all other dependencies from a test of a composite Streamlet and replacing them with more predictable, simpler Streamlets.
    \item ``Stubbing'' an unimplemented dependency; even if the purpose is not to perform a unit test, it may help to temporarily substitute a dependency, to verify a larger design.
    \item Simulating a dependency which cannot otherwise (efficiently) run in software; e.g., if a dependency would normally draw data from a hardware component (such as memory, device storage, or a sensor), replacing it with something which reads or produces data in software enables the overall design to be tested.
    \item Verifying that a dependency is being used; e.g., by substituting the random number generator Streamlet, it is possible to assert that it transferring a specific number is actually propagated to the output of the dependent Streamlet.
\end{itemize}

The syntax for this functionality can be fairly simple, e.g.:

\begin{lstlisting}[basicstyle=\ttfamily\scriptsize]
// substitute(<streamlet_declaration>, <replacement implementation>)
substitute(streamlet_decl, "/test/path/");
substitute(streamlet_decl, !an_intrinsic);
substitute(streamlet_decl, { input--output; });
\end{lstlisting}

\textit{Where} such substitution statements are allowed is to be determined based on actual implementation. As it will depend on the complexity of performing a substitution across different target languages. Should it prove very simple in most relevant languages, it can be performed per test - otherwise, it may make more sense per test file, or at the level of the entire (test) project.

\section{Setting up Subjects}\label{sec:setup}


The last property to address when creating tests is appropriate setup steps for a subject Streamlet. Even when considering tests are simulated, it is still likely that a test may involve a reset procedure before or during assertions; a designer may also want to verify reset behavior itself.

Additionally, Streamlets will have one or more domains consisting of a clock and reset signal each, which may be different. The Tydi specification does not place constraints on the reset signal, other than requiring that the \textit{ready} and \textit{valid} signals are released during a reset. As such, the reset signal can have any sensitivity and synchronicity, and a reset may take any number of cycles.

As such, reset/setup syntax needs to minimally account for the following properties:
\begin{itemize}
    \item Whether the reset signal is sensitive on a positive or negative edge.
    \item Whether the reset signal needs to be held for a certain number of cycles.
    \item When accounting for multiple domains, whether each domain's reset signal behaves differently, and whether they can be reset simultaneously.
\end{itemize}

Specific configuration of clock signals and frequency is less important, but may prove helpful when testing components which explicitly require different clock speeds to operate correctly. Initially, clock behavior is best left to annotations (described in Section \ref{subsec:annotations}), as the actual implementation of a clock may differ between target languages.

How to specifically represent these properties has not been fully evaluated, and more considerations may arise from actual implementation. However, in the absence of such an implementation and based on the previously established properties, the following syntax can be proposed:

\begin{lstlisting}[basicstyle=\ttfamily\scriptsize,caption={``Arranging'' subject streamlets, by driving their reset signals.},label={lst:setup}]
domain 'a { reset: low };
domain 'b { reset: low };
domain 'c { reset: high };
domain 'd { reset: high };

subject1 = streamlet_def_name1<'a, 'b>;
subject2 = streamlet_def_name2<'c>;
subject3 = streamlet_def_name2<'d>;

process arrange1 = sequence arrange1_sequence {
  {
    reset('a, 3);
  }, {
    wait('a, 1);
  }, {
    reset('b, 1);
    reset('c, 1);
  }
};

process arrange2 = {
  reset('d, 1);
};

"sequence label": sequence sequence_name {
  "initial setup": {
    arrange1; // Note that nested sequences are allowed.
    arrange2; // arrange2 will occur in parallel to all of arrange1
  }, "test stage": {
    // Perform some assertions
  }, "partial reset": {
    arrange_name2; // The contents of arrange_name2 are inserted into this scope.
    reset('a, 3);
  }, ...
}
\end{lstlisting}

To elaborate on this syntax, and address some properties not (clearly) included in this example:

\begin{itemize}
    \item \lstinline{domain}s are defined with \lstinline{reset: low} or \lstinline{reset: high}, indicating whether the reset signal needs to be held low or high. Other properties may be added if they are useful and generally applicable in defining domains (e.g., (relative) clock speed).
    \item \lstinline{process}es are stored sequences or operations, to be easily reused. These are useful for arranging subjects, but may also be used for other kinds of test organization.
    \item \lstinline{reset(<1>, <2>)} is a pre-defined process with two parameters: \begin{enumerate}
        \item Which \textit{reset signal (domain)} to drive. The reset signal will be inverted from its default state.
        \item How many cycles to \textit{hold} the reset signal. (Currently assumed to be an integer $\geq1$.)
    \end{enumerate}
    \item \lstinline{wait(<1>, <2>)} is a pre-defined process, with two parameters: \begin{enumerate}
        \item Which \textit{clock signal (domain)} to wait relative to.
        \item How many cycles to \textit{wait} for. (Currently assumed to be an integer $\geq1$.)
    \end{enumerate}
    \item \lstinline{reset(...)} and \lstinline{wait(...)} may only be used in the context of a sequence. If a domain is being reset or waited on, any subject streamlets which depend on that domain cannot be asserted on. Likewise, resets and waits can be applied in parallel with other resets and waits (e.g., waiting on two domains at the same time, moving to the next stage if both have passed), but cannot be applied to the same domain in the same scope.
    \item As in structural implementations, Streamlet definitions with only the \textit{default} domain can still be assigned a domain on instantiation. Tests require that domains are explicitly declared.
\end{itemize}



\chapter{Implementation}\label{ch:implementation}


As discussed in the Methodology Section (\ref{sec:methodology}), in order to demonstrate the intermediate representation's capabilities and evaluate various approaches, a prototype toolchain was implemented\footnote{\url{https://github.com/matthijsr/til-vhdl}} over the course of the thesis. This toolchain consists of a query system for storing and retrieving the IR's declarations and expressions on-demand, a preliminary grammar and parser which stores its results in the query system, and a backend which uses the query system and emits VHDL.

\section{Query System}\label{sec:query_system}

The first component of the prototype toolchain is the query system for storing and computing information of the IR. The decision to use a query system rather than more traditional passes of compilation was inspired by ongoing work on the Rust compiler \cite{rustcompilerteamqueries2021} and implemented using the Salsa framework \cite{salsa-rssalsa2022}. The advantage of such a system is that information can be retrieved or computed on-demand, and the results of previously executed queries are automatically stored, and only re-computed when their dependencies change.



The query system currently performs the following tasks:

\begin{itemize}
    \item \textbf{Storing information} --- The query system stores types, Interfaces, Streamlets and Implementations. The query system also tracks Namespaces, Projects, and the declarations therein, but those declarations are ultimately stored as identifiers of the query system's database.
    \item \textbf{Validation} --- The query system is responsible for validating definitions against the Tydi specification, well before a backend is able to extract any information. For example: \begin{itemize}
        \item \textit{Name}s used as identifiers for ports, Streamlets and declarations must be formatted correctly, according to the Tydi specification (as mentioned in Section \ref{sec:background_tydi}). Likewise, it ensures that identifiers are unique where relevant (such as in ports of Interfaces).
        \item All ports of an Interface with explicitly named \textit{domain}s must be assigned a domain.
        \item \textit{Link}s to behavior must be correctly formatted paths.
        \item Streamlet \textit{instances} in a structural implementation are correctly assigned their domains, and assigned default domains if the parent Streamlet has a default domain.
        \item \textit{Connection}s between ports of interfaces must have compatible directions, types, and domains, as explained in Section \ref{subsec:interface_compat}.
        \item Once a structural implementation is being stored, all ports of all instances and the parent Streamlet must have been connected.
    \end{itemize}
    \item \textbf{All Streamlets} --- Regardless of how they are organized in namespaces, the query system is able to retrieve all declared Streamlets and automatically set appropriate \textit{Path Name}s based on the namespace they were a part of.
    \item \textbf{Physical streams} --- In order to represent the logical Stream definitions used for ports in hardware, they must be converted to physical streams, as described in Section \ref{subsec:physical_streams}. The query system performs this conversion, and also tracks which logical types the physical streams themselves originally related to.
\end{itemize}

Another use-case for the query system is the high-level assertions described in Section \ref{sec:assertions}; converting abstract streams of data on a logical Stream into appropriate, generic calls to the signals that make up its physical streams. Through these functions, a backend would only need to implement the methods for addressing physical streams in order to support these complex, abstract assertions.

While these are still a work in progress, Section \ref{sec:spec_poc} showcases how the query system is already capable of taking abstract transfers of data structures and converting them into appropriate addressing of a physical stream. As explained in Section \ref{subsec:assertions_future}, the query system can take on the bulk of the work implementing high-level assertions once a backend implements the specifics for addressing individual physical streams.

\section{Grammar and Parser}\label{sec:grammar_parser}

While the query system is effectively an implementation of the IR in its own right, text-based representations are more portable and can allow for more flexible expressions. Furthermore, a purpose-built language reduces the amount of scaffolding required when testing complete projects in the IR, as compared to setting up the query system manually.

To this end, the prototype toolchain also features a simple grammar (referred to as Tydi Intermediate Language, or \textit{TIL}) and parser, implemented using Chumsky \cite{barrettochumsky2022}. Using the parser, a project expressed in TIL can be stored in the query system. TIL also served as a more stable target for a front-end, computation-oriented language (called Tydi-lang) which was being developed in parallel with the IR by Yongding Tian, as mentioned in Section \ref{sec:problem_statement}.

\subsection{Parsers}


Before designing a grammar, it was necessary to determine which libraries were available to parse the intermediate language. The initial requirements for such libraries were relatively simple:

\begin{enumerate}
    \item The library needs to target Rust, as this is what the query system was written in. Adding an interfacing pass between another language and Rust would not be a productive use of time.
    \item The parsing method needs to support lexing (tokenization) and (integrate with) some form of evaluation, in addition to conventional syntax parsing (producing an abstract syntax tree). The goal is to avoid needing to rely on multiple different parser libraries.
    \item Defining a grammar in the parser needs to be well-documented, ideally with examples provided as part of the documentation, or through other users' projects. As building a parser and defining a grammar is not the primary goal of this thesis, it should not require too much time.
\end{enumerate}

Based on these requirements, \textit{crates.io}'s list of most downloaded ``grammar'' crates\footnote{\url{https://crates.io/keywords/grammar?sort=downloads}} and a cursory search through \textit{Y Combinator} and the Rust subreddit (e.g., \cite{python2019, rodyamirovwhat2020, zestererchumsky2021}), it was possible to produce a shortlist of potential candidates based on other users' experiences. This shortlist is included here as Table \ref{tab:parsers}.

\begin{table}[h]
    \centering
    \begin{tabular}{l|l}
\textbf{Name} & \textbf{Kind}                                                              \\ \hline
lalrpop \cite{lalrpop2022}       & LR(1) (\textbf{L}eft-to-right, \textbf{R}ightmost derivation in reverse, \textbf{1} lookahead symbol) \\
lrpar \cite{grammar2022}         & LR(1)                                                                      \\
nom \cite{couprienom2022}           & combinator                                                                 \\
chumsky \cite{barrettochumsky2022}       & combinator                                                                 \\
rust-peg \cite{mehallparsing2022}      & PEG (Parsing Expression Grammar)                                           \\
pest \cite{pest2022}         & PEG                                                                       
\end{tabular}
    \caption{A list of parser libraries evaluated for this project}
    \label{tab:parsers}
\end{table}

Of these, \textit{nom} and \textit{Chumsky} directly provide (macro) functions to build and combine parsers in Rust. \textit{LALRPOP}, \textit{lrpar} and \textit{rust-peg} all allow users to define grammars in a separate syntax, which is subsequently converted into Rust code to be referenced; \textit{lrpar} is somewhat notable for using the existing \textit{Yacc} syntax, rather than a custom one. \textit{pest} also relies on an external grammar definition, but only exposes pre-defined functions, rather than generated ones to be imported.

In order to quickly determine which parser library was best suited for quickly defining a grammar and parser, each library's documentation (and possible examples) were followed to the point it was possible to define and evaluate a simple programming language. Of these, \textit{lalrpop}, \textit{Chumsky} and \textit{nom}'s documentation was easiest to follow, featuring clear tutorials and ample examples, though \textit{nom}'s largely stopped short of parsing programming languages. \textit{Chumsky}'s separate example parsers were not immediately functional, due to the methods they relied on receiving breaking changes in the interim, but were relatively easy to repair after finishing the tutorial and using the IDE to automatically suggest changes. The fact that \textit{Chumsky} directly employs Rust functions was ultimately what resulted in a decision in its favor, as this is what allowed it to integrate with the IDE's (Visual Studio Code with \textit{rust-analyzer}) existing analysis and suggestion capabilities.

This direct integration with Rust also meant that a parser built in Chumsky could directly use existing types and functionality built for the query system, such as using the \texttt{Name} constructor to determine and store valid identifiers, or simple use-cases such as re-using the existing \textit{Direction} enumeration for ``Forward'' and ``Reverse''. This benefit extends to the evaluation pass, which amounts to more Rust functions to interpret the abstract syntax tree; this meant that rather than building custom functionality for tracking identifiers and validating statements, it was possible to directly store declarations in the query system, and rely on its errors and validation.

\begin{figure}
    \centering
    \includegraphics[width=0.98\linewidth]{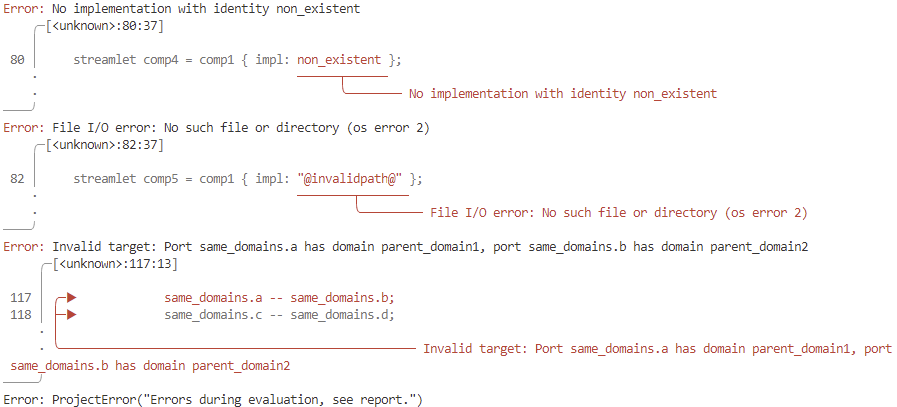}
    \caption{Errors from the evaluation pass, rendered in the terminal by \textit{Ariadne}.}
    \label{fig:Ariadne_error}
\end{figure}

A final, interesting but non-essential quality of Chumsky is that its errors (using \textit{spans} of character positions) are easily rendered by its sister project, \textit{Ariadne} \cite{barrettoariadne2022}. \textit{Ariadne} allows a compiler/parser to emit labelled and color-coded errors to the terminal, not unlike Rust's own \textit{rustc} compiler. The quality of these errors will ultimately depend on the implementation, but chumksy's error recovery strategies and reporting is quite flexible. Figure \ref{fig:Ariadne_error} illustrates how errors emitted by the parser are human-readable, how Chumsky's error recovery allows for multiple errors to be detected and reported within the same file, and how spans can be propagated even to the evaluation pass.

Theoretically, it would be possible to use the spans and abstract syntax tree emitted by the parsers built in Chumsky to build a \textit{language server} for an IDE (specifically Visual Studio Code's Language Server Protocol), to provide support for in-line color-coding of syntax, linting of errors, and other analysis. There is no straightforward path to building such a service, however, and so no effort was made towards it beyond an initial cursory exploration of the possibilities.


\subsection{Grammar}


TIL features expressions for declaring namespaces, types, Interfaces, Streamlets and Streamlet implementations, as well as some syntax sugar for subsetting Streamlets into interfaces. This grammar has been fully implemented in the prototype toolchain, in that it can also be emitted to VHDL using the backend described in the next subsection.

\textbf{Namespaces} are simple containers for other declarations, their only innate property is their name, which can be expressed as a \textit{path}. Note that paths in this context are purely abstract, and do not reflect any hierarchy in the grammar or IR itself, they can simply be used to \textit{communicate} hierarchy to a backend, and/or propagate it from a front-end.

\includegraphics[width=0.5\linewidth]{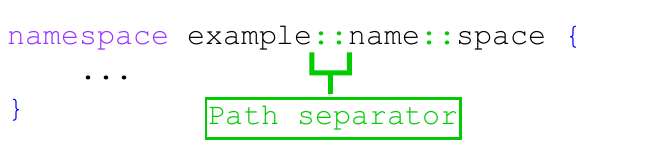}

The \textbf{types} described in Section \ref{sec:background_tydi} can be declared using the \textit{type} keyword, an identifier, and an expression. Type expressions either reference these identifiers, or directly describe the type's properties.

\includegraphics[width=0.6\linewidth]{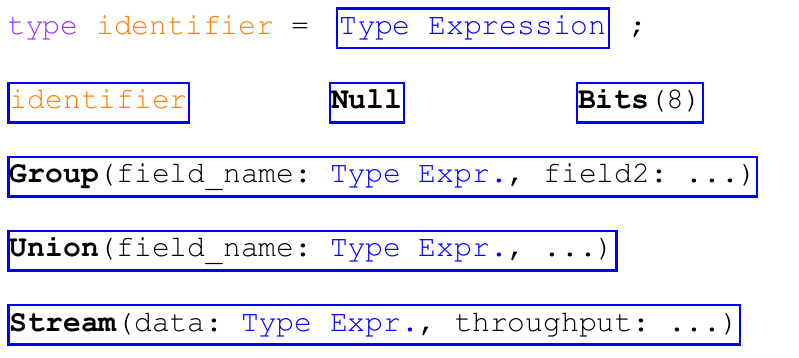}

\textbf{Interfaces}, as described in Section \ref{subsec:interfaces} are collections of ports and (clock and reset) domains. They can be separately declared with an identifier, to enable reuse.

\includegraphics[width=0.8\linewidth]{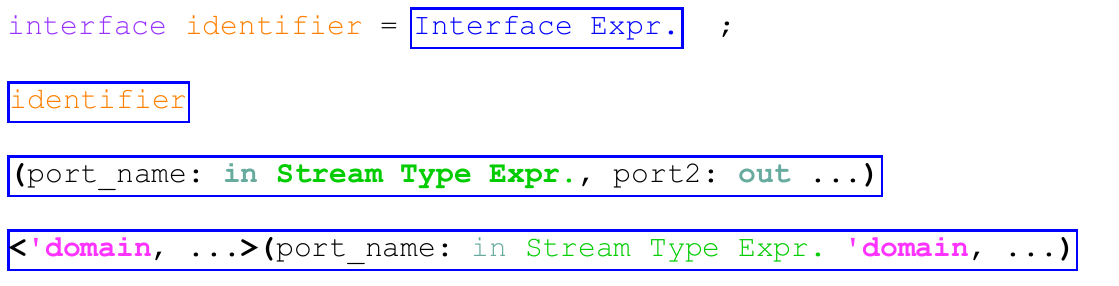}

There are two kinds of \textbf{implementations}, \textit{links} to behavior, and \textit{structural} implementations which connect Streamlets declared in the IR. This is elaborated on in Section \ref{sec:component_implementation}. Links simply use double-quotes to enclose a path to a directory, while structural implementations are scopes with two kinds of statements: One to create a Streamlet \textit{instance} and connect the Interface's domains, and another to connect ports between instances and/or the enclosing Streamlet.

\includegraphics[width=0.8\linewidth]{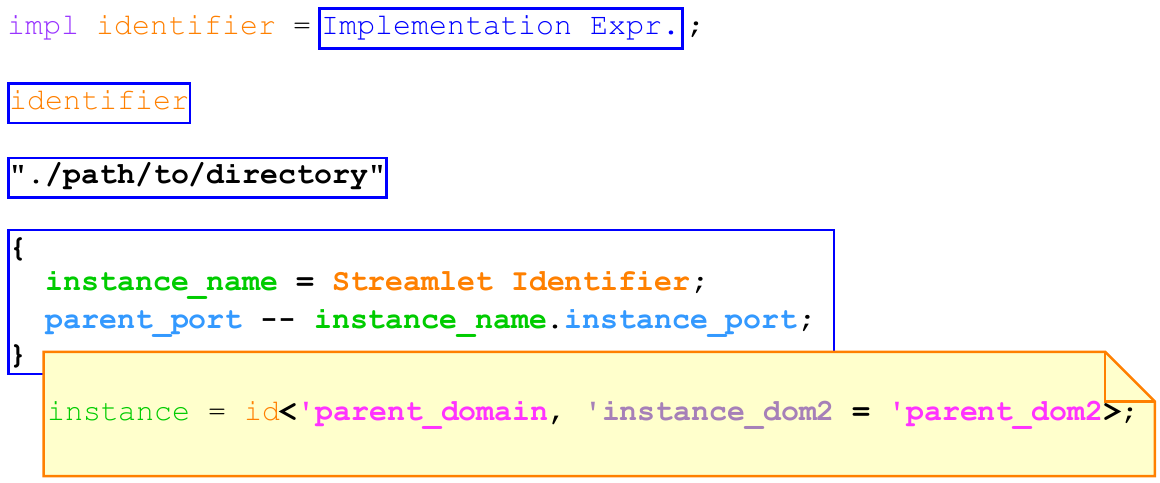}

\textbf{Streamlets} are a combination of the expressions above, and consist of an Interface and optionally an implementation. These are intended to be the output of a backend.

\includegraphics[width=0.75\linewidth]{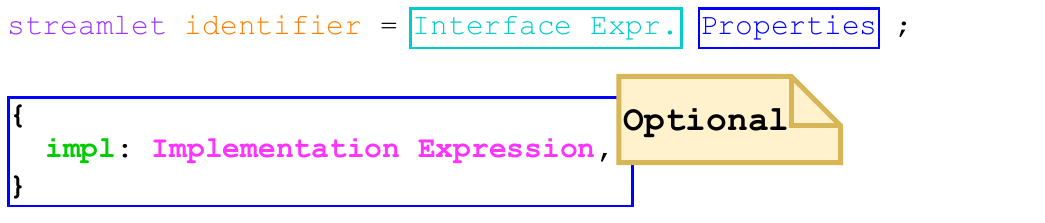}

Finally, \textbf{Documentation} is expressed by enclosing text with \textit{\#} signs, and must precede their subject, as shown in Listing \ref{lst:docex}. As explained in Section \ref{subsec:interfaces}, documentation is distinct from comments in that it is an actual property of Streamlets, ports, and implementations.

\begin{lstlisting}[basicstyle=\ttfamily,caption={Documentation Example},label={lst:docex}]
#documentation (optional)#
streamlet comp1 = (
    // This is a comment
    a: in stream,
    b: out stream,
    #this is port
documentation#
    c: in stream2,
    d: out stream2,
);
\end{lstlisting}

For a complete example of TIL, see Listing \ref{lst:full_til_example}.

\subsection{Parser Implementation}

The overall TIL parser performs the following passes:
\begin{enumerate}
    \item Lexing; converting the initial text file(s) into tokens. \begin{itemize}
        \item The primary token categories (in order of parsing priority) are: Documentation, versions (e.g., \textit{1.0.1}), numbers, path strings, operators, control characters, keywords, and identifiers.
        \item Anything not interpreted as a token (mainly whitespace) is treated as padding, and ignored. This includes comments. (Comments start with \texttt{//}, or are enclosed by \texttt{///} when multi-line.)
        \item Identifiers and keywords are continuous strings of characters, not separated by whitespace, operators or control characters.
        \item Documentation is any character (except \texttt{\#}), enclosed by \texttt{\#}s.
        \item The lexer pass recovers by simply skipping to the next input and retrying once all subsequent inputs have been parsed. This is not especially robust (i.e., it may not detect multiple errors), but simple to implement.
    \end{itemize}
    \item If the lexer pass succeeds, its resulting tokens are parsed to an abstract syntax tree (AST). \begin{itemize}
        \item The root node of any syntax tree is the \textit{namespace}; a namespace contains any number of statements.
        \item Statements in the current version of TIL are only declarations, but may be expanded to include imports and potentially certain intrinsic functions.
        \item There are four kinds of definitions to be declared (in order of parsing priority): Types, implementations, interfaces, and streamlets.
        \item Definitions themselves are expressions, and may use an identifier to reference a prior declaration.
        \item Identifier tokens are further combined into Path Names (Names separated by \texttt{::}), (port/field) labels (Names followed by a \texttt{:}) and domain names (Names preceded by a \texttt{'}). Though whether each parser is applied depends on context; e.g., labels only exist in ports lists of interfaces, while domain names only exist in domain lists of interfaces and domain assignments.
        \item Structural implementation definition bodies, despite being part of an expression, parse a list of statements. Structural body statements are Streamlet instantiations or port connections.
        \item This expression parsing can recover from some errors by looking for the next delimiter (e.g., if an error occurs parsing an interface \texttt{( ... )}, it will simply look for the closing \texttt{)}) and storing this as an \texttt{Error} node.
    \end{itemize}
    \item If the AST parsing pass succeeds, all declarations are \textit{evaluated}. \begin{itemize}
        \item Identifiers are converted to Names and Path Names, using the existing \texttt{try\_} constructors built for the query system, which consume an arbitrary string and then validate whether they match the Tydi requirements for identifiers.
        \item Each declaration is immediately stored into the query system's database during evaluation.
        \item The \textit{interned} identifiers of each declaration are stored in a HashMap for that type of declaration, using the identifer as a key. While the query system already ensures that identifiers are unique and exist, this method allows for better error recovery and error messages. (And while this was not implemented, pairing the original declaration with a span could allow for error messages which directly reference the previous declaration.)
        \item When evaluation of a declaration fails, this pass recovers by storing an \texttt{EvalError} node. Subsequent declarations are still evaluated, and all EvalErrors are reported (and rendered) once all declarations have been evaluated.
        \item Structural definition body statements are also evaluated during this pass (as part of the implementation expression/declaration evaluation), errors are still recovered as part of the complete declaration, however. (I.e., only the first error in a structural body will be reported, and all other statements are ignored.)
    \end{itemize}
\end{enumerate}

\begin{figure}
    \centering
    \includegraphics[width=0.98\linewidth]{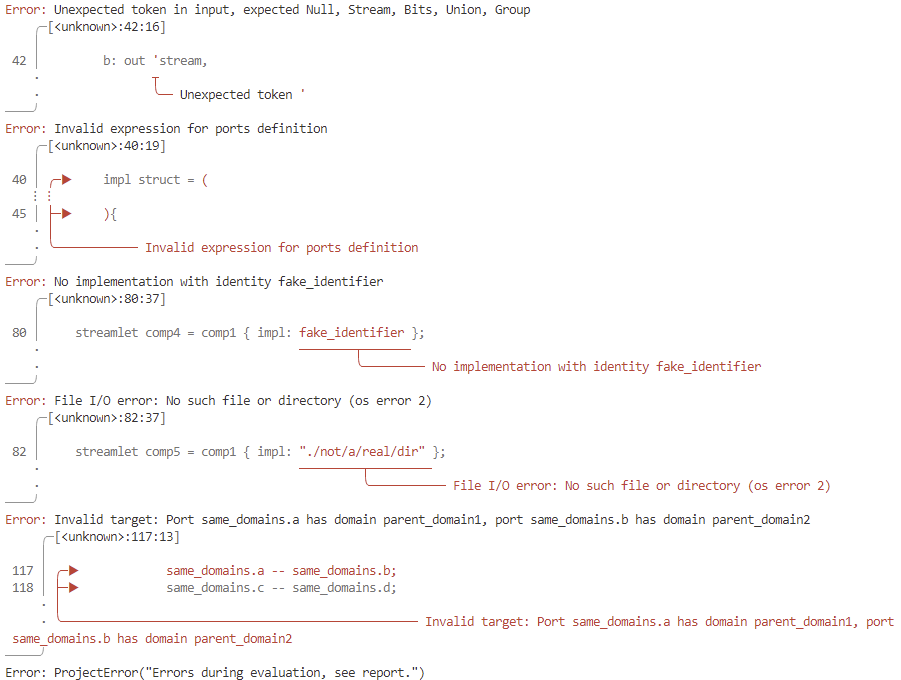}
    \caption{Errors simultaneously reported from both the abstract syntax tree parsing pass and evaluation pass.}
    \label{fig:multiple_pass_errors}
\end{figure}

Note that while in this implementation, each pass is only executed if the previous succeeded, this is not actually a requirement of Chumsky or the parsers implemented in it. Provided a pass has sufficiently robust error recovery, performing the next pass is a viable way of reporting more comprehensive errors. Figure \ref{fig:multiple_pass_errors} illustrates what happens when evaluation is allowed to occur despite AST parsing errors (this can be achieved by removing the initial \texttt{return Err} from the parser library's \texttt{into\_query\_storage} function). This is an optimistic scenario, however, as some AST errors (especially those involving delimiters) will still prevent all subsequent parsing.

\section{VHDL Backend}\label{sec:vhdl_backend}

In order to verify that the IR could actually be compiled to a hardware description, a VHDL backend was incorporated into the prototype toolchain. As all concepts expressed in the IR would need to be emitted to VHDL, this helps explore which properties are necessary or helpful for targeting hardware.

VHDL was chosen as the target because it is well-supported by multiple toolchains for both synthesis and simulation, and simply because its syntax was personally most familiar. Similar methods as those for emitting VHDL can be employed when emitting other hardware description languages, such as Verilog, FIRRTL and LLHD.

The ``passes'' used when emitting to VHDL in this example backend are intentionally very simple (for instance, while namespaces could correspond to their own VHDL packages, all namespaces are instead combined into a single package), though they do leverage the the query system's ability to incrementally compute and retrieve information:
\begin{enumerate}
    \item The ``all streamlets'' query described in Section \ref{sec:query_system} is used to retrieve all the Streamlet declarations in the project.
    \item For each Streamlet, the Streams that make up its Interface are split into physical streams, of which the signals are converted into ports. These ports make up a component with a unique name based on the Streamlet declaration and the namespace in which it was declared. These components are added to a single VHDL package.
    \item For each Streamlet, an architecture declaration is either imported or generated, as discussed in the next subsections.
\end{enumerate}

\subsection{Components and Organization}

As mentioned in the preface, each Streamlet is converted to a unique \textit{component} definition, regardless of how it is implemented. Taking the following namespace and Streamlet declaration, as an example:

\begin{lstlisting}[language=TIL,basicstyle=\ttfamily\scriptsize,caption={A simple namespace containing a Streamlet declaration},label={lst:comp1_example}]
namespace my::example::space {
    type stream = Stream (
        data: Bits(8),
        dimensionality: 0,
        synchronicity: Sync,
        complexity: 4,
    );

    #Streamlet documentation#
    streamlet comp1 = (
        a: in stream,
        b: out stream,
        #Port
documentation#
        c: in stream,
        d: out stream,
    );
}
\end{lstlisting}

We see that there is a namespace called \textit{my::example::space} and a single Streamlet declaration called \textit{comp1}. When the VHDL backend imports all Streamlets, the namespace is combined with the declaration identifier to form a unique identifier for the whole project, \textit{my::example::space::comp1}. As multiple underscores are not valid in Tydi's \textit{Name}s, we can safely use two underscores to represent the path separators, making for \lstinline{my__example__space__comp1_com}. The \texttt{\_com} suffix being a holdover from the original Tydi VHDL interface generator, which used it to distinguish \textit{canonical} representations of interfaces compared to the \textit{``fancy''} equivalents which used record types for better readability, as implementing similar functionality in the new backend is recommended as future work (as described in Section \ref{sec:readability}).

Likewise, the signals that make up the physical stream(s) split from each port receives a prefix based on that physical stream's name, for clarity. E.g., the \textit{data} signal of port \texttt{a} becomes \texttt{a\_data}. This results in the following package and component definition:

\begin{lstlisting}[language=VHDL,basicstyle=\ttfamily\scriptsize,caption={The TIL from Listing \ref{lst:comp1_example} converted to a VHDL package.},label={lst:comp1_vhdl}]
package proj is

  -- Streamlet documentation
  component my__example__space__comp1_com
    port (
      clk : in std_logic;
      rst : in std_logic;
      a_valid : in std_logic;
      a_ready : out std_logic;
      a_data : in std_logic_vector(7 downto 0);
      b_valid : out std_logic;
      b_ready : in std_logic;
      b_data : out std_logic_vector(7 downto 0);
      -- Port
      -- documentation
      c_valid : in std_logic;
      c_ready : out std_logic;
      c_data : in std_logic_vector(7 downto 0);
      d_valid : out std_logic;
      d_ready : in std_logic;
      d_data : out std_logic_vector(7 downto 0)
    );
  end component;

end proj;
\end{lstlisting}

Note that the documentation from Listing \ref{lst:comp1_example} is converted into comments in Listing \ref{lst:comp1_vhdl}. Documentation on Streamlet declarations is added above the respective VHDL component and entity declarations, documentation on a structural implementation is added above the architecture declaration, and documentation on the (IR) ports of Interfaces is added above the sets of (VHDL) ports that make up the logical Stream.

The \textit{clock} and \textit{reset} signals of the component are simply called \textit{clk} and \textit{rst}, as the Streamlet's interface only features the unnamed \textit{default} domain. Should it feature explicitly named domains, however, such as \lstinline{<'domain_name>}, this will take the form \lstinline{domain_name__clk} and \lstinline{domain_name__rst} instead. The relation between ports and specific domains is only tracked within the IR, as there is no way to reflect this property in VHDL directly.

The architecture definitions are stored in separate \texttt{.vhd} files named after the produced components, e.g., \texttt{my\_\_example\_\_space\_\_comp1.vhd}. This output is emitted to the same directory as the package definition. Each architecture also imports the package it was declared in by default, which simplifies the use of other components in structural implementations:

\begin{lstlisting}[language=VHDL,basicstyle=\ttfamily\scriptsize]
library work;
use work.proj.all;
\end{lstlisting}

\subsection{Linked Implementations}

Linked implementations are expressed as paths to a directory, as explained in section \ref{subsec:linked_impls}. For example:

\begin{lstlisting}[language=TIL,basicstyle=\ttfamily\scriptsize]
streamlet comp2 = comp1 {
    impl: "./vhdl_dir"
};
\end{lstlisting}

How to use these paths is to be decided by each backend: The current VHDL backend simply checks whether a file matching the naming scheme specified before exists in the directory (\texttt{my\_\_}\allowbreak\texttt{example\_\_}\allowbreak\texttt{space\_\_}\allowbreak\texttt{comp2.vhd}, in this case), and generates an empty architecture at that location if one does not exist. Then, the file is directly copied to the output. Note that as Streamlets are independently converted to component definitions for the package, any linked implementation must match the generated definition exactly for the project (and other uses of the Streamlet) to work correctly.

\subsection{Structural Implementations}

Structural implementations represent the bulk of the VHDL backend's computation, as these involve defining a (non-empty) architecture. The statements and properties the backend must implement are as follows:

\begin{enumerate}
    \item Instantiating Streamlets (\lstinline{a = comp1}); the backend must implement creating a named instance of a Streamlet, to be used throughout the rest of the implementation.
    \item Connecting ports (\lstinline{a.a -- a.b}); the backend must allow for ports of both instances and the parent Streamlet (made up of one or more physical streams and therefore multiple signals) to be connected.
    \item Assigning domains to instances (\lstinline{a = comp<'a, 'b>}); the backend must assign the appropriate clock and reset signals from the parent streamlet to an instantiated child Streamlet.
\end{enumerate}

To demonstrate how the backend implements this functionality, we will define a simple example, extending the previous listings:

\begin{lstlisting}[language=TIL,basicstyle=\ttfamily\scriptsize,caption={Declaring a Streamlet with multiple domains, and a structural implementation.},label={lst:struct_example}]
streamlet domains_only = <'a, 'b, 'c>();

streamlet comp3 = <'x, 'y>(
    q: in stream 'x,
    r: out stream 'x,
) {
    impl: {
        dom_ex = domains_only<'x, 'y, 'y>;
        inst = comp2<'x>;
        q -- inst.a;
        r -- inst.b;
        inst.c -- inst.d;
    }
};
\end{lstlisting}

To start, instances of Streamlets can be represented as simple \textit{port mapping}s of their respective component in the architecture. The name of the instance can be reflected as a label, e.g.:

\begin{lstlisting}[language=VHDL,basicstyle=\ttfamily\scriptsize]
dom_ex: my__example__space__domains_only_com port map( ...
\end{lstlisting}

As ports of instances need to be connected to both the parent Streamlet and other instances, the backend first defines a set of signals matching each port, allowing them to be connected in different parts of the architecture, rather than during port mapping. These signals are given unique names by suffixing their name with the instance's name, e.g.: \lstinline{inst__a_valid}

Domain assignments are defined directly on instantiation, however, and always draw from the parent Streamlet's domains. This means the clock and reset signals can be assigned directly on port mapping:

\begin{lstlisting}[language=VHDL,basicstyle=\ttfamily\scriptsize]
inst: my__example__space__comp2_com port map(
  clk => x__clk,
  rst => x__rst,
  ...
\end{lstlisting}

Finally, one last property to account for is that while physical streams have a single overall ``direction'' in Tydi itself, they are still made up of signals with different directions. In particular, the \textit{ready} signal will always be reversed relative to the other signals; as such, its assignment must be reversed in the resulting VHDL:

\begin{lstlisting}[language=VHDL,basicstyle=\ttfamily\scriptsize]
inst__a_valid <= q_valid;
q_ready <= inst__a_ready;
inst__a_data <= q_data;
r_valid <= inst__b_valid;
inst__b_ready <= r_ready;
r_data <= inst__b_data;
\end{lstlisting}

The full architecture (and TIL namespace) can be found in Appendix \ref{app:full_vhdl_example}.

\subsection{Additional and Future Functionality}

While structural implementations are the most complex fully-implemented feature of the backend, they do not reflect the full functionality implemented over the course of the thesis. For reference, the VHDL backend actually consists of two \textit{crate}s (Rust libraries): \textit{til-vhdl} and \textit{vhdl}. The latter is a library which is solely focused on programmatically defining and \textit{validating} VHDL, independent from Tydi or the IR, while the former simply uses it to convert the IR into VHDL.

These libraries were split to ensure that the concerns of \textit{generating correct VHDL syntax} and of \textit{converting the IR to VHDL} could be separated. As a result, the \textit{vhdl} library is capable of generating various VHDL statements, expressions and properties which were not used by \textit{til-vhdl}, such as:

\begin{itemize}
    \item \textit{Processes}, which were employed by the proof-of-concept for high-level assertions described in Section \ref{sec:spec_poc}, along with...
    \item \textit{Assertions}, which can compare values/signals and report an error message if they do not match.
    \item \textit{Types} other than \textit{std\_logic} and \textit{std\_logic\_vector}; \textit{vhdl} actually supports booleans, \textit{time}s, and arbitrary \textit{array} and \textit{record} types, along with constant expressions of \textit{severity}.
    \item \textit{Constant expressions} and \textit{relations}, as shown in Section \ref{sec:spec_poc}, the \textit{vhdl} library is also able to emit constant expressions of bits and bit vectors, though it can do the same for arrays, records, booleans and times (in various units). It can combine these as relations using using various operators (equality, greater/less than, logical operators, etc.), which themselves form (associative, boolean) relations.
    \item \textit{Imports}; while the current implementation only requires the \textit{ieee} and project package imports, the vhdl library can easily track multiple imports and prevent duplicate imports.
    \item Support for \textit{variables} and \textit{constants} in addition to \textit{signals}.
\end{itemize}

The reason the \textit{vhdl} library has support for these is because, while \textit{til-vhdl} often only required a small subset of VHDL, it was not significantly more difficult to implement a more complete and correct set of the relevant VHDL syntax. As a result, future functionality of \textit{til-vhdl} can be implemented more easily.

\section{Example}\label{sec:impl_example}

As an end-to-end example of the toolchain, the repository \cite{matthijsrtil2022} features a \texttt{demo-cmd} command-line application project which incorporates the parser, query system and VHDL backend. This project has been verified to compile and run using the Rust compiler \texttt{rustc}\footnote{\url{https://doc.rust-lang.org/cargo/getting-started/installation.html}} version 1.61.0 on both Windows 10 and Ubuntu 20.04 (through Windows Subsystem for Linux 2). To try the demonstration yourself, follow the following steps:

\begin{enumerate}
    \item Clone the repository (e.g., \texttt{git clone https://github.com/matthijsr/til-vhdl.git})
    \item Switch to the \texttt{demo-cmd} directory (\texttt{cd demo-cmd} from the repository's root, or \texttt{cd ./til-vhdl/demo-cmd/} if you have just cloned the repository)
    \item Build the application with \texttt{cargo build}, this will also install any dependencies.
    \item Run the application with \texttt{cargo run ./til\_samples/paper\_example.til ./output}, where \texttt{./til\_samples/paper\_example.til} is the input TIL file, and \texttt{./output} is the output directory.
    \item Once compilation has succeeded, there should be a \texttt{proj} directory in the chosen output directory, inside of this directory are the resulting \texttt{.vhd} VHDL architecture definitions and package. (As projects were not implemented in TIL, ``proj'' is used as the default project name.)
\end{enumerate}

\begin{figure}[h]
    \centering
    \includegraphics[width=0.98\linewidth]{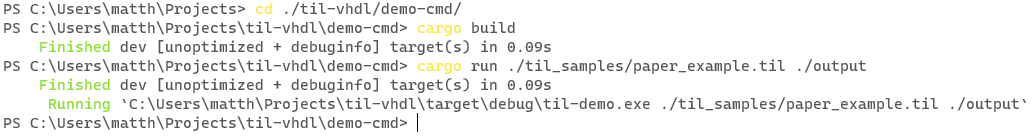}
    \caption{Successfully building and running the \texttt{demo-cmd} example application.}
    \label{fig:windows_demo_cmd}
\end{figure}

Once the application has been set up, users are free to try other TIL files and output directories. For example, the \texttt{./til\_samples/evaluation\_axi.til} file was used for the evaluation in Section \ref{sec:effort}. Users may write their own TIL files, or modify the existing ones. (For instance, to try out error reporting by changing \texttt{same\_domains = dom\_example<'parent\_domain1, 'parent\_domain1>;} to \texttt{same\_domains = dom\_example<'parent\_domain1, 'parent\_domain\textbf{2}>;}.)

\section{Partial Implementations}\label{sec:impl_partial}


As noted in their respective sections, work on some features had begun, but did not reach a complete or satisfactory state within the timeframe of the thesis. To summarize:

\begin{itemize}
    \item (Section \ref{sec:spec_poc}) Transaction-level assertions were partially implemented at the physical stream level, but lack support for logical streams or other features discussed in the chapter overall.
    \item (Section \ref{sec:project_structure}) Imports of declarations from other sources, and minimal support for projects have been implemented on the query system, but lack support from the parser.
\end{itemize}




\chapter{Evaluation}\label{ch:evaluation}


The prototype toolchain was developed not just as a demonstration, but also to test different approaches and verify their effectiveness. This section contains the evaluation of the prototype toolchain and its features.

\section{Tydi Specification}\label{sec:tydi_spec}

As a result of explicitly translating the Tydi specification to code, a few oversights and contradictions in the specification came to light. Fortunately, it was possible to discuss the intent with the designers of the specification (Johan Peltenburg, Jeroen van Straten, and Matthijs Brobbel), who were originally part of the ABS group. As such, in addition to submitting these as issues on the specification's GitHub, I was able to directly propose and utilize (interim) solutions for the purposes of my toolchain, or determine the original intent.\footnote{Also tracked on this project's repository, here: \url{https://github.com/matthijsr/til-vhdl/issues/81}} The following subsections describe the issues I have found and their proposed and/or interim solutions, their contents are adapted from the submitted reports.

\subsection{Directly nested Streams which must both be retained}\label{subsec:issue_nested}
\stitle{Report} \url{https://github.com/abs-tudelft/tydi/issues/221}

\stitle{Background} When a Stream contains another Stream as its \textit{data}, the \textit{Split} function\footnote{\url{https://abs-tudelft.github.io/tydi/specification/logical.html\#split-function}} assigns both the parent and child streams ``$\varnothing$'' (empty name), and employs ``flattening'' to combine their \textit{throughput}, \textit{synchronicity}, \textit{dimensionality} and \textit{direction}.

When a Stream has no element-manipulating data (\textit{data} is either Null or a Stream) and no \textit{user} property, it is discarded from the result. In effect, this creates a new physical stream with the original child Stream's \textit{data}, combined with the parent Stream's properties.

\stitle{Issue} When \textit{keep} ($x$) is true and/or \textit{user} ($T_u$) is non-Null, the parent Stream must be retained.

If a parent Stream has $x={true}$ and/or a non-Null $T_u$ property, both Streams are still assigned ``$\varnothing$'', but the parent Stream will conflict with the child Stream. Implementing the result of the Split function as a map in code, this means that either the child Stream simply replaces the parent Stream altogether (thereby losing the parent Stream's user property), or the Split function fails. Additionally, this does not account for child Streams having a \textit{synchronicity} which flattens its \textit{last} signal on the assumption that the (now non-existent) parent stream will drive the \textit{last} bits for its enclosing dimensions.

However, this behavior is not described in the specification, it only specifies that the names resulting from the Split function ``are case-insensitively \textbf{unique}, emptyable strings consisting of letters, numbers, and/or underscores, not starting or ending in an underscore, and not starting with a digit'' (emphasis mine).

\stitle{Interim solution} In the toolchain's implementation, the Split function will fail when it encounters a situation where two physical Streams have identical names.

\stitle{Proposed solutions} Explicitly specify that it is illegal for nested Streams (Streams which only have another Stream type as their \textit{data}) to have a \textit{keep} and/or \textit{user} property on more than one of these Streams, and ``flattening'' should incorporate the singular user property into the resulting physical stream.

Alternatively, Streams should have a non-empty \textit{name} property, or directly-nested children with retained parents should automatically receive one (e.g., \texttt{data} or \texttt{child}). This will avoid conflicts in the result of the Split function, regardless of whether these names are unique. (As nested Streams will simply have the parent and child Stream names joined in a ``Path Name'', as is currently the case for field names in Groups and Unions.)

\subsection{Significance of Strobe and Index Signals}

\stitle{Report} \url{https://github.com/abs-tudelft/tydi/issues/223}

\stitle{Background} Tydi provides three different signals to indicate whether the lanes of a \textit{data} signal are active during a transfer. At \textit{complexity} $C\geq7$, the strobe (\textit{strb}) signal can encode the validity of every lane independently. At \textit{complexity} $C<7$, the start index (\textit{stai}) and end index (\textit{endi}) encode which \textit{range} of lanes is active instead; as both are 0-indexed, however, such Streams also have a single \textit{strb} bit to indicate whether \textit{all} of the transfer's data lanes are inactive.

When a Stream with $C<7$ is a source to a Stream with $C\geq7$, it will drive all \textit{strb} bits simultaneously from its single \textit{strb} bit, and the higher-complexity sink must instead interpret the \textit{stai} and \textit{endi} signals.

\stitle{Issue} While Tydi's \textit{data} signal specification\footnote{\url{https://abs-tudelft.github.io/tydi/specification/physical.html\#data-signal-description}} indicates that the \textit{stai} and \textit{endi} signals are redundant when $C\geq7$, it is not entirely clear whether they are still \textit{significant}.

Since the \textit{strb} signal still exists at $C<7$ as a single bit to indicate that all of transfer's data lanes are inactive, considering the inverse, \textit{stai} and \textit{endi} must be significant despite \textit{strb} indicating that ``all lanes'' are active. Otherwise, these signals would be unable to encode lane activity. At the same time, while \textit{stai} and \textit{endi} are insignificant when all \textit{strb} bits are driven low, it is not clear whether this applies when some \textit{strb} bits are high and some are low.

As an example of these conflicts, consider the following examples:
\begin{enumerate}
    \item ${stai}=0$, ${endi}=0$ and ${strb}={"010"}$
    \item ${stai}=0$, ${endi}=0$ and ${strb}={"101"}$
\end{enumerate}

In both instances, the start- and end-indices indicate that only the first lane is active. However, in the first instance the \textit{strb} indicates that the first lane is inactive, but the second lane is active. In the second instance, the \textit{strb} does indicate that he first lane is active, but also indicates that the third lane is active.

\stitle{Proposed solution} The start- and end index signals should only be significant when \textit{all} strobe signal bits are driven high. This ensures lower-complexity sources can still connect to higher-complexity sinks, but does not allow for sources with $C\geq7$ to create confusing transfers.

\subsection{Transferring Empty Outer Sequences at Lower Complexities}

\stitle{Report} \url{https://github.com/abs-tudelft/tydi/issues/224}

\stitle{Background} The specification for the \textit{last} signal\footnote{\url{https://abs-tudelft.github.io/tydi/specification/physical.html\#last-signal-description}} notes different constraints for complexities $C<4$ and at $C<8$, with the intent of placing certain requirements on sources transferring sequences. The constraints at $C<4$ are as follows:
\begin{enumerate}
    \item ``It is illegal to assert a \textit{last} bit for dimension $j$ without also asserting the last bits for dimensions $j′<j$ in the same lane.''
    \item ``It is illegal to assert the \textit{last} bit for dimension 0 when the respective data lane is inactive, except for empty sequences.''
\end{enumerate}

The intention of these rules is to prevent source Streams with $C<4$ from postponing \textit{last} flags of outer dimensions to subsequent transfers. For instance, when transferring \textit{[ [ data, data ] ]}, the source cannot first perform a transfer with all data and ${last}={"01"}$ (last in dimension 0), followed by an empty transfer with ${last}={"10"}$ (last in dimension 1).

\stitle{Issue} The first constraint prevents Streams with $C<4$ from transferring empty outer sequences. That is to say, they cannot perform a transfer \textit{[ ]} (${last}={"10"}$), and may only transfer \textit{[ [ ] ]} (${last}={"11"}$) instead. As a result, the example sequence from the \textit{last} signal specification cannot actually be transferred at lower complexities:
\begin{lstlisting}[basicstyle=\ttfamily]
["Hello", "World"], ["Tydi", "is", "nice"], [""], []
\end{lstlisting}

Complexity is meant to be a property affecting \textit{how} data can be transferred and how Streams are physically implemented: These constraints mean that the complexity property also affects what kind of data can be transferred.

\stitle{Proposed solution} The first rule for $C<4$ (requiring \textit{last} in dimensions $j′<j$) should be amended with an exception for empty sequences, just like the second rule. This ensures \textit{complexity} does not affect what kinds of data can be transferred, while still ensuring \textit{last} flags cannot be postponed at lower complexities.

\subsection{Indicating Inactive Lanes at Lower Complexities}

\stitle{Report} \url{https://github.com/abs-tudelft/tydi/issues/226}

\stitle{Background} The specification for signal omission\footnote{\url{https://abs-tudelft.github.io/tydi/specification/physical.html\#signal-omission}} places the following constraints on the start index (\textit{stai}), end index (\textit{endi}) and strobe (\textit{strb}) signals which govern whether element (\textit{data}) lanes in a transfer are active, based on complexity $C$, number of element lanes $N$, and dimensionality $D$:
\begin{enumerate}
    \item \textit{endi} is contingent on $(C\geq5 \lor D\geq1)∧N>1$
    \item \textit{stai} is contingent on $C\geq6 ∧ N>1$
    \item \textit{strb} is contingent on $C\geq7 \lor D\geq1$
\end{enumerate}

If these constraints are not met, a physical stream is unable to indicate whether individual element lanes are inactive. As these constraints are part of the signal omission specification, it is implied that such streams have no need to do so.

\stitle{Issue} When a Stream has properties $C<5 ∧ D = 0 ∧ {throughput} > 1$, its physical implementation will have multiple element lanes, but lack the ability to indicate whether they are inactive. This means each transfer \textit{must} consist of exactly $N$ elements

\stitle{Proposed solutions} This constraint is either an oversight, or should be clarified to be an actual requirement for lower complexity Streams, rather than a physical implementation detail:

\begin{enumerate}
    \item If it is an oversight, the requirement for \textit{endi} being contingent on $(C\geq5 \lor D\geq1) ∧ N>1$ should be changed to being solely contingent on $N>1$.
    \item If it is intentional, and physical streams should be able to transfer arbitrary sets of elements, the number of element lanes $N$ being greater than $1$ should be contingent on $D>0 \lor C\geq5$.
    \item If it is intentional overall, the requirement for Streams with complexity $C<5$, $D=0$ and \textit{throughput} $t>1$ to transfer only sets of elements equal to or divisible by $\ceil{t}$ should be specified as part of the logical Stream specification, as well.
\end{enumerate}

\subsection{Minor Inconsistencies}

\stitle{Reports} \begin{enumerate}
    \item \url{https://github.com/abs-tudelft/tydi/issues/222}
    \item \url{https://github.com/abs-tudelft/tydi/issues/225}
\end{enumerate}

\stitle{Background, Issue and Solution} These contradictions are relatively minor inconsistencies related to phrasing and constraints conflicting over multiple separate requirements:

\begin{enumerate}
    \item The signaling specification states that when $C<8$, the \textit{last} bits for lanes $0$ through $N-2$ must be driven low, suggesting that the singular \textit{last} value applies to lane $N-1$. One of the constraints for $C<4$ mentions that ``It is illegal to assert the \textit{last} bit for dimension 0 when \textbf{the respective data lane} is inactive, except for empty sequences.'' (emphasis mine). As a result, sequences which do not align with the number of element lanes (and do not have a start index signal, so must align to the first lane) would not be able to assert \textit{last}. \begin{itemize}
        \item The solution is to change the phrasing for the $C<4$ rule to refer to \textit{transfer} data, rather than a data lane.
    \end{itemize}
    \item The specification inconsistently refers to the strobe (\textit{strb}) signal encoding whether individual data lanes are active, with some constraints suggesting this is only the case at $C\geq8$, while others suggest this applies to $C\geq7$. \begin{itemize}
        \item This is simply an oversight: $C\geq7$ must allow \textit{strb} to encode individual lane activity, as otherwise $C=7$ is identical in functionality to $C=6$. ($C=8$ adds the ability to encode a \textit{last} value per lane over $C=7$.)
    \end{itemize}
\end{enumerate}

\section{Readability}\label{sec:readability}

\subsection{Readable Output}

As the IR relies on other languages to express functionality, it will generally be necessary for the descriptions a backend \textit{does} generate to be readable by designers, barring a frontend emitting both the IR and the behavioral descriptions. To this end, the IR exposes ``documentation'' to backends, enabling designers to propagate some intent to component templates and interfaces. The prototype VHDL backend propagates this documentation as comments, and generates indented VHDL with port and signal names derived from the TIL port and field names.

There is one area in which much information and readability is lost, however: The physical streams emitted by the VHDL backend feature standard \textit{data} and \textit{user} signals as bit vectors, meaning that the names of element fields of Groups and Unions are lost. As described in Section \ref{subsec:intrinsics}, the Tydi documentation describes alternative ways to represent physical streams to retain this information. For instance, Groups and Unions could be expressed as record types in VHDL, multiple element lanes as arrays of the base type, and even physical streams themselves could be collected into records (split into separate records for up and downstream signals). These are not only useful for implementation, but can also provide more information when simulating a design.

In fact, the \textit{Implementations} section of the original Tydi paper \cite{peltenburgtydi2020} assumes that designers would prefer such a solution, and illustrates that automatically generating such records from Tydi logical types would greatly reduce the number of lines of code designers would need to write. To better enable such alternative representations, making changes to the IR to require type identifiers, rather than storing only the official properties of logical types may prove beneficial, as described in Section \ref{subsec:interface_compat}. Doing so would allow a backend to generate alternative representations with meaningful type names, which could then be directly reused by multiple interfaces, albeit at the expense of the ability to directly connect physically compatible types.

\subsection{Type Identifiers}

The initial approach towards the IR was to stay very close to the Tydi specification itself, and avoid any added or divergent functionality. As the Tydi specification did not feature identifiers for types, this would be diverging from the specification. Even if these identifiers were only tracked by the IR itself and not propagated to the backend, using them to determine ``compatibility'' appeared to be too opinionated.

However, as the previous section and Section \ref{subsec:interface_compat} reflect, identifiers being a property of types can yield significant benefits. Moreover, insisting that types must be compatible based on their definition, and not their identifier, is opinionated in its own way. As such, there are a number of different approaches to implementing type identifiers:

\begin{enumerate}
    \item Whether identifiers affect compatibility is \textit{configurable}; the existing TIL syntax remains unchanged, but the query system's functions for determining compatibility are configurable, and may either behave as it currently does, or require identical identifiers as well. This allows either the backend or frontend to specify which behavior is preferred.
    \item Create explicit distinctions between \textit{anonymous} types, \textit{named} types, and \textit{aliases} of types. This enables the frontend (or designer) to choose which behavior is preferred on a per-project or per-type level, and would be implemented as follows: \begin{itemize}
        \item \textit{Anonymous} types are the current type definitions, they are not declared, but used in declarations or directly on Interface ports: \lstinline{Bits(8)}
        \item \textit{Named} types modify the properties of a type, setting the name; when checking whether types are compatible, their names must also match:\begin{itemize}
            \item \lstinline{type byte = Bits(8);}
            \item \lstinline{type char = byte;}
            \item \lstinline{char != byte}
        \end{itemize}
        \item \textit{Aliases} are how current type declarations work, they are identifiers for types, but do not modify the type's properties, and are only tracked as part of namespaces: \begin{itemize}
            \item \lstinline{alias char = byte;}
            \item \lstinline{char == byte}
            \item \lstinline{alias reg = Bits(8);}
            \item \lstinline{reg == Bits(8)} and \lstinline{reg != char}
        \end{itemize}
    \end{itemize}
    \item Types \textit{must} have identifiers. This is the most divergent approach from both the current IR and of the Tydi specification, but does have a number of merits: \begin{itemize}
        \item It guarantees that backends have unique identifiers for ``fancy'' types, ensuring they are also compatible in the target language (e.g., record types in VHDL).
        \item By giving Streams explicit names, there can be no conflicts between physical stream names, such as those described in Section \ref{subsec:issue_nested}.
    \end{itemize}
\end{enumerate}

As an aside, it is worth noting that technically, frontends can already implement a kind of identifier-based compatibility between types, by declaring every type as a \lstinline{Group} with one field (e.g., \break\lstinline{Group(byte: Bits(8))}). Though this puts the onus of tracking the uniqueness of identifiers on said frontends. More importantly, the fact that Tydi already supports compatibility restrictions based solely on identifiers calls the accuracy of anonymous types enforcing ``physical compatibility'' into question.




\section{Hardware Description Effort}\label{sec:effort}


The goal of the IR is to describe streams carrying complex data structures more effectively than conventional HDLs. As such, while ``lines of code'' is not an especially relevant metric for an IR overall, it can be applied to the amount of effort required to express interfaces and connections. To evaluate the IR's effectiveness in this regard, Tydi equivalents of the AXI4-Stream \cite{armlimitedamba2010} and AXI4 \cite{armlimitedintroduction2021} interface standards were declared in TIL.

\begin{table}[]
    \centering
    \tiny
    \begin{tabular}{lll}
\hline
\multicolumn{1}{c}{\textbf{Signal}} & \multicolumn{1}{c}{\textbf{Source}} & \multicolumn{1}{c}{\textbf{Description}}                                                                                                                                                                                                                                                                     \\ \hline
\textbf{ACLK}                       & Clock source                        & The global clock signal. All signals are sampled on the rising edge of ACLK.                                                                                                                                                                                                                                 \\ \hline
\textbf{ARESETn}                    & Reset source                        & The global reset signal. ARESETn is active-LOW.                                                                                                                                                                                                                                                              \\ \hline
\textbf{TVALID}                     & Master                              & \begin{tabular}[c]{@{}l@{}}TVALID indicates that the master is driving a valid transfer.\\  \\ A transfer takes place when both TVALID and TREADY are asserted.\end{tabular}                                                                                                                                 \\ \hline
\textbf{TREADY}                     & Slave                               & TREADY indicates that the slave can accept a transfer in the current cycle.                                                                                                                                                                                                                                  \\ \hline
\textbf{TDATA{[}(8n-1):0{]}}        & Master                              & TDATA  is the primary payload that is used to provide the data that is passing  across the interface. The width of the data payload is an integer  number of bytes.                                                                                                                                          \\ \hline
\textbf{TSTRB{[}(n-1):0{]}}         & Master                              & TSTRB is the byte qualifier that indicates whether the content of the associated byte of TDATA is processed as a data byte or a position byte.                                                                                                                                                               \\ \hline
\textbf{TKEEP{[}(n-1):0{]}}         & Master                              & \begin{tabular}[c]{@{}l@{}}TKEEP is the byte qualifier that indicates whether the content of the associated byte of TDATA is processed as part of the data stream.\\  \\ Associated bytes that have the TKEEP byte qualifier deasserted are null bytes and can be removed from the data stream.\end{tabular} \\ \hline
\textbf{TLAST}                      & Master                              & TLAST indicates the boundary of a packet.                                                                                                                                                                                                                                                                    \\ \hline
\textbf{TID{[}(i-1):0{]}}           & Master                              & TID is the data stream identifier that indicates different streams of data.                                                                                                                                                                                                                                  \\ \hline
\textbf{TDEST{[}(d-1):0{]}}         & Master                              & \textbf{TDEST provides routing information for the data stream.}                                                                                                                                                                                                                                             \\ \hline
\textbf{TUSER{[}(u-1):0{]}}         & Master                              & \textbf{TUSER is user defined sideband information that can be transmitted alongside the data stream.}                                                                                                                                                                                                       \\ \hline
\end{tabular}
    \caption{The AXI4-Stream signal specification, source: \cite{armlimitedamba2010}}
    \label{tab:axi4stream_spec}
\end{table}

Table \ref{tab:axi4stream_spec} shows the signal specification of AXI4-Stream, while Listing \ref{lst:axi4streamtil} shows how it was implemented as a Tydi Stream for the purposes of this evaluation, along with the resulting (VHDL) signals in Listing \ref{lst:axi4streamvhdl}. AXI4 was spread over 5 Streams for Address Write, Write Data, Write Response, Address Read, and Read Data. Appendix \ref{app:axi4spec} shows the full signal specification of AXI4, while Appendix \ref{app:axi4til} shows the full TIL definitions of both AXI4-Stream- and AXI4-equivalent Streams, and the resulting VHDL component.

\begin{minipage}{0.95\linewidth}
\begin{lstlisting}[basicstyle=\ttfamily,caption={An AXI4-Stream-equivalent interface in TIL.},label={lst:axi4streamtil},escapechar=@]
type axi4stream = Stream (
    data: Union (
        data: Bits(8),
        null: Null, // Equivalent to TSTRB
    ),
    throughput: 128.0, // Data bus width
    dimensionality: 1, // Equivalent to TLAST
    synchronicity: Sync,
    complexity: 7, // Tydi's strobe is equivalent to TKEEP
    user: Group (
        TID: Bits(8),
        TDEST: Bits(4),
        TUSER: Bits(1),
    ),
);

@\textcolor{gray}{streamlet example = (}@
    axi4stream: in axi4stream,
\end{lstlisting}
\end{minipage}

\begin{minipage}{0.95\linewidth}
\begin{lstlisting}[basicstyle=\ttfamily,caption={Result of Listing \ref{lst:axi4streamtil} in VHDL.},label={lst:axi4streamvhdl}]
axi4stream_valid : in std_logic;
axi4stream_ready : out std_logic;
axi4stream_data : in std_logic_vector(1151 downto 0);
axi4stream_last : in std_logic;
axi4stream_stai : in std_logic_vector(6 downto 0);
axi4stream_endi : in std_logic_vector(6 downto 0);
axi4stream_strb : in std_logic_vector(127 downto 0);
axi4stream_user : in std_logic_vector(12 downto 0);
\end{lstlisting}
\end{minipage}


Once a Stream type has been declared, it can be easily reused for any number of ports, and ports only require one statement (\lstinline{port_a -- port_b;}) to connect, which is far fewer than the signals which make up a stream (or AXI4 channel). Table \ref{tab:loc_vs_signals} illustrates this difference: The AXI4-Stream equivalent requires a single Stream overall, while AXI4 requires a Stream per channel, and can be either split across multiple ports, or combined into a Group with Reverse Streams for the Read Data and Response channels, depending on the use case. Both result in identical physical streams, but using multiple ports allows for them to be connected to different Streamlets if necessary.

\begin{table}[ht]
\centering
\begin{tabular}{l|ll}
                          & \textbf{Type Declaration} & \textbf{Interface} \\ \hline
AXI4 equiv. (TIL)         & 48*              & 5         \\
AXI4 equiv. (TIL, Group)  & 59*              & 1         \\
AXI4 equiv. (VHDL)        & -                & 28        \\
AXI4                      & -                & 44        \\ \hline
AXI4-Stream equiv. (TIL)  & 15*              & 1         \\
AXI4-Stream equiv. (VHDL) & -                & 8         \\
AXI4-Stream               & -                & 9        
\end{tabular}
\caption{Lines of code to represent an interface in TIL, compared to the resulting number of signals in VHDL or for an equivalent interface standard. *Only required once.}
\label{tab:loc_vs_signals}
\end{table}

As an aside, these AXI4 and AXI4-Stream definitions are areas where the type parameters discussed in Section \ref{subsec:type_parameters} could be applied very effectively. A type could define the basic requirements for an AXI4(-Stream)-equivalent interface, while using type parameters to set the variable properties of AXI4(-Stream), such as data bus width.

\section{Parser}\label{sec:eval_parser}

As development of a text-based grammar and parser was secondary to development of the query system and VHDL backend, it did not receive as much attention, and work and research started later in the course of the thesis overall. Despite these limitations, the parser developed using \textit{Chumsky} was satisfactory overall; it was possible to very quickly and relatively easily define a grammar and build a parser which translated to the majority of concepts expressed in the query system. Hence, this section will summarize the specific merits and demerits of Chumsky, based on experience using it to build the TIL parser, as well as recommend potential improvements to the TIL parser.

\subsection{Merits of Chumsky}

Overall, Chumsky was easy to work with, and can certainly be recommended for continued work on a TIL parser, or any other projects which might require a domain-specific language parser built in Rust. To summarize the specific advantages:

\begin{enumerate}
    \item It is (comparatively) easy to work with, featuring a short but descriptive tutorial\footnote{\url{https://github.com/zesterer/chumsky/blob/master/tutorial.md}} and many built-in parser functions, such as \texttt{delimited\_by} to indicate grammar is enclosed by specific symbols, and \texttt{foldr} and \texttt{foldl} for \textit{folding} right- and left recursive grammar into nested expressions.
    \item As parser definitions are fully native to Rust, they integrate well with IDEs (detecting errors, suggesting functions, providing documentation hints) and with other functionality written in Rust (such as functions and types originally created for the query system).
    \item Parsers can be split over multiple functions or variables, making them reusable and easy to organize over files and directories.
    \item It features built-in error-recovery strategies to use with parsers, enabling it to generate partial ASTs and/or evaluate more of a file, even when errors are encountered. This is not necessarily unique, but is made very accessible.
    \item While not directly part of Chumsky, its sister project \textit{Ariadne} \cite{barrettoariadne2022} can be used to render error reports in the terminal in a clear, color-coded way.
\end{enumerate}

\subsection{Issues Using Chumsky}

While Chumsky proved to be a good fit for the project, and provides many useful features, it does have a number of issues to account for:

\begin{enumerate}
    \item Chumsky is explicitly not designed to be a high-performance parser, as its repository description notes, ``Chumsky focuses on high-quality errors and ergonomics over performance.''\footnote{\url{https://github.com/zesterer/chumsky\#performance}} While parsing speed was not an issue for the TIL parser's implementation, it could scale poorly to larger, multi-file projects, and may be ill-suited to other kinds of parsers.
    \item While being able to include \textit{spans} of character positions in parsed lexical tokens and nodes of the abstract syntax tree is very useful for error reporting, there does not appear to be a way to quickly remove them from all nodes/tokens. This is inconvenient when attempting to write unit tests against the parser to assert its output is correct - needing to either manually include the expected spans in the comparison, or write a custom method to remove them. As different parser stages also expect to receive tokens with spans, this also convolutes their input when testing.
    \item Chumsky is not portable between different languages: While its parser definitions being native to Rust was marked as an advantage, this also means that the TIL parser cannot be easily translated to a different language altogether. Using independent grammar definitions can reduce the effort of implementing parsers in different languages, although of the parser libraries evaluated, only \textit{lrpar} \cite{grammar2022} employed a syntax (\textit{Yacc}) for which libraries exist in other languages.
    \item In the event a parser composed of multiple different parsers fails (notably due to stack overflows), it is extremely difficult to find out through debugging or error messages what part of the code caused the issue. This is due to the expanded code resulting from Chumsky's inline and macro functions being difficult to trace back to the source. This is not unique to Chumsky, however, as tracing the source of a stack overflow is rather difficult regardless.
\end{enumerate}

As a note on the last issue, stack overflows were consistently resolved by splitting up larger parsers into multiple functions (to better reason about their operation), and by removing potential sources of ambiguity. One such source of ambiguity causing stack overflows in the TIL parser was that originally, the AST parser would attempt to parse any kind of definition expression (i.e., whether it was an identifier, type, interface, implementation or Streamlet) after a declaration, and select the appropriate ones afterward. Splitting up the parsers for the definition expressions into separate functions, and only combining them with the declaration parser where they were appropriate resolved the stack overflows.


\subsection{Recommendations for TIL Parser}\label{subsec:parser_recs}

The following aspects of the TIL parser can be improved, or may be useful additions (note, these are not additions to TIL itself):

\begin{enumerate}
    \item More robust error recovery strategies applied to all passes, e.g.: \begin{itemize}
        \item The lexer pass currently uses \texttt{skip\_then\_retry}, it may be possible to use \texttt{skip\_until} for certain constructs, or create a custom strategy.
        \item The AST pass has difficulty with unclosed delimiters, it should be possible to create a recovery strategy which skips until the next declaration keyword.
        \item The evaluation pass does not recover inside structural implementation definitions, and instead recovers the entire declaration. Instead, an error recovery strategy on every structural implementation statement should be implemented.
    \end{itemize}
    \item Attempt to produce more helpful errors; e.g, when a duplicate identifier error occurs, also point towards the previous declaration.
    \item As a larger change/addition, attempt to emit the parser's results to a language server (protocol), so an IDE can provide syntax highlighting and error linting.
\end{enumerate}




\chapter{Conclusion}\label{ch:conclusion}

\section{Conclusions and Summary}


This thesis presents an IR for defining interfaces and integrating components using the Tydi specification. The prototype toolchain used to evaluate and demonstrate the ideas in this thesis features the ability to efficiently express Tydi interfaces and connect components using a simple grammar, and emit these as VHDL components and architectures.

Of note is the ability to p ropagate high-level, abstract properties such as documentation down from the IR (and any potential front-end) to the target language, to improve readability and more easily verify its outputs. As an extension of this, emitting alternative representations for Tydi's interfaces to retain type information could improve readability further. The thesis outlines potential changes to the IR to better enable this, in particular the addition of identifiers as properties of types.

This thesis also proposes the use of and a potential syntax for high-level assertions against interfaces defined in the IR, and a partial proof-of-concept for such tests was implemented in conjunction with the prototype toolchain. Alongside the high-level assertions, the limitations of such tests were discussed, and a potential solution was proposed in the form of ``test Streamlets'' and substitutions of Streamlet implementations. Then, the requirements for setting up individual Streamlets for testing were described, in order to aid future work.

As part of the overall evaluation, a number of inconsistencies in the Tydi specification were also identified and reported. The identification of these issues was a direct result of implementing the Tydi specification programmatically.

Overall, the work done as part of this thesis has been effective in demonstrating and testing the limits of an intermediate representation and toolchain specifically for composing components using the Tydi specification. However, there are many avenues to improve both the IR and toolchain, as summarized in the next section.

\section{Recommendations for Future Work}


Based on the findings of the thesis, we can make the following recommendations for future work:

\begin{itemize}
    \item Implement a framework for typed, transaction-level assertions as described in Section \ref{sec:assertions}.
    \item Document the workings of the IR and any ancillary components outside of this thesis and the published paper as part of the open-source repository.
    \item Define a number of intrinsics, and attempt to emit them through a backend, as described in Section \ref{subsec:intrinsics}. Having the ability to verify their functionality through the transaction-level assertions would be ideal, as this would also allow them to be verified over different backends.
    \item Add support for other language features, such as type parameters (\ref{subsec:type_parameters}), generation (\ref{subsec:generation}) and annotations (\ref{subsec:annotations}).
    \item Consider adding support for Streams without a clock domain, enabling asynchronous transfers as discussed in Section \ref{subsec:interfaces}. Do note that this will also require modifications to the Tydi specification.
    \item Implement and use ``projects'' in both the query system and TIL, as described in Section \ref{sec:project_structure}.
    \item Add support for ``substituting'' implementations of Streamlets for the purposes of testing, as described in Section \ref{subsec:substitution}.
    \item Consider means for setting up ``subjects'' of tests correctly, given potentially unusual reset behavior and other requirements, as described in Section \ref{sec:setup}.
    \item Make identifiers a (potentially optional) component of type compatibility, as described in Section \ref{sec:readability}.
    \item As also described in Section \ref{sec:readability}, emit interfaces which better reflect their original logical type definitions.
    \item Make improvements to the TIL parser, as described in Section \ref{subsec:parser_recs}.
    \item Should performance become a concern, establish a ``benchmark'' TIL project and attempt to measure the speed with which the TIL parser and query system + backend perform their tasks. Of particular interest is whether the query system is effectively storing and reusing previous queries.
    \item Once the query system, parser, and other components have been appropriately iterated on and fully documented, publish them as crates on \textit{crates.io}, so that others may use them more easily. Also ensure that their respective dependencies are subsequently converted to dependencies on the published crates, rather than their relation in the repository.
\end{itemize}

A number of these recommendations are also being tracked as issues on the \texttt{til-vhdl} repository: \url{https://github.com/matthijsr/til-vhdl/issues}

\appendix

\chapter{Complete TIL Example}

\begin{lstlisting}[basicstyle=\ttfamily\scriptsize,caption={A full TIL namespace with explanations as comments.},label={lst:full_til_example}]
namespace my::example::space {
    // Type declarations
    type byte = Bits(8);
    // Type expressions can be identifers or in-line declarations
    type select = Union(val: byte, empty: Null);
    type rgb = Group(r: select, g: select, b: select);
    // Streams have many properties, but some are optional
    type stream = Stream (
        data: rgb,
        throughput: 2.0, // 1.0 by default
        dimensionality: 0,
        synchronicity: Sync,
        complexity: 4,
        direction: Forward, // Forward by default
        user: Null, // Null by default
        keep: false, // false by default
    );
    type stream2 = stream;

    // A streamlet declaration
    #documentation (optional)#
    streamlet comp1 = (
        // Ports are *name* : *direction* *stream expression*
        a: in stream,
        b: out stream,
        # port documentation #
        c: in stream2,
        d: out stream2,
    );

    // An independent interface declaration
    interface iface1 = (a: in stream, b: out stream);

    #streamlet documentation
newline documentation#
    streamlet comp2 = iface1;
    
    // Implementation declarations
    #This is implementation documentation.#
    impl struct = (
        a: in stream,
        b: out stream,
        c: in stream2,
        d: out stream2,
    ){
        // Ports can be connected with --
        a -- b;

        // Streamlet instances are declared with
        // *instance name* = *streamlet name*
        a = comp1;
        b = comp1;

        // Ports on streamlet instances can be addressed with .
        a.a -- b.b;
        a.b -- b.a;
        
        // Ports on instances can also be connected to local ports
        c -- a.c;
        d -- b.d;
        a.d -- b.c;
    };

    // Linked implementations are paths enclosed by double quotes.
    impl link = comp1 "./vhdl_dir";

    streamlet comp3 = comp1 {
        impl:
        #This is implementation documentation, too.#
        {
            p1 = comp2;
            p2 = comp2;
            a -- p1.a;
            b -- p1.b;
            c -- p2.a;
            d -- p2.b;
        }
    };

    streamlet comp4 = comp1 { impl: struct };

    streamlet comp5 = comp1 { impl: "./vhdl_dir" };

    // 'domains represent combined clock and reset domains, and how they relate
    // to a port's stream.
    streamlet dom_example = <
     'domain1,
     'domain2,
    >(
        a: in stream 'domain1, 
        b: out stream 'domain2,
        c: in stream 'domain2, 
        d: out stream 'domain1,
    );

    streamlet blank_doms = <'a, 'b, 'c>();

    // In the above example, the domains of ports a and b are different, making them incompatible
    // despite having the same type. a and d, and b and c can be connected, however.
    //
    // However, a structural implementation can assign the same domain twice,
    // making a and b, and c and d compatible again.
    streamlet struct_dom_example = <
      'parent_domain1,
      'parent_domain2,
    > () {
        impl: {
            different_domains = dom_example<'parent_domain1, 'parent_domain2>;

            // Try changing these to <'parent_domain1, 'parent_domain2>
            // to see what happens when domains don't match.
            same_domains = dom_example<'parent_domain1, 'parent_domain1>;

            different_domains.a -- different_domains.d;
            different_domains.b -- different_domains.c;

            same_domains.a -- same_domains.b;
            same_domains.c -- same_domains.d;

            // For clarity, when assigning domains it's also possible to specify
            // which domain of the instance is being assigned to, rather than using their order.
            explicit_doms = blank_doms<'c = 'parent_domain1, 'a = 'parent_domain2, 'b = 'parent_domain2>;

            // It's also possible to mix named assignments with ordered assignments,
            // provided the named assignments succeed all ordered assignments.
            mixed_assignments = blank_doms<'parent_domain2, 'c = 'parent_domain1, 'b = 'parent_domain2>;
        }
    };

    // When a parent interface has no explicit domains, it is instead given a "default" domain.
    // This default domain is also automatically assigned to any instances which do have explicit domains.
    streamlet default_domains = (
        a: in stream,
        b: out stream,
        c: in stream,
        d: out stream
    ) {
        impl: {
            explicit_domains_instance = dom_example;

            explicit_domains_instance.a -- a;
            explicit_domains_instance.b -- b;
            explicit_domains_instance.c -- c;
            explicit_domains_instance.d -- d;
        }
    };
}
\end{lstlisting}

\chapter{VHDL Backend Example}\label{app:full_vhdl_example}

\begin{lstlisting}[language=TIL,basicstyle=\ttfamily\scriptsize,caption={The full TIL namespace defined over the course of Section \ref{sec:vhdl_backend}},label={lst:full_til_vhdl}]
namespace my::example::space {
    type stream = Stream (
        data: Bits(8),
        dimensionality: 0,
        synchronicity: Sync,
        complexity: 4,
    );

    #Streamlet documentation#
    streamlet comp1 = (
        a: in stream,
        b: out stream,
        #Port
documentation#
        c: in stream,
        d: out stream,
    );

    streamlet comp2 = comp1 {
        impl: "./vhdl_dir"
    };

    streamlet domains_only = <'a, 'b, 'c>();

    streamlet comp3 = <'x, 'y>(
        q: in stream 'x,
        r: out stream 'x,
    ) {
        impl: {
            dom_ex = domains_only<'x, 'y, 'y>;
            inst = comp2<'x>;
            q -- inst.a;
            r -- inst.b;
            inst.c -- inst.d;
        }
    };
}
\end{lstlisting}

\begin{lstlisting}[language=TIL,basicstyle=\ttfamily\scriptsize,caption={The VHDL architecture output by the VHDL backend for \textit{comp3} of Listing \ref{lst:full_til_vhdl}},label={lst:full_struct_vhdl}]
library ieee;
use ieee.std_logic_1164.all;

library work;
use work.proj.all;

entity my__example__space__comp3_com is
  port (
    x__clk : in std_logic;
    x__rst : in std_logic;
    y__clk : in std_logic;
    y__rst : in std_logic;
    q_valid : in std_logic;
    q_ready : out std_logic;
    q_data : in std_logic_vector(7 downto 0);
    r_valid : out std_logic;
    r_ready : in std_logic;
    r_data : out std_logic_vector(7 downto 0)
  );
end my__example__space__comp3_com;

architecture my__example__space__comp3 of my__example__space__comp3_com is
  signal inst__a_valid : std_logic;
  signal inst__a_ready : std_logic;
  signal inst__a_data : std_logic_vector(7 downto 0);
  signal inst__b_valid : std_logic;
  signal inst__b_ready : std_logic;
  signal inst__b_data : std_logic_vector(7 downto 0);
  signal inst__c_valid : std_logic;
  signal inst__c_ready : std_logic;
  signal inst__c_data : std_logic_vector(7 downto 0);
  signal inst__d_valid : std_logic;
  signal inst__d_ready : std_logic;
  signal inst__d_data : std_logic_vector(7 downto 0);
begin
  dom_ex: my__example__space__domains_only_com port map(
    a__clk => x__clk,
    a__rst => x__rst,
    b__clk => y__clk,
    b__rst => y__rst,
    c__clk => y__clk,
    c__rst => y__rst
  );
  inst: my__example__space__comp2_com port map(
    clk => x__clk,
    rst => x__rst,
    a_valid => inst__a_valid,
    a_ready => inst__a_ready,
    a_data => inst__a_data,
    b_valid => inst__b_valid,
    b_ready => inst__b_ready,
    b_data => inst__b_data,
    c_valid => inst__c_valid,
    c_ready => inst__c_ready,
    c_data => inst__c_data,
    d_valid => inst__d_valid,
    d_ready => inst__d_ready,
    d_data => inst__d_data
  );
  inst__a_valid <= q_valid;
  q_ready <= inst__a_ready;
  inst__a_data <= q_data;
  r_valid <= inst__b_valid;
  inst__b_ready <= r_ready;
  r_data <= inst__b_data;
  inst__c_valid <= inst__d_valid;
  inst__d_ready <= inst__c_ready;
  inst__c_data <= inst__d_data;
end my__example__space__comp3;
\end{lstlisting}

\chapter{AXI4 Specification}\label{app:axi4spec}

The following tables were taken from \cite{armlimitedintroduction2021}.

\begin{table}[h]
    \centering
    \small
\begin{tabular}{ll}
\hline
\multicolumn{1}{c}{\textbf{Write Address (AW) channel signals}}             & \multicolumn{1}{c}{\textbf{AXI version}}                          \\ \hline
AWVALID                                                                     & AXI3 and AXI4                                                     \\ \hline
AWREADY                                                                     & AXI3 and AXI4                                                     \\ \hline
AWADDR{[}31:0{]}                                                            & AXI3 and AXI4                                                     \\ \hline
AWSIZE{[}2:0{]}                                                             & AXI3 and AXI4                                                     \\ \hline
AWBURST{[}1:0{]}                                                            & AXI3 and AXI4                                                     \\ \hline
AWCACHE{[}3:0{]}                                                            & AXI3 and AXI4                                                     \\ \hline
AWPROT{[}2:0{]}                                                             & AXI3 and AXI4                                                     \\ \hline
AWID{[}x:0{]}                                                               & AXI3 and AXI4                                                     \\ \hline
\begin{tabular}[c]{@{}l@{}}AWLEN{[}3:0{]}\\  \\ AWLEN{[}7:0{]}\end{tabular} & \begin{tabular}[c]{@{}l@{}}AXI3 only\\  \\ AXI4 only\end{tabular} \\ \hline
\begin{tabular}[c]{@{}l@{}}AWLOCK{[}1:0{]}\\  \\ AWLOCK\end{tabular}        & \begin{tabular}[c]{@{}l@{}}AXI3 only\\  \\ AXI4 only\end{tabular} \\ \hline
AWQOS{[}3:0{]}                                                              & AXI4 only                                                         \\ \hline
AWREGION{[}3:0{]}                                                           & AXI4 only                                                         \\ \hline
AWUSER{[}x:0{]}                                                             & AXI4 only                                                         \\ \hline
\end{tabular}
    \caption{Write Address}
    \label{tab:axi4aw}
\end{table}

\begin{table}[h]
    \centering
    \small
\begin{tabular}{ll}
\hline
\multicolumn{1}{c}{\textbf{Write Data (W) channel signals}} & \multicolumn{1}{c}{\textbf{AXI version}} \\ \hline
WVALID                                                      & AXI3 and AXI4                            \\ \hline
WREADY                                                      & AXI3 and AXI4                            \\ \hline
WLAST                                                       & AXI3 and AXI4                            \\ \hline
WDATA{[}x:0{]}                                              & AXI3 and AXI4                            \\ \hline
WSTRB{[}x:0{]}                                              & AXI3 and AXI4                            \\ \hline
WID{[}x:0{]} {]}                                            & AXI3 only                                \\ \hline
WUSER{[}x:0{]}                                              & AXI4 only                                \\ \hline
\end{tabular}
    \caption{Write Data}
    \label{tab:axi4w}
\end{table}

\begin{table}[h]
    \centering
    \small
\begin{tabular}{ll}
\hline
\multicolumn{1}{c}{\textbf{Write response (B) channel signals}} & \multicolumn{1}{c}{\textbf{AXI version}} \\ \hline
BWVALID                                                         & AXI3 and AXI4                            \\ \hline
BWREADY                                                         & AXI3 and AXI4                            \\ \hline
BRESP{[}1:0{]}                                                  & AXI3 and AXI4                            \\ \hline
BID{[}x:0{]}                                                    & AXI3 and AXI4                            \\ \hline
BUSER{[}x:0{]}                                                  & AXI4 only                                \\ \hline
\end{tabular}
    \caption{Write response}
    \label{tab:axi4b}
\end{table}

\begin{table}[h]
    \centering
    \small
\begin{tabular}{ll}
\hline
\multicolumn{1}{c}{\textbf{Read Address (AR) channel signals}}              & \multicolumn{1}{c}{\textbf{AXI version}}                          \\ \hline
ARVALID                                                                     & AXI3 and AXI4                                                     \\ \hline
AREADY                                                                      & AXI3 and AXI4                                                     \\ \hline
ARADDR{[}31:0{]}                                                            & AXI3 and AXI4                                                     \\ \hline
ARSIZE{[}2:0{]}                                                             & AXI3 and AXI4                                                     \\ \hline
ARBURST{[}1:0{]}                                                            & AXI3 and AXI4                                                     \\ \hline
ARCACHE{[}3:0{]}                                                            & AXI3 and AXI4                                                     \\ \hline
ARPROT{[}2:0{]}                                                             & AXI3 and AXI4                                                     \\ \hline
ARID{[}x:0{]}                                                               & AXI3 and AXI4                                                     \\ \hline
\begin{tabular}[c]{@{}l@{}}ARLEN{[}3:0{]}\\  \\ ARLEN{[}7:0{]}\end{tabular} & \begin{tabular}[c]{@{}l@{}}AXI3 only\\  \\ AXI4 only\end{tabular} \\ \hline
\begin{tabular}[c]{@{}l@{}}ARLOCK{[}1:0{]}\\  \\ ARLOCK\end{tabular}        & \begin{tabular}[c]{@{}l@{}}AXI3 only\\  \\ AXI4 only\end{tabular} \\ \hline
ARQOS{[}3:0{]}                                                              & AXI4 only                                                         \\ \hline
ARREGION{[}3:0{]}                                                           & AXI4 only                                                         \\ \hline
ARUSER{[}x:0{]}                                                             & AXI4 only                                                         \\ \hline
\end{tabular}
    \caption{Read Address}
    \label{tab:axi4ar}
\end{table}

\begin{table}[h]
    \centering
    \small
\begin{tabular}{ll}
\hline
\multicolumn{1}{c}{\textbf{Read Data (R) channel signals}} & \multicolumn{1}{c}{\textbf{AXI version}} \\ \hline
RVALID                                                     & AXI3 and AXI4                             \\ \hline
RREADY                                                     & AXI3 and AXI4                             \\ \hline
RLAST                                                      & AXI3 and AXI4                            \\ \hline
RDATA{[}x:0{]}                                             & AXI3 and AXI4                            \\ \hline
RRESP{[}1:0{]}                                             & AXI3 and AXI4                            \\ \hline
RID{[}x:0{]}                                               & AXI3 and AXI4                            \\ \hline
RUSER{[}x:0{]}                                             & AXI4 only                                \\ \hline
\end{tabular}
    \caption{Read Data}
    \label{tab:axi4r}
\end{table}

\chapter{AXI4 TIL Definition and VHDL Output}\label{app:axi4til}

\begin{lstlisting}[basicstyle=\ttfamily\scriptsize,caption={The full TIL definition of AXI4-Stream and AXI4},label={lst:full_axi4_til}]
namespace evaluation {
    type axi4stream = Stream (
        data: Union (
            data: Bits(8),
            null: Null, // Equivalent to TSTRB
        ),
        throughput: 128.0, // Data bus width
        dimensionality: 1, // Equivalent to TLAST
        synchronicity: Sync,
        complexity: 7, // Tydi's strobe is equivalent to TKEEP
        user: Group (
            TID: Bits(8),
            TDEST: Bits(4),
            TUSER: Bits(1),
        ),
    );

    type axi4_address = Stream (
        data: Group (
            ADDR: Bits(32),
            SIZE: Bits(3),
            BURST: Bits(2),
            CACHE: Bits(4),
            PROT: Bits(3),
            ID: Bits(4),
            LEN: Bits(8),
            LOCK: Bits(1),
            QOS: Bits(4),
            REGION: Bits(4),
        ),
        dimensionality: 0,
        synchronicity: Sync,
        complexity: 1,
        user: Bits(4),
    );
    type axi4_write_data = Stream (
        data: Bits(8),
        throughput: 256.0, // Max transfers
        dimensionality: 1, // Equivalent to LAST
        synchronicity: Sync,
        complexity: 7, // Adds a strobe
        user: Bits(4),
    );
    type axi4_read_data = Stream (
        data: Bits(8),
        throughput: 256.0, // Max transfers
        dimensionality: 1, // Equivalent to LAST
        synchronicity: Sync,
        complexity: 7, // Adds a strobe
        user: Group (
            RESP: Bits(2),
            ID: Bits(4),
            USER: Bits(4),
        ),
    );
    type axi4_response = Stream (
        data: Group (
            RESP: Bits(2),
            ID: Bits(4),
        ),
        dimensionality: 0,
        synchronicity: Sync,
        complexity: 1,
        user: Bits(4),
    );

    type axi4 = Stream (
        data: Group (
            AW: axi4_address,
            W: Stream (
                data: Bits(8),
                throughput: 256.0, // Max transfers
                dimensionality: 1, // Equivalent to LAST
                synchronicity: Sync,
                complexity: 7, // Adds a strobe
                user: Bits(4),
            ),
            B: Stream (
                direction: Reverse,
                data: Group (
                    RESP: Bits(2),
                    ID: Bits(4),
                ),
                dimensionality: 0,
                synchronicity: Sync,
                complexity: 1,
                user: Bits(4),
            ),
            AR: axi4_address,
            R: Stream (
                direction: Reverse,
                data: Bits(8),
                throughput: 256.0, // Max transfers
                dimensionality: 1, // Equivalent to LAST
                synchronicity: Sync,
                complexity: 7, // Adds a strobe
                user: Group (
                    RESP: Bits(2),
                    ID: Bits(4),
                    USER: Bits(4),
                ),
            ),
        ),
        dimensionality: 0,
        synchronicity: Sync,
        complexity: 1,
    );

    streamlet example = (
        axi4stream: in axi4stream,
        axi4_aw: out axi4_address,
        axi4_w: out axi4_write_data,
        axi4_b: in axi4_response,
        axi4_ar: out axi4_address,
        axi4_r: in axi4_read_data,
        axi4: out axi4,
    );
}
\end{lstlisting}

\begin{lstlisting}[language=VHDL,basicstyle=\ttfamily\scriptsize,caption={The full VHDL component using the AXI4-Stream and AXI4 equivalent Tydi Streams},label={lst:full_axi4_vhdl}]
  component evaluation__example_com
    port (
      clk : in std_logic;
      rst : in std_logic;
      axi4stream_valid : in std_logic;
      axi4stream_ready : out std_logic;
      axi4stream_data : in std_logic_vector(1151 downto 0);
      axi4stream_last : in std_logic;
      axi4stream_stai : in std_logic_vector(6 downto 0);
      axi4stream_endi : in std_logic_vector(6 downto 0);
      axi4stream_strb : in std_logic_vector(127 downto 0);
      axi4stream_user : in std_logic_vector(12 downto 0);
      axi4_aw_valid : out std_logic;
      axi4_aw_ready : in std_logic;
      axi4_aw_data : out std_logic_vector(64 downto 0);
      axi4_aw_user : out std_logic_vector(3 downto 0);
      axi4_w_valid : out std_logic;
      axi4_w_ready : in std_logic;
      axi4_w_data : out std_logic_vector(2047 downto 0);
      axi4_w_last : out std_logic;
      axi4_w_stai : out std_logic_vector(7 downto 0);
      axi4_w_endi : out std_logic_vector(7 downto 0);
      axi4_w_strb : out std_logic_vector(255 downto 0);
      axi4_w_user : out std_logic_vector(3 downto 0);
      axi4_b_valid : in std_logic;
      axi4_b_ready : out std_logic;
      axi4_b_data : in std_logic_vector(5 downto 0);
      axi4_b_user : in std_logic_vector(3 downto 0);
      axi4_ar_valid : out std_logic;
      axi4_ar_ready : in std_logic;
      axi4_ar_data : out std_logic_vector(64 downto 0);
      axi4_ar_user : out std_logic_vector(3 downto 0);
      axi4_r_valid : in std_logic;
      axi4_r_ready : out std_logic;
      axi4_r_data : in std_logic_vector(2047 downto 0);
      axi4_r_last : in std_logic;
      axi4_r_stai : in std_logic_vector(7 downto 0);
      axi4_r_endi : in std_logic_vector(7 downto 0);
      axi4_r_strb : in std_logic_vector(255 downto 0);
      axi4_r_user : in std_logic_vector(9 downto 0);
      axi4__AW_valid : out std_logic;
      axi4__AW_ready : in std_logic;
      axi4__AW_data : out std_logic_vector(64 downto 0);
      axi4__AW_user : out std_logic_vector(3 downto 0);
      axi4__W_valid : out std_logic;
      axi4__W_ready : in std_logic;
      axi4__W_data : out std_logic_vector(2047 downto 0);
      axi4__W_last : out std_logic;
      axi4__W_stai : out std_logic_vector(7 downto 0);
      axi4__W_endi : out std_logic_vector(7 downto 0);
      axi4__W_strb : out std_logic_vector(255 downto 0);
      axi4__W_user : out std_logic_vector(3 downto 0);
      axi4__B_valid : in std_logic;
      axi4__B_ready : out std_logic;
      axi4__B_data : in std_logic_vector(5 downto 0);
      axi4__B_user : in std_logic_vector(3 downto 0);
      axi4__AR_valid : out std_logic;
      axi4__AR_ready : in std_logic;
      axi4__AR_data : out std_logic_vector(64 downto 0);
      axi4__AR_user : out std_logic_vector(3 downto 0);
      axi4__R_valid : in std_logic;
      axi4__R_ready : out std_logic;
      axi4__R_data : in std_logic_vector(2047 downto 0);
      axi4__R_last : in std_logic;
      axi4__R_stai : in std_logic_vector(7 downto 0);
      axi4__R_endi : in std_logic_vector(7 downto 0);
      axi4__R_strb : in std_logic_vector(255 downto 0);
      axi4__R_user : in std_logic_vector(9 downto 0)
    );
  end component;
\end{lstlisting}

\printbibliography

\end{document}